\documentclass{elsart}

\usepackage[square,comma]{natbib}
\usepackage{graphicx}
\usepackage{pxfonts}
\usepackage{amssymb}

\begin{document}

\begin{frontmatter}



\title{Transverse coherence properties of X-ray beams in third-generation synchrotron radiation sources}


\author{Gianluca Geloni}
\author{Evgeni Saldin}
\author{Evgeni Schneidmiller}
\author{and Mikhail Yurkov}

\address{Deutsches Elektronen-Synchrotron (DESY), Hamburg,
Germany}

\begin{abstract}
This article describes a complete theory of spatial coherence for
undulator radiation sources. Current estimations of coherence
properties often assume that undulator sources are
quasi-homogeneous, like thermal sources, and rely on the
application of the van Cittert-Zernike theorem for calculating the
degree of transverse coherence. Such assumption is not adequate
when treating third generation light sources, because the vertical
(geometrical) emittance of the electron beam is comparable or even
much smaller than the radiation wavelength in a very wide spectral
interval that spans over four orders of magnitude (from $0.1
\mathrm{\AA}$ up to $10^3 \mathrm{\AA}$). Sometimes, the so-called
Gaussian-Schell model, that is widely used in statistical optics
in the description of partially-coherent sources, is applied as an
alternative to the quasi-homogeneous model. However, as we will
demonstrate, this model fails to properly describe coherent
properties of X-ray beams from non-homogeneous undulator sources.
As a result, a more rigorous analysis is required. We propose a
technique, based on statistical optics and Fourier optics, to
explicitly calculate the cross-spectral density of an undulator
source in the most general case, at any position after the
undulator. Our theory, that makes consistent use of dimensionless
analysis, allows relatively easy treatment and physical
understanding of many asymptotes of the parameter space, together
with their region of applicability. Particular emphasis is given
to the asymptotic situation when the horizontal emittance is much
larger than the radiation wavelength, and the vertical emittance
is arbitrary. This case is practically relevant for third
generation synchrotron radiation sources.
\end{abstract}


\end{frontmatter}

% main text

\clearpage
\section{\label{sec:intro} Introduction}

In recent years, continuous evolution of synchrotron radiation
(SR) sources has resulted in a dramatic increase of brilliance
with respect to older designs. Among the most exciting properties
of third generation facilities of today is a high flux of coherent
X-rays \cite{EDGA}. The availability of intense coherent X-ray
beams has triggered the development of a number of new
experimental techniques based on coherence properties of light
such as scattering of coherent X-ray radiation, X-ray photon
correlation spectroscopy and phase contrast imaging. The
interested reader may find an extensive reference list to these
applications in \cite{PETR}.

Characterization of transverse coherence properties of SR at a
specimen position is fundamental for properly planning, conducting
and analyzing experiments involving above-mentioned techniques.
These objectives can only be met once coherence properties of
undulator sources are characterized and then propagated along the
photon beamline up to the specimen.

This paper is mainly dedicated to a description of transverse
coherence properties of SR sources. Therefore, it constitutes the
first step for tracking coherence properties through optical
elements, which can be only done when the source is quantitatively
described. Solution of the problem of evolution of radiation
properties in free-space is also given. This can be used to
characterize radiation at the specimen position when there are no
optical elements between the source and the specimen.

Since SR is a random process, the description of transverse
coherence properties of the source and its evolution should be
treated in terms of probabilistic statements. Statistical optics
\cite{GOOD, MAND, NEIL} presents most convenient tools to deal
with fluctuating electromagnetic fields. However, it was mainly
developed in connection with polarized thermal light that is
characterized by quite specific properties. Besides obeying
Gaussian statistics, polarized thermal light has two other
specific characteristics allowing for major simplifications of the
theory: stationarity and homogeneity (i.e. perfect incoherence) of
the source \cite{GOOD, MAND}.

As we will see, SR obeys Gaussian statistics, but stationarity and
homogeneity do not belong to SR fields. Thus, although the
language of statistical optics must be used to describe SR
sources, one should avoid a-priori introduction of an incorrect
model. In contrast to this, up to now it has been a widespread
practice to assume that undulator sources are perfectly incoherent
(homogeneous) and to draw conclusions about transverse coherence
properties of undulator light based on these assumptions
\cite{PETR}.

In particular, in the case of thermal light, transverse coherence
properties of the radiation can be found with the help of the
well-known van Citter-Zernike (VCZ) theorem. However, for third
generation light sources either planned or in operation, the
horizontal electron beam (geometrical)
emittance\footnote{Emittances are measured in m $\cdot$ rad.
However, radians are a dimensionless unit. Therefore, we can
present emittances measured in meters. This is particularly useful
in the present paper, since we need to compare emittances with
radiation wavelength.} is of order of $1 \div 3$ nm while the
vertical emittance is bound to the horizontal through a coupling
factor $\zeta \sim 10^{-2}$, corresponding to vertical emittances
of order $0.1 \div 0.3~ \mathrm{\AA}$. As we will show, these
facts imply that the VCZ theorem cannot be applied in the vertical
direction, aside for the hard X-ray limit at wavelengths shorter
than $0.1 ~\mathrm{\AA}$. Similar remarks hold for future sources,
like Energy Recovery Linac (ERL)-based spontaneous radiators. ERL
technology is expected to constitute the natural evolution of
today SR sources and to provide nearly fully diffraction-limited
sources in the $1 \AA$-range, capable of three order of magnitudes
larger coherent flux compared to third generation light sources
\cite{EDGA}. Horizontal and vertical emittance will be of order of
$0.3 \AA$, which rules out the applicability of the VCZ theorem.

The need for characterization of partially coherent undulator
sources is emphasized very clearly in reference \cite{HOWE}.
However, studying spatial-coherence properties of undulator
sources is not a trivial task. Difficulties arise when one tries
to simultaneously include the effect of intrinsic divergence of
undulator radiation, and of electron beam size and divergence.

An attempt to find the region of applicability of the VCZ theorem
for third generation light sources is reported in \cite{TAKA}.
Authors of \cite{TAKA} conclude that the VCZ theorem can only be
applied when SR sources are close to the diffraction limit. We
will see that the VCZ theorem can only be applied in the opposite
case.

A model to describe partially-coherent sources, the
Gaussian-Schell model, is widely used in statistical optics. Its
application to SR is described in \cite{COI1,COI2,COI3}. The
Gaussian-Schell model includes non-homogeneous sources, because
the typical transverse dimension of the source can be comparable
or smaller than the transverse coherence length. While
\cite{COI1,COI2,COI3} are of general theoretical interest, they do
not provide a satisfactory approximation of third generation SR
sources. In fact, as we will see, undulator radiation has very
specific properties that cannot be described in terms of
Gaussian-Schell model.

The first treatment of transverse coherence properties from SR
sources accounting for the specific nature of undulator radiation
and anticipating operation of third generation SR facilities is
given, in terms of Wigner distribution, in \cite{KIM2,KIM3}.
References \cite{KIM2,KIM3} present the most general algorithm for
calculating properties of undulator radiation sources.

In the present paper we base our theory on the characterization of
the cross-spectral density of the system. The cross-spectral
density is merely the statistical correlation function of the
radiation field at two different positions on the observation
plane at a given frequency. It is equivalent to the Wigner
distribution, the two quantities being related by Fourier
transformation. Based on cross-spectral density we developed a
comprehensive theory of third generation SR sources in the
space-frequency domain, where we exploited the presence of small
or large parameters intrinsic in the description of the system.
First, we took advantage of the particular but important situation
of perfect resonance, when the field from the undulator source can
be presented in terms of analytical functions. Second, we
exploited the practical case of a Gaussian electron beam, allowing
further analytical simplifications. Third, we took advantage of
the large horizontal emittance (compared with the radiation
wavelength) of the electron beam, allowing separate treatments of
horizontal and vertical directions. Finally, we studied asymptotic
cases of our theory for third generation light sources with their
region of applicability. In particular, we considered both
asymptotes of a small and a large vertical emittance compared with
the radiation wavelength, finding surprising results. In the case
of a small vertical emittance (smaller than the radiation
wavelength) we found that the radiation is not diffraction limited
in the vertical direction, an effect that can be ascribed to the
influence of the large horizontal emittance on the vertical
coherence properties of radiation. In the case of a large vertical
emittance we described the quasi-homogeneous source in terms of a
non-Gaussian model, which has higher accuracy and wider region of
applicability with respect to a simpler Gaussian model. In fact,
the non-Gaussian model proposed in our work accounts for
diffraction effects. Nonetheless, in order to solve the
propagation problem in free-space, a geometrical optics approach
can still be used.

We organize our work as follows. First, a general algorithm for
the calculation of the field correlation function of an undulator
source is given in Section \ref{sec:due}. Such algorithm is not
limited to the description of third generation light sources, but
can also be applied, for example, to ERL sources. In Section
\ref{sec:main} we specialize our treatment to third generation
light sources exploiting a large horizontal emittance compared to
the radiation wavelength, and studying the asymptotic limit for a
small vertical emittance of the electron beam. In the following
Section \ref{sec:quas} we consider the particular case of
quasi-homogeneous undulator sources, and discuss the applicability
range of the van VCZ theorem. Finally, in Section \ref{sec:conc}
we come to conclusions.

\section{\label{sec:due} Cross-spectral density of an undulator source}

\subsection{\label{sub:def} Thermal light and Synchrotron Radiation: some concepts and definitions}

The great majority of optical sources emits thermal light. Such is
the case of the sun and the other stars, as well as of
incandescent lamps. This kind of radiation consists of a large
number of independent contributions (radiating atoms) and is
characterized by random amplitudes and phases in space and time.
Its electromagnetic field can be conveniently described in terms
of statistical optics that has been intensively developed during
the last few decades \cite{GOOD, MAND}.

\begin{figure}
\begin{center}
\includegraphics*[width=140mm]{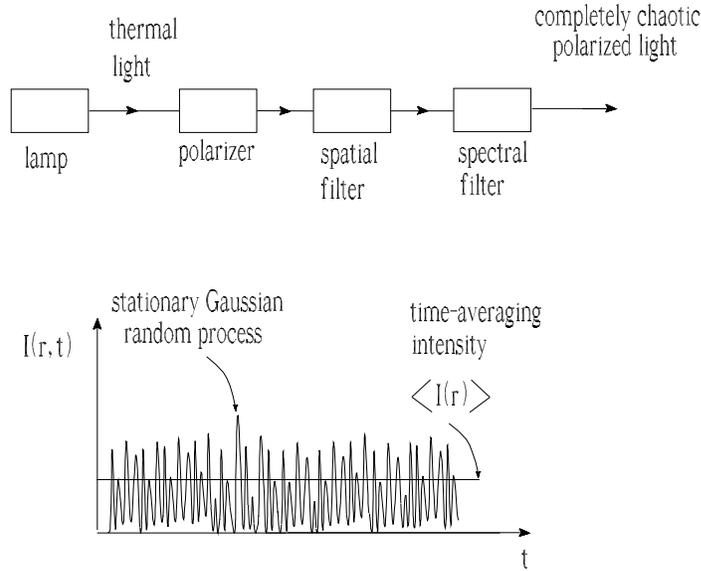}% Here is how to import EPS art
\caption{\label{thermal} Light intensity from an incandescent lamp
driven by a constant electric current. A statistically stationary
wave has an average that does not vary with time.}
\end{center}
\end{figure}
Consider the light emitted by a thermal source passing through a
polarization analyzer, as in Fig. \ref{thermal}. Properties of
polarized thermal light are well-known in statistical optics, and
are referred to as properties of completely chaotic, polarized
light \cite{GOOD, MAND}. Thermal light is a statistical random
process and statements about such process are probabilistic
statements. Statistical processes are handled using the concept of
statistical ensemble, drawn from statistical mechanics, and
statistical averages are performed over an ensemble, or many
realizations, or outcomes of the statistical process under study.
Polarized thermal sources obey a very particular kind of random
process in that it is Gaussian and stationary\footnote{Ergodic
too. From a qualitative viewpoint, a given random process is
ergodic when all ensemble averages can be substituted by time
averages. Ergodicity is a stronger requirement than stationarity
\cite{GOOD,MAND}.}. Moreover, they are homogeneous.

The properties of Gaussian random processes are well-known in
statistical optics. For instance, the real and imaginary part of
the complex amplitudes of the electric field from a polarized
thermal source have Gaussian distribution, while the instantaneous
radiation power fluctuates in accordance with the negative
exponential distribution. It can be shown \cite{GOOD} that a
linearly filtered Gaussian process is also a Gaussian random
process. As a result, the presence of a monochromator and a
spatial filter as in the system depicted in Fig. \ref{thermal} do
not change the statistics of the signal. Finally, higher order
field correlation functions can be found in terms of the second
order correlation. This dramatically simplifies the description of
the random process.

Stationarity is a subtle concept. There are different kinds of
stationarity. However, for Gaussian processes different kinds of
stationarity coincide \cite{GOOD,MAND}. In this case, stationarity
means that all ensemble averages are independent of time. It
follows that a necessary condition for a certain process to be
stationary is that the signal last forever. Yet, if a signal lasts
much longer than the short-scale duration of the field
fluctuations (its coherence time $\tau_c$)  and it is observed for
a time much shorter than its duration $\sigma_T$, but much longer
than $\tau_c$ it can be reasonably considered as everlasting and
it has a chance to be stationary as well, as in the case of
thermal light.

Finally, thermal sources are homogeneous. The field is correlated
on the transverse direction over the possible minimal distance,
which is of order of a wavelength. This means that the radiation
intensity at the source remains practically unvaried on the scale
of a correlation length. One can say, equivalently, that the
source is homogeneous. Homogeneity is the equivalent, in the
transverse direction, of stationarity and implies a constant
ensemble-averaged intensity along the transverse direction.

\begin{figure}
\begin{center}
\includegraphics*[width=140mm]{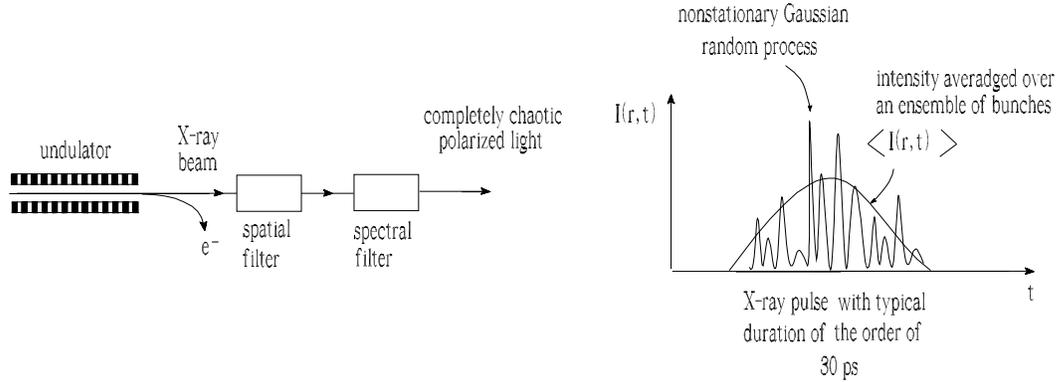}% Here is how to import EPS art
\caption{\label{SR} The intensity of an X-ray beam from a SR
source. A statistically non-stationary wave has a time-varying
intensity averaged over an ensemble of bunches.}
\end{center}
\end{figure}

Consider now a SR source, as depicted\footnote{Radiation at the
detector consists of a carrier modulation of frequency $\omega$
subjected to random amplitude and phase modulation. The Fourier
decomposition of the radiation contains frequencies spread about
the monochromator bandwidth: it is not possible, in practice, to
resolve the oscillations of the radiation fields which occur at
the frequency of the carrier modulation. Therefore, for comparison
with experimental results, we average the theoretical results over
a cycle of oscillations of the carrier modulation.} in Fig.
\ref{SR}. Like thermal light, also SR is a random process. In
fact, relativistic electrons in a storage ring emit SR passing
through bending magnets or undulators. The electron beam shot
noise causes fluctuations of the beam density which are random in
time and space from bunch to bunch. As a result, the radiation
produced has random amplitudes and phases. Moreover, in Section
\ref{sub:seco} we will demonstrate that SR fields obey Gaussian
statistics. Statistical properties satisfied by single-electron
contributions (elementary phasors) to the total SR field are
weaker than those satisfied by single-atom contributions in
thermal sources. Thus, the demonstration that thermal light obeys
Gaussian statistics cannot be directly applied to the SR case, and
some condition should be formulated on the parameter space to
define the region where SR is indeed a Gaussian random process. We
will show this fact with the help of Appendix A of \cite{OURU}.

In contrast with thermal light, SR is intrinsically
non-stationary, because it presents a time-varying
ensemble-averaged intensity on the temporal scale of the duration
of the X-ray pulse generated by a single electron bunch. For this
reason, in what follows the averaging brackets $\langle ...
\rangle$ will always indicate the ensemble average over bunches
(and not a time average).
Finally, SR sources are not completely incoherent, or homogeneous.
In fact, there is a close connection between the state of
coherence of the source and the angular distribution of the
radiant intensity. A thermal source that is correlated over the
minimal possible distance (which is of order of the wavelength) is
characterized by a radiant intensity distributed over a solid
angle of order $2 \pi$. This is not the case of SR light, that is
confined within a narrow cone in the forward direction. The high
directionality of SR rules out the possibility of description in
terms of thermal light. As we will see, depending on the
situation, SR may or may not be described by a quasi-homogeneous
model, where sources are only locally coherent over a distance of
many wavelengths but the linear dimension of the source is much
larger than the correlation distance\footnote{Note that the high
directionality of SR is not in contrast with the poor coherence
which characterizes the quasi-homogeneous limit.}.

In spite of differences with respect to the simpler case of
thermal light, SR fields can be described in terms of statistical
optics. However, statistical optics was developed in relation with
thermal light emission. Major assumptions typical of this kind of
radiation like, for example, stationarity, are retained in
textbooks \cite{GOOD,MAND}. Thus, usual statistical optics
treatment must be modified in order to deal with SR problems. We
will reduce the problem of characterization of transverse
coherence properties of undulator sources in the space-frequency
domain to the calculation of the correlation of the field produced
by a single electron with itself. This correlation is known as
cross-spectral density. The non-stationarity of the process
imposes some (practically non restrictive) condition on the
parameter space region where a treatment based on cross-spectral
density can be applied to SR phenomena. Such condition involves
the length of the electron bunch, the number of undulator periods
and the radiation wavelength. Given the fact that the electron
bunch length can vary from about $10$ ps for third generation SR
sources to about $100$ fs  for ERL sources, understanding of the
region of applicability this condition is a-priori not obvious.
All these subjects will be treated in the next Section
\ref{sub:seco}.

\subsection{\label{sub:seco} Second-order correlations in
space-frequency domain}

In SR experiments with third generation light sources, detectors
are limited to about $100$ ps time resolution. Therefore, they
cannot resolve a single X-ray pulse in time domain, whose duration
is about $30$ ps. They work, instead, by counting the number of
photons at a certain frequency over an integration time longer
than the pulse. It is therefore quite natural to consider signals
in the frequency domain. With this in mind, let ${\bar{E}}_b(z,
\vec{r}_{{}}, \omega)$ be a fixed polarization component of the
Fourier transform of the electric field at location $(z,
\vec{r}_{{}})$, in some cartesian coordinate system, and frequency
$\omega$ by a given collection of electromagnetic sources. We will
often name it, slightly improperly, "the field". Subscript "b"
indicates that the field is generated by the entire bunch.
${\bar{E}}_b(z, \vec{r}_{{}}, \omega)$ is linked to the time
domain field ${{E}}_b(z, \vec{r}_{{}}, t)$ through the Fourier
transform

\begin{equation}
\bar{{E}}_b(\omega) = \int_{-\infty}^{\infty} dt {{E}}_b(t) \exp(i
\omega t)~, ~~~~~{{E}}_b(t) = \frac{1}{2\pi}
\int_{-\infty}^{\infty} d\omega \bar{{E}}_b(\omega) \exp(-i \omega
t)~. \label{ftran}
\end{equation}
We will be interested in the case of an ultra relativistic
electron beam going through a certain magnetic system, an
undulator in particular. In this case $z$ is the observation
distance along the optical axis of the undulator and
$\vec{r}_{{}}$ fixes the transverse position of the observer. The
contribution of the $k$-th electron to the field depends on the
transverse offset $\vec{l}_{k}$ and deflection angles
$\vec{\eta}_{k}$ that the electron has at some reference point on
the optical axis $z$, e.g. the center of the undulator. Moreover,
the arrival time $t_k$ at the center of the undulator has the
effect of multiplying the field by a phase factor $\exp{(i\omega
t_k)}$, i.e. the time-domain electric field is retarded by a time
$t_k$. At this point we do not need to explicitly specify the
dependence on offset and deflection. The total field can be
written as

\begin{equation}
{\bar{E}}_b\left(z,\vec{r}_{{}},\omega\right)=\sum_{k=1}^{N_e}
\bar{E}\left(\vec{\eta}_k,\vec{l}_k,z,\vec{r}_{{}},\omega\right)
\exp{(i\omega t_k)} ~, \label{total}
\end{equation}
where $\vec{\eta}_k,\vec{l}_k$ and $t_k$ are random variables and
$N_e$ is the number of electrons in the bunch. Note that $\bar{E}$
in Eq. (\ref{total}) is a complex quantity, that can be written as
$\bar{E} = \mid\bar{E}_k\mid \exp(i \phi_k)$. It follows that the
SR field pulse at fixed frequency and position is a sum of a many
phasors, one for each electron, of the form $\bar{E} \exp{(i\omega
t_k)} = \mid\bar{E}_{k}\mid \exp(i \phi_k + i \omega t_k)$.

Elementary phasors composing the sum obey three statistical
properties, that are satisfied in SR problems of interest. First,
random variables $t_k$ are statistically independent of each
other, and of variables $\vec{\eta}_k$ and $\vec{l}_k$. Second,
random variables $\mid\bar{E}_{k}\mid$ (at fixed frequency
$\omega$), are identically distributed for all values of $k$, with
a finite mean $\langle \mid\bar{E}_{k}\mid \rangle$ and a finite
second moment $\langle \mid\bar{E}_{k}\mid^2 \rangle $. These two
assumptions follows from the properties of shot noise in a storage
ring, which is a fundamental effect related with quantum
fluctuations. Third, we assume that the electron bunch duration
$\sigma_T$ is large enough so that $\omega \sigma_T \gg 1$: under
this assumption, phases $\omega t_k$ can be regarded as uniformly
distributed on the interval $(0, 2\pi)$. The assumption $\omega
\sigma_T \gg 1$ exploits the first large parameter of our theory,
and is justified by the fact that $\omega$ is the undulator
resonant frequency, which is high enough, in practical cases of
interest, to guarantee that $\omega \sigma_T \gg 1$ for any
realistic choice of $\sigma_T$. Based on the three previously
discussed properties, and with the help of the central limit
theorem, it can be demonstrated\footnote{The proof follows from a
slight generalization of Section 2.9 in \cite{GOOD}. Namely, it
can be shown by direct calculation that real and imaginary part of
the total phasor ${\bar{E}}_b$ are uncorrelated, with zero mean
and equal variance. Then, using the central limit theorem, we
conclude that ${\bar{E}}_b$ is a circular complex Gaussian random
variable (at fixed $z$, $\vec{r}_{{}}$ and $\omega$).} that the
real and the imaginary part of $\bar{E}_b$ are distributed in
accordance to a Gaussian law.

As a result, SR is a non-stationary Gaussian random process.

Because of this, higher-order correlation functions can be
expressed in terms of second-order correlation functions with the
help of the moment theorem \cite{GOOD}. As a result, the knowledge
of the second-order field correlation function in frequency
domain, $\Gamma_\omega$, is all we need to completely characterize
the signal from a statistical viewpoint. The following definition
holds:

\begin{equation}
\Gamma_{\omega}\left(z,\vec{r}_{{} 1},\vec{r}_{{}
2},\omega_1,\omega_2\right) = \left<
{\bar{E}}_b\left(z,\vec{r}_{{}
1},\omega_1\right){\bar{E}}^*_b\left(z,\vec{r}_{{}
2},\omega_2\right) \right>~, \label{gamma}
\end{equation}
where brackets $<...>$ indicate ensemble average over electron
bunches. For any given function $w\left(\vec{\eta}_k, \vec{l}_k,
t_k\right)$, the ensemble average is defined as

\begin{eqnarray}
\left<w\left(\vec{\eta}_k, \vec{l}_k, t_k\right)\right> = \int
d\vec{\eta}_k \int d\vec{l}_k \int_{-\infty}^{\infty} d t_k
w\left(\vec{\eta}_k, \vec{l}_k, t_k\right) P\left(\vec{\eta}_k,
\vec{l}_k, t_k\right) \label{ensembledef}~,
\end{eqnarray}
where integrals in $d\vec{l}_k$ and $d\vec{\eta}_k$ span over all
offsets and deflections, and $P=P(\vec{\eta}_k, \vec{l}_k, t_k)$
indicates the probability density distribution in the joint random
variables $\vec{\eta}_k$, $\vec{l}_k$, and $t_k$. Note that, since
all electrons have the same probability of arrival around a given
offset, deflection, and time, $P$ is independent of $k$. Moreover,
already discussed independence of $t_k$ from $\vec{l}_k$ and
$\vec{\eta}_k$ allows to write $P$ as

\begin{eqnarray}
P\left(\vec{\eta}_k, \vec{l}_k, t_k\right) =
f_\bot\left(\vec{l}_k,\vec{\eta}_k\right) f(t_k)~.
\label{independence}
\end{eqnarray}
Here $f$ is the longitudinal bunch profile of the electron beam,
while $f_\bot$ is the transverse phase-space distribution.

Substituting Eq. (\ref{total}) in Eq. (\ref{gamma}) one has

\begin{eqnarray}
\Gamma_{\omega} = \left<\sum_{m,n=1}^{N_e} \bar{E}
\left(\vec{\eta}_m ,\vec{l}_m , z ,\vec{r}_{{} 1}, \omega_1\right)
\bar{E}^* \left(\vec{\eta}_n,\vec{l}_n, z,\vec{r}_{{} 2},
\omega_2\right) \exp{[i(\omega_1 t_m- \omega_2 t_n)]} \right> .\cr
&& \label{gamma2}
\end{eqnarray}
Expansion of Eq. (\ref{gamma2}) gives

\begin{eqnarray}
&&\Gamma_{\omega} = \sum_{m=1}^{N_e} \left\langle \bar{E}
\left(\vec{\eta}_m,\vec{l}_m,z,\vec{r}_{{} 1}, \omega_1\right)
\bar{E}^*\left(\vec{\eta}_m,\vec{l}_m,z,\vec{r}_{{} 2},
\omega_2\right) \exp{[i( \omega_1- \omega_2) t_m]} \right\rangle
\cr && + \sum_{m\ne n} \left\langle
\bar{E}\left(\vec{\eta}_m,\vec{l}_m,z,\vec{r}_{{} 1},
\omega_1\right) \exp{(i \omega_1
t_m)}\right\rangle\left\langle\bar{E}^*
\left(\vec{\eta}_n,\vec{l}_n,z,\vec{r}_{{} 2}, \omega_2\right)
\exp{(- i \omega_2 t_n)} \right\rangle.\cr && \label{gamma3}
\end{eqnarray}
With the help of Eq. (\ref{ensembledef}) and Eq.
(\ref{independence}), the ensemble average $\langle \exp{(i \omega
t_k)} \rangle$ can be written as the Fourier transform of the
longitudinal bunch profile function $f$, that is

\begin{equation}
\left\langle \exp{(i \omega t_k)} \right\rangle =
\int_{-\infty}^{\infty} d t_k f(t_k) \exp(i\omega t_k) \equiv
\bar{f}(\omega)~. \label{FTlong}
\end{equation}
Using Eq. (\ref{FTlong}), Eq. (\ref{gamma3}) can be written as

\begin{eqnarray}
\Gamma_{\omega} &&= \sum_{m=1}^{N_e} \bar{f}( \omega_1- \omega_2)
\left \langle \bar{E} \left(\vec{\eta}_m,\vec{l}_m, z, \vec{r}_{{}
1}, \omega_1\right) \bar{E}^*
\left(\vec{\eta}_m,\vec{l}_m,z,\vec{r}_{{} 2}, \omega_2\right)
\right\rangle \cr && + \sum_{m\ne n} \bar{f}( \omega_1)\bar{f}(-
\omega_2)  \left\langle
\bar{E}\left(\vec{\eta}_m,\vec{l}_m,z,\vec{r}_{{} 1},
\omega_1\right) \right\rangle
\left\langle\bar{E}^*\left(\vec{\eta}_n, \vec{l}_n,z,\vec{r}_{{}
2}, \omega_2\right) \right\rangle ~, \label{gamma4}
\end{eqnarray}
where $\bar{f}^*(\omega_2)=\bar{f}(- \omega_2)$ because ${f}$ is a
real function. When the radiation wavelengths of interest are much
shorter than the bunch length we can safely neglect the second
term on the right hand side of Eq. (\ref{gamma4}) since the form
factor product $\bar{f}( \omega_1) \bar{f}(- \omega_2)$ goes
rapidly to zero for frequencies larger than the characteristic
frequency associated with the bunch length: think for instance, at
a centimeter long bunch compared with radiation in the Angstrom
wavelength range\footnote{When the radiation wavelength of
interested is comparable or longer than the bunch length, the
second term in Eq. (\ref{gamma4}) is dominant with respect to the
first, because it scales with the number of particles
\textit{squared}: in this case, analysis of the second term leads
to a treatment of coherent synchrotron radiation phenomena (CSR).
In this paper we will not be concerned with CSR and we will
neglect the second term in Eq. (\ref{gamma4}), assuming that the
radiation wavelength of interest is shorter than the bunch length.
Also note that $\bar{f}( \omega_1- \omega_2)$ depends on the
\textit{difference} between $ \omega_1$ and $ \omega_2$, and the
first term cannot be neglected.}. Therefore we write

\begin{eqnarray}
\Gamma_{\omega} &&= \sum_{m=1}^{N_e} \bar{f}( \omega_1- \omega_2)
\left \langle \bar{E} \left(\vec{\eta}_m,\vec{l}_m, z, \vec{r}_{{}
1}, \omega_1\right) \bar{E}^*
\left(\vec{\eta}_m,\vec{l}_m,z,\vec{r}_{{} 2}, \omega_2\right)
\right\rangle \cr && = N_e \bar{f}( \omega_1- \omega_2) \left
\langle \bar{E}\left(\vec{\eta},\vec{l}, z, \vec{r}_{{} 1},
\omega_1\right) \bar{E}^* \left(\vec{\eta},\vec{l},z,\vec{r}_{{}
2}, \omega_2\right) \right\rangle  ~. \label{gamma5}
\end{eqnarray}
As one can see from Eq. (\ref{gamma5}) each electron is correlated
just with itself: cross-correlation terms between different
electrons was included in the second term on the right hand side
of Eq. (\ref{gamma4}), that has been dropped.

If the dependence of $\bar{E}$ on $ \omega$ is slow enough,
$\bar{E}$ does not vary appreciably on the characteristic scale of
$\bar{f}$. Thus, we can substitute
$\bar{E}^*(\vec{\eta},\vec{l},z,\vec{r}_{{} 2}, \omega_2)$ with
$\bar{E}^*(\vec{\eta},\vec{l},z,\vec{r}_{{} 2}, \omega_1)$ in Eq.
(\ref{gamma5}).
\begin{figure}
\begin{center}
\includegraphics*[width=110mm]{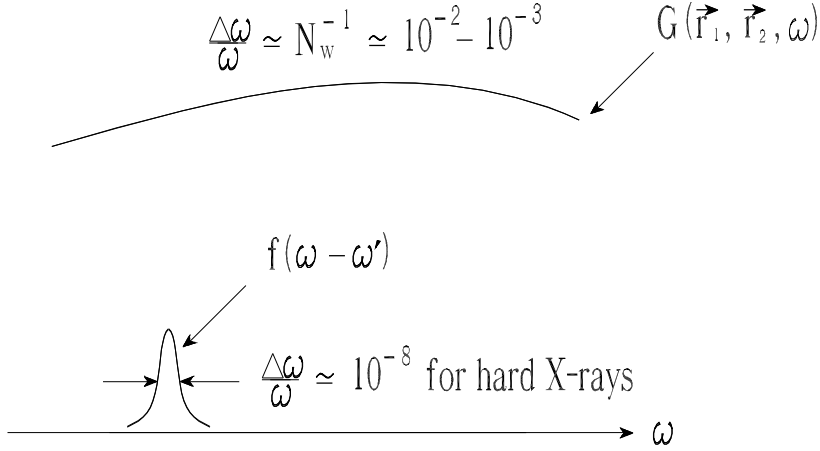}% Here is how to import EPS art
\caption{\label{rela} Schematic illustration of the relative
frequency dependence of the spectral correlation function
$\bar{f}(\omega-\omega')$ and of the cross-spectral density
function (the cross-power spectrum) $G(z,\vec{r}_{{}
1},\vec{r}_{{} 2}, \omega)$ of SR at points $\vec{r}_{{} 1}$ and
$\vec{r}_{{} 2}$ at frequency $\omega$.}
\end{center}
\end{figure}
The situation is depicted in Fig. \ref{rela}. On the one hand, the
characteristic scale of $\bar{f}$ is given by $1/\sigma_T$, where
$\sigma_T$ is the characteristic bunch duration. On the other
hand, the minimal possible bandwidth of undulator radiation is
achieved on axis and in the case of a filament beam. It is peaked
around the resonant frequency ${\omega_r} = {2 \gamma_z^2
c}/{\lambdabar_w}$ ($\lambda_w$ being the undulator period and
$\gamma_z$ the average longitudinal Lorentz factor) and amounts to
$\omega_r/N_w$, $N_w\gg 1$ being the number of undulator periods
(of order $10^2 - 10^3$). Since $\omega_r/N_w$ is a minimum for
the radiation bandwidth, it should be compared with $1/\sigma_T$.
For instance, at wavelengths of order $1 \AA$, $N_w \sim 10^3$ and
$\sigma_T \sim 30$ ps (see \cite{PETR}), one has $\omega_r/N_w
\sim 2\cdot 10^{16}$ Hz, which is much larger than $1/\sigma_T
\sim 3\cdot 10^{10}$ Hz. From this discussion follows that, in
practical situations of interest, we can simplify Eq.
(\ref{gamma5}) to

\begin{eqnarray}
\Gamma_{\omega}(z,\vec{r}_{{} 1},\vec{r}_{{} 2}, \omega_1,
\omega_2) = N_e \bar{f}( \omega_1- \omega_2) G\left(z,\vec{r}_{{}
1},\vec{r}_{{} 2}, \omega_1\right)~, \label{gamma6prima}
\end{eqnarray}
where

\begin{equation}
G(z,\vec{r}_{{} 1},\vec{r}_{{} 2}, \omega) \equiv \left\langle
\bar{E} \left(\vec{\eta},\vec{l},z,\vec{r}_{{} 1}, \omega\right)
\bar{E}^*\left(\vec{\eta},\vec{l},z,\vec{r}_{{} 2}, \omega\right)
\right\rangle~.\label{coore}
\end{equation}
Eq. (\ref{gamma6prima}) fully characterizes the system under study
from a statistical viewpoint. Correlation in frequency and space
are expressed by two separate factors. In particular, spatial
correlation is expressed by the cross-spectral density
function\footnote{Note, however, that $G$ depends on $\omega$.}
$G$. In other words, we are able to deal separately with spatial
and spectral part of the correlation function in space-frequency
domain under the non-restrictive assumption $\omega_r/N_w \gg
1/\sigma_T$.

Eq. (\ref{gamma6prima}) is the result of our theoretical analysis
of the second-order correlation function in the space-frequency
domain. We can readily extend this analysis to the case when the
observation plane is behind a monochromator with transfer function
$T(\omega)$. In this case, Eq. (\ref{gamma6prima}) modifies to

\begin{eqnarray}
\Gamma_{\omega}\left(z,\vec{r}_{{} 1},\vec{r}_{{} 2}, \omega_1,
\omega_2\right) = N_e \bar{f}( \omega_1- \omega_2)
G\left(z,\vec{r}_{{} 1},\vec{r}_{{} 2}, \omega_1\right)
T(\omega_1) T^*(\omega_2)~. \label{gamma6}
\end{eqnarray}
It is worth to note that, similarly to Eq. (\ref{gamma6prima}),
also Eq. (\ref{gamma6}) exhibits separability of correlation
functions in frequency and space.

From now on we will be concerned with the calculation of the
cross-spectral density $G(z,\vec{r}_{{} 1},\vec{r}_{{} 2},
\omega)$, independently of the shape of the remaining factors on
the right hand side of Eq. (\ref{gamma6}).

\begin{figure}
\begin{center}
\includegraphics*[width=140mm]{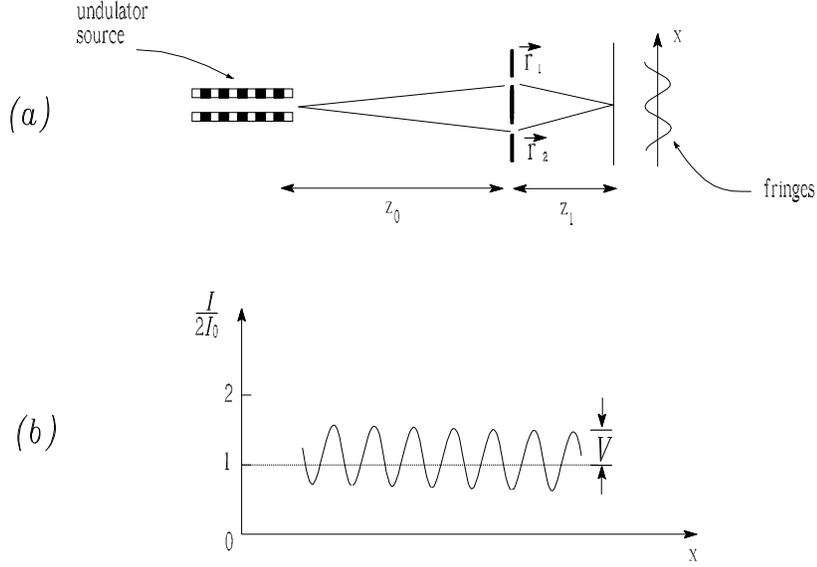}% Here is how to import EPS art
\caption{\label{ed12c} Measurement of the cross-spectral density
of a undulator source. (a) Young's double-pinhole interferometer
demonstrating the coherence properties of undulator radiation. The
radiation beyond the pinholes must be spectrally filtered by a
monochromator or detector (not shown in figure). (b) The fringe
visibility of the resultant interference pattern.}
\end{center}
\end{figure}
Before proceeding we introduce, for future reference, the notion
of spectral degree of coherence $g$, that can be presented as a
function of $\vec{r}_{{} 1}$ and $\vec{r}_{{} 2}$:

\begin{equation}
g\left(\vec{r}_{{} 1},\vec{r}_{{} 2}\right) =
\frac{{G}\left(\vec{r}_{{} 1},\vec{r}_{{} 2}\right)}{\left\langle
\left |\bar{E}\left(\vec{r}_{{} 1}\right)\right|^2\right \rangle
^{1/2} \left \langle \left|\bar{E}\left(\vec{r}_{{}
2}\right)\right|^2\right\rangle^{1/2}} ~. \label{normfine}
\end{equation}
Consider Fig. \ref{ed12c}, depicting a Young's double-pinhole
interferometric measure. Results of Young's experiment vary with
$\vec{r}_{{} 1}$ and $\vec{r}_{{} 2}$. The modulus of the spectral
degree of coherence, $|g|$ is related with the fringe visibility
of the interference pattern. In particular, the relation between
fringes visibility $V$ and $g$ is given by

\begin{eqnarray}
V = 2 \frac{ \left<\left|\bar{E}\left(\vec{r}_{{}
1}\right)\right|^2\right>^{1/2}
\left<\left|\hat{E}\left(\vec{r}_{{}
2}\right)\right|^2\right>^{1/2}}{
\left<\left|\hat{E}\left(\vec{r}_{{}1}\right)\right|^2\right>+
\left<\left|\hat{E}\left(\vec{r}_{{}2}\right)\right|^2\right>}
\left|g\left(\vec{r}_{{}1},\vec{r}_{{}2}\right)\right|~.
\label{Vfrin}
\end{eqnarray}
Phase of $g$ is related to the position of the fringes instead.
Thus, spectral degree of coherence and cross-spectral density are
related with both amplitude \textit{and} position of the fringes
and are physically measurable quantities that can be recovered
from a Young's interference experiment.

\subsection{\label{sub:temp} Relation between space-frequency and
space-time domain}

This paper deals with transverse coherence properties of SR
sources in the space-frequency domain. However, it is interesting
to briefly discuss relations with the space-time domain, and
concepts like quasi-stationarity, cross-spectral purity and
quasi-monochromaticity that are often considered in literature
\cite{GOOD,MAND}.

First, the knowledge of $\Gamma_{\omega}$ in frequency domain is
completely equivalent to the knowledge of the mutual coherence
function \cite{WOLF}:

\begin{equation}
\Gamma_{t}(z,\vec{r}_{{} 1},\vec{r}_{{} 2},t_1,t_2) = \left<
{{E}}(z,\vec{r}_{{} 1},t_1){{E}}^*(z,\vec{r}_{{} 2},t_2) \right>~.
\label{gammatime}
\end{equation}
Next to $\Gamma_t$, the complex degree of coherence is defined as

\begin{eqnarray}
\gamma_t(z,\vec{r}_{{} 1},\vec{r}_{{} 2},t_1,t_2)  =
\frac{\Gamma_{t}(z,\vec{r}_{{} 1},\vec{r}_{{}
2},t_1,t_2)}{\left[\Gamma_{t}(z,\vec{r}_{{} 1},\vec{r}_{{}
1},t_1,t_1) \Gamma_{t}(z,\vec{r}_{{} 2},\vec{r}_{{}
2},t_2,t_2)\right]^{1/2} }~.\label{comdeg}
\end{eqnarray}
The presence of a monochromator (see Eq. (\ref{gamma6})) is
related with a bandwidth of interest $\Delta \omega_\mathrm{m}$,
centered around a given frequency $\omega_o$ (typically, the
undulator resonant frequency). Then, $T(\omega)$ is peaked around
$\omega_o$ and rapidly goes to zero as we move out of the range $(
\omega_o-\Delta \omega_\mathrm{m}/2, \omega_o+\Delta
\omega_\mathrm{m}/2)$. Using Eq. (\ref{gamma6}) we write the
mutual coherence function as

\begin{eqnarray}
\Gamma_t(z,\vec{r}_{{} 1},\vec{r}_{{} 2},t_1,t_2) &=&
\frac{N_e}{(2\pi)^2} \int_{-\infty}^{\infty} d\omega_1
\int_{-\infty}^{\infty} d\omega_2 \bar{f}(\omega_1 - \omega_2)
T(\omega_1) T^*(\omega_2) \cr && \times G(z,\vec{r}_{{}
1},\vec{r}_{{} 2},\omega_2) \exp{(-i\omega_1 t_1)} \exp{(i\omega_2
t_2)} ~.\label{trasfgammabreak}
\end{eqnarray}
If the characteristic bandwidth  of the monochromator, $\Delta
\omega_\mathrm{m}$, is large enough so that $T$ does not vary
appreciably on the characteristic scale of $\bar{f}$, i.e. $\Delta
\omega_\mathrm{m} \gg 1/\sigma_T$, then
$T(\omega_1)T^*(\omega_2)\bar{f}(\omega_1- \omega_2)$ is peaked at
$\omega_1-\omega_2=0$. In this case the process is
quasi-stationary. With the help of new variables $\Delta \omega =
\omega_1 - \omega_2$ and $\bar{\omega}=(\omega_1+\omega_2)/2$, we
can simplify Eq. (\ref{trasfgammabreak}) accounting for the fact
that $\bar{f}(\omega_1- \omega_2)$ is strongly peaked around
$\Delta \omega=0$. In fact we can consider $T(\omega_1)
T^*(\omega_2) G(z,\vec{r}_{{} 1},\vec{r}_{{} 2},\omega_1) \simeq
|T(\bar{\omega})|^2 G(z,\vec{r}_{{} 1},\vec{r}_{{} 2},
\bar{\omega})$, so that we can integrate separately in $\Delta
\omega$ and $\bar{\omega}$ to obtain

\begin{eqnarray}
\Gamma_t(z,\vec{r}_{{} 1},\vec{r}_{{} 2},t_1,t_2) &=&
\frac{N_e}{(2\pi)^2} \int_{-\infty}^{\infty} d\Delta \omega
~\bar{f}(\Delta \omega) \exp{\left(-i \Delta{\omega}
\bar{t}\right)} \cr && \times \int_{-\infty}^{\infty}
d\bar{\omega} ~|T(\bar{\omega})|^2 G(z,\vec{r}_{{} 1},\vec{r}_{{}
2},\bar{\omega}) \exp{\left[-i \bar{\omega} \Delta t
\right]}~.\label{trasfgammabreak2}
\end{eqnarray}
where $\bar{t} = (t_1+t_2)/2$ and $\Delta t = t_1 - t_2$. This
means that the mutual coherence function can be factorized as
$\Gamma_t(\bar{t}, \Delta t) = f(\bar{t}) \widetilde{G}_t(\Delta
t)$. In the case no monochromator is present, $\widetilde{G}_t$
coincides with the Fourier transform of the cross-spectral
density, and the correspondent correlation function
$\Gamma_\omega(\bar{\omega},\Delta \omega)$ has been seen to obey
Eq. (\ref{gamma5}). Therefore:

$(a_1)$ The temporal correlation function $\widetilde{G}_t(\Delta
t)$ and the spectral density distribution of the source
$|T(\bar{\omega})|^2 G(\bar{\omega})$ form a Fourier pair.

$(b_1)$ The intensity distribution of the radiation pulse
$f(\bar{t})$ and the spectral correlation function $\bar{f}(\Delta
\omega)$ form a Fourier pair.

\begin{figure}
\begin{center}
\includegraphics*[width=140mm]{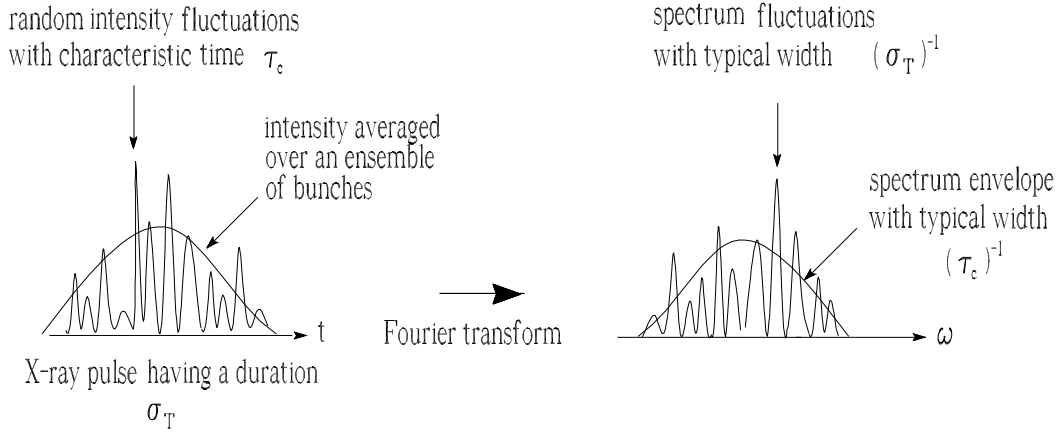}% Here is how to import EPS art
\caption{\label{avecarr} Reciprocal width relations of Fourier
transform pairs.}
\end{center}
\end{figure}
Statement $(a_1)$ can be regarded as an analogue, for
quasi-stationary sources, of the well-known Wiener-Khinchin
theorem, which applies to stationary sources and states that the
temporal correlation function and the spectral density are a
Fourier pair.  Since there is symmetry between time and frequency
domains, an inverse Wiener-Khinchin theorem must also hold, and
can be obtained by the usual Wiener-Khinchin theorem by exchanging
frequencies and times. This is simply constituted by statement
$(b_1)$. Intuitively, statements $(a_1)$ and $(b_1)$ have their
justification in the reciprocal width relations of Fourier
transform pairs (see Fig. \ref{avecarr}).

It should be stressed that statistical optics is almost always
applied in the stationary case \cite{GOOD,MAND}. Definitions of
$\Gamma_t$ and $\gamma_t$ are also usually restricted to such
case. The case of a stationary process corresponds to the
asymptote of a Dirac $\delta$-function for the spectral
correlation function $\bar{f}$. The inverse Wiener-Khinchin
theorem applied to this asymptotic case would imply an infinitely
long radiation pulse, i.e. an infinitely long electron beam. In
contrast to this, the width of $\bar{f}$ is finite, and
corresponds to a finite width of $f$ of about $30$ ps. Thus, one
cannot talk about stationarity. However, when $\Delta
\omega_\mathrm{m} \gg 1/\sigma_T$, the spectral width of the
process (i.e. the width of $|T|^2G$ in $\omega$) is much larger
than the width of $\bar{f}$. In this case, the process is
quasi-stationary. The situation changes completely if a
monochromator with a bandwidth $\Delta \omega_\mathrm{m} \lesssim
1/\sigma_T$ is present. In this case, Eq. (\ref{trasfgammabreak2})
cannot be used anymore, and one is not allowed to treat the
process as quasi-stationary. In the large majority of the cases
monochromator characteristics are not good enough to allow
resolution of $\bar{f}$. There are, however, some special cases
when $\Delta \omega_\mathrm{m} \lesssim 1/\sigma_T$. For instance,
in \cite{NOST} a particular monochromator is described with a
relative resolution of $10^{-8}$ at wavelengths of about $1 \AA$,
or $\omega_o \sim 2 \cdot 10^{19}$ Hz. Let us consider, as in
\cite{NOST}, the case of radiation pulses of $32$ ps duration.
Under the already accepted assumption $1/\sigma_T \ll
\omega_o/N_w$, we can identify the radiation pulse duration with
$\sigma_T$. Then we have $\Delta \omega_\mathrm{m} \sim 2 \cdot
10^{11}$ Hz  which is of order of $2 \pi/\sigma_T \sim 2 \cdot
10^{11}$ Hz: this means that the monochromator has the capability
of resolving $\bar{f}$.

Cases discussed up to now deal with radiation that is not
cross-spectrally pure \cite{MAN2} (or \cite{MAND}, paragraph
4.5.1). In fact, the absolute value of the spectral degree of
coherence $g$ is a function of $\omega$. Moreover, as remarked in
\cite{COI1,COI2,COI3}, the spectrum of undulator radiation depends
on the observation point. This fact can also be seen in the time
domain from Eq. (\ref{trasfgammabreak}) or Eq.
(\ref{trasfgammabreak2}), because the complex degree of coherence
$\gamma_t$ cannot be split into a product of temporal and spatial
factors. However, if we assume $\Delta \omega_\mathrm{m} N_w /
\omega_o \ll 1$ (that is usually true), $G(z,\vec{r}_{{}
1},\vec{r}_{{} 2},\omega)$ $ = G(z,\vec{r}_{{} 1},\vec{r}_{{}
2},\omega_o)$ is a constant function of frequency within the
monochromator line, i.e. it is independent on the frequency
$\omega$. As a result $\Gamma_t$ in Eq. (\ref{trasfgammabreak})
can be split in the product of a temporal and a spatial factor and
therefore, in this case, light is cross-spectrally pure:

\begin{equation}
\Gamma_t(z_o,\vec{r}_{{} o1},\vec{r}_{{} o2},t_1,t_2) = N_e
\widetilde{g}_t(t_1,t_2) G\left(z,\vec{r}_{{} 1},\vec{r}_{{}
2},\omega_o\right)~, \label{CScasegen}
\end{equation}
where $\widetilde{g}_t(t_1,t_2)$ is defined by

\begin{eqnarray}
\widetilde{g}_t = \frac{1}{(2\pi)^2} \int_{-\infty}^{\infty}
d\omega_1 \int_{-\infty}^{\infty} d\omega_2 \bar{f}(\omega_1 -
\omega_2) T(\omega_1) T^*(\omega_2) \exp{[i(\omega_2 t_2-\omega_1
t_1)]} ~.\cr &&\label{gtilda}
\end{eqnarray}
Note that, for instance, in the example considered before $\Delta
\omega_\mathrm{m} /\omega_o \simeq 10^{-8}$ and $N_w \simeq 10^2
\div 10^3$, i.e. $\Delta \omega_\mathrm{m} N_w / \omega_o \ll 1$.

It is important to remark that, since we are dealing with the
process in the space-frequency domain, whether the light is
cross-spectrally pure or not is irrelevant concerning the
applicability of our treatment, because we can study the
cross-spectral density $G$ for any frequency component.

Finally, for the sake of completeness, it is interesting to
discuss the relation between $G$ and the mutual intensity function
as usually defined in textbooks \cite{GOOD, MAND} in
\textit{quasimonochromatic} conditions. The assumption $\Delta
\omega_\mathrm{m} \gg 1/\sigma_T$ describes a quasi-stationary
process. In the limit $\sigma_T \longrightarrow \infty$ we have a
stationary process. Now letting $\Delta \omega_m \longrightarrow
0$ slowly enough so that $\Delta \omega_\mathrm{m} \gg
1/\sigma_T$,  Eq. (\ref{trasfgammabreak2}) remains valid while
both $\bar{f}(\Delta \omega)$ and $|T(\bar{\omega})|^2$ become
approximated better and better by Dirac $\delta$-functions,
$\delta(\Delta\omega)$ and $\delta(\bar{\omega}-\omega_o)$,
respectively. Then, $\Gamma_t \sim G \exp [-i \omega_o
(t_1-t_2)]$. Aside for an unessential factor, depending on the
normalization of $\bar{f}$ and $|T({\omega})|^2$, this relation
between $\Gamma_t$ and $G$ allows identification of the mutual
intensity function with $G$ as in \cite{GOOD,MAND}. In this case,
light is obviously cross-spectrally pure.

\subsection{\label{sub:fiel} Undulator field by a single particle with offset and deflection}

In order to give an explicit expression for the cross-spectral
density of undulator radiation, we first need an explicit
expression for $\bar{E} \left(\vec{\eta},\vec{l},z,\vec{r}_{{}},
\omega\right)$, the field contribution from a single electron with
given offset and deflection. This can be obtained by solving
paraxial Maxwell's equations in the space-frequency domain with a
Green's function technique. We refer the reader to \cite{OURF},
where it was also shown that paraxial approximation always applies
for ultrarelativistic systems with $1/\gamma^2\ll 1$, $\gamma$
being the relativistic Lorentz factor. Since paraxial
approximation applies, the envelope of the field
$\widetilde{E}(\omega) = \bar{E}(\omega) \exp{(-{i\omega z}/{c})}$
is a slowly varying function of $z$ on the scale of the wavelength
$\lambda$, and for the sake of simplicity will also be named "the
field". In Eq. (17) of reference \cite{OURF} an expression for
$\vec{\widetilde{E}}(z, \vec{r},\omega)$ generated by an electron
moving along a generic trajectory $\vec{r}(z)$ was found. Working
out that equation under the resonance approximation for the case
of a planar undulator where $\vec{r}_{}(z) = {K}/({\gamma k_w})
\cos{(k_w z)} \vec{e}_x + \vec{l} +z \vec{\eta}$, $\vec{e}_x$
being the unit vector in the $x$-direction, yields the
horizontally polarized field

\begin{eqnarray}
\tilde{E}_{C}&=& \frac{(-e) K \omega   A_{JJ}}{2 c^2  \gamma}
\int_{-L_w/2}^{L_w/2}  \frac{d z'}{{z}-{z}'} \exp \left\{i
\left[\left(C+\frac{\omega \left|\vec{\eta}\right|^2}{2
c}\right){z}' + \frac{\omega\left(\vec{r}-\vec{l}-\vec{\eta}z'
\right)^2 }{2 c(z-z')}\right] \right\}~ .\cr &&
\label{undunormfin00}
\end{eqnarray}
Here we defined the detuning parameter $C = \omega
/(2{\gamma}_z^2c)-k_w = ({\Delta\omega}/{\omega_r}) k_w$, where
$\omega = \omega_r + \Delta \omega$. Thus, $C$ specifies "how
much" $\omega$ differs from the fundamental resonance frequency
${\omega_r} = {2 \gamma_z^2 c}/{\lambdabar_w}$. The subscript "C"
in $\widetilde{E}_C$ indicates that this expression is valid for
arbitrary detuning parameter. Moreover, $\lambdabar_w \equiv 1/k_w
= \lambda_w/(2\pi)$, $\lambda_w$ being the undulator period;
$\gamma_z \equiv \gamma/(1+K^2/2)$; $K=(\lambda_w e H_w) / (2 \pi
m_\mathrm{e} c^2)$ is the undulator parameter, $m_\mathrm{e}$
being the electron mass and $H_w$ being the maximum of the
magnetic field produced by the undulator on the $z$ axis. Finally,
$L_w$  is the undulator length and $A_{JJ} \equiv
J_0[{K^2}/({4+2K^2})] - J_1[ {K^2}/({4+2K^2})]$, $J_n$ indicating
the Bessel function of the first kind of order $n$. It should be
stressed that Eq. (\ref{undunormfio}) was derived under the
resonance approximation meaning that the large parameter $N_w \gg
1$ was exploited, together with conditions $\Delta \omega/\omega_r
\ll 1$ and $C+\omega r^2/(2 c z^2) \ll k_w$, meaning that we are
looking at frequencies near the fundamental and angles within the
main lobe of the directivity diagram. Moreover, the reader should
keep in mind that no focusing elements are accounted for in the
undulator. This fact is intrinsically related to the choice of
$\vec{r}(z)$ done above.

Further algebraic manipulations  (see Appendix B of \cite{OURU})
show that Eq. (\ref{undunormfin00}) can be rewritten as:

\begin{eqnarray}
\widetilde{E}_C &=& \frac{(-e) K \omega  A_{JJ}}{2 c^2 \gamma}
\int_{-L_w/2}^{L_w/2} \frac{d z'}{z-z'} \cr && \times \exp
\left\{i \left[C z' +
\frac{\omega\left(\vec{r}-\vec{l}~~\right)^2}{2 c z} +\frac{\omega
z z'}{2c (z-z')}
\left(\frac{\vec{r}}{z}-\frac{\vec{l}}{z}-\vec{\eta}\right)^2
\right] \right\} ~.\cr && \label{undunormfio}
\end{eqnarray}
In this paper we will make a consistent use of dimensional
analysis, which allows one to classify the grouping of dimensional
variables in a way that is most suitable for subsequent study.
Normalized units will be defined as

\begin{eqnarray}
&&\hat{E} = -\frac{2 c^2  \gamma}{K \omega e  A_{JJ}}
\widetilde{E}~, \cr && \vec{\hat{\eta}} =\vec{{\eta}}
\sqrt{\frac{\omega L_w}{c}}  ~,\cr &&\hat{C} = L_w C = 2 \pi N_w
\frac {\omega-\omega_r}{\omega_r}~,\cr&&\vec{\hat{r}}_{{} }
={\vec{r}}_{{}} \sqrt{{\omega \over{L_w c}}}~,\cr&&\vec{\hat{l}}
={\vec{l}}\sqrt{{\omega \over{L_w c}}}~,\cr&&
\hat{z}={z\over{L_w}}~.\label{Cnorm}
\end{eqnarray}
Moreover,  for any distance $\hat{z}$, we introduce
$\vec{\hat{\theta}} = \vec{\hat{r}}/\hat{z}$. The algorithm for
calculating the cross-spectral density will be formulated in terms
of dimensionless fields. Therefore we re-write Eq.
(\ref{undunormfio}) as

\begin{eqnarray}
\hat{E}_C=  \exp\left[\frac{i
\hat{z}}{2}\left|\vec{\hat{\theta}}-\frac{\vec{\hat{l}}}{\hat{z}}\right|^2\right]
\Psi_{C} \left(\hat{z},\hat{C},
\left|\vec{\hat{\theta}}-\frac{\vec{\hat{l}}}{\hat{z}}
-\vec{\hat{\eta}}\right|\right)~, \label{undunormfin}
\end{eqnarray}
where

\begin{eqnarray}
\Psi_{C}(\hat{z},\hat{C},\alpha) \equiv \int_{-1/2}^{1/2}
\frac{d\hat{z}'}{\hat{z}-\hat{z}'} \exp \left\{i
\left[\hat{C}\hat{z}' +\frac{\hat{z}\hat{z}' \alpha^2
}{2(\hat{z}-\hat{z}')}\right] \right\} ~.\label{psig}
\end{eqnarray}
A physical picture of the evolution of the field along the $z$
direction was given in reference \cite{OURF}, where Fourier optics
ideas were used to develop a formalism ideally suited for the
analysis of any SR problem. In that reference, the use of Fourier
optics led to establish basic foundations for the treatment of SR
fields, and in particular of undulator radiation, in terms of
laser beam optics. Radiation from an ultra-relativistic electron
can be interpreted as radiation from a virtual source, which
produces a laser-like beam. In principle, such virtual source can
be positioned everywhere down the beam, but there is a particular
position where it is similar, in many aspects, to the waist of a
laser beam. In the case of an undulator this location is the
center of the insertion device. A virtual source located at that
position ("the" virtual source) exhibits a plane wavefront.
Therefore, it is completely specified by a real-valued amplitude
distribution of the field (see Eq. (34) of \cite{OURF}). This
amplitude can be derived from the far zone field distribution.
Free-space propagation from the virtual source through the near
zone and up to the far-zone, can be performed with the help of the
Fresnel formula:

\begin{equation}
{ \hat{E}}( {z},\vec{r}_{}) = \frac{i }{2 \pi ( \hat{z}-
\hat{z}_s)} \int d \vec{\hat{r}'}_{{}}~
\hat{E}(\hat{z}_s,\vec{\hat{r}'}) \exp{\left[\frac{i
\left|{\vec{\hat{r}}}-\vec{ \hat{r}'}\right|^2}{2  (\hat{z}-
\hat{z}_s)}\right]}~, \label{fieldpropback}
\end{equation}
where the integral is performed over the transverse plane, and
$z_s$ is the virtual source position down the beamline.

\begin{figure}
\begin{center}
\includegraphics*[width=110mm]{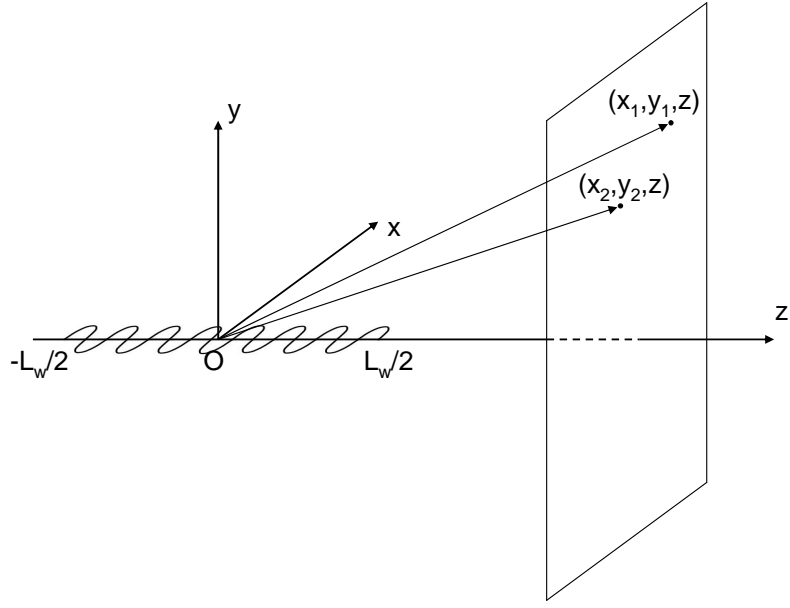}% Here is how to import EPS art
\caption{\label{geo} Illustration of the undulator geometry and of
the observation plane. }
\end{center}
\end{figure}

\begin{figure}
\begin{center}
\includegraphics*[width=110mm]{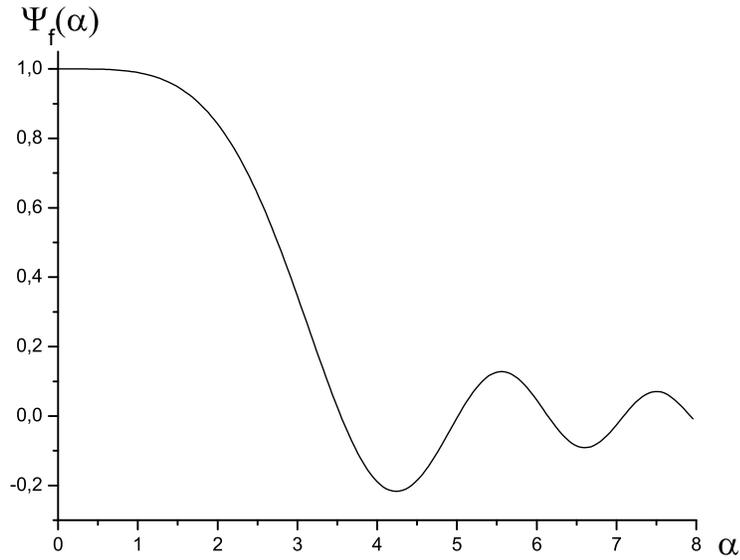}% Here is how to import EPS art
\caption{\label{psif} Universal function $\Psi_f(\alpha)$, used to
calculate the far-zone radiation field of a single electron at the
fundamental harmonic at perfect resonance.}
\end{center}
\end{figure}

\begin{figure}
\begin{center}
\includegraphics*[width=110mm]{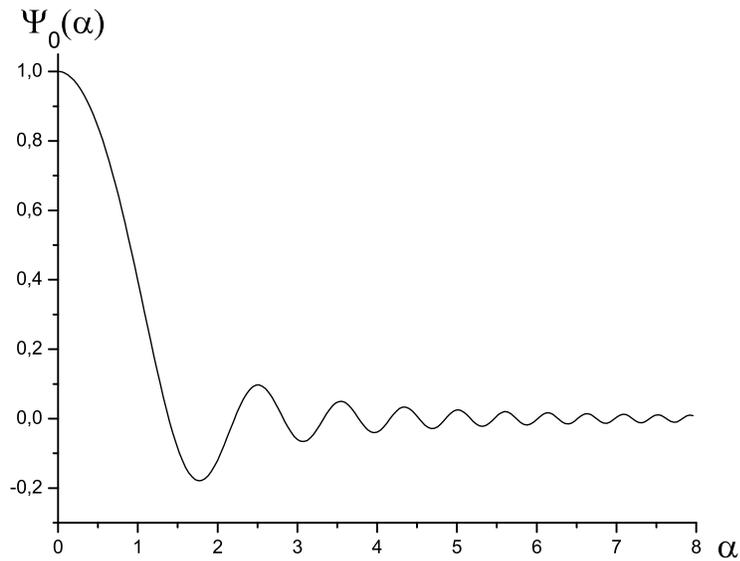}% Here is how to import EPS art
\caption{\label{psio} Universal function $\Psi_0(\alpha)$, used to
calculate the radiation field of a single electron at the
fundamental harmonic at perfect resonance on the virtual source
plane.}
\end{center}
\end{figure}

These considerations were applied in \cite{OURF} to the case of
undulator radiation under the applicability region of the resonant
approximation. With reference to Fig. \ref{geo}, we let $z=0$ be
the center of the undulator. Thus, the position of the virtual
source is fixed in the center of the undulator too, $z_s=0$. For
simplicity, the resonance condition with the fundamental harmonic
was assumed satisfied, i.e. $\omega = \omega_r$. For this case, an
analytical description of undulator radiation was provided.
The horizontally-polarized field produced by a single electron
with offset $\vec{\hat{l}}$ and deflection $\vec{\hat{\eta}}$ in
the far-zone (i.e. at  $\hat{z} \gg 1$) can be represented by the
scalar quantity\footnote{Note that for a particle moving on axis,
at $\vec{\hat{l}}=0$ and $\vec{\hat{\eta}}=0$ the quadratic phase
$\exp[i \hat{z} \hat{\theta}^2/2]$ in Eq. (\ref{undurad4}) is
indicative of a spherical wavefront in paraxial approximation on
the observation plane. When $\vec{\hat{l}}$ is different from
zero, the laser-like beam is shifted, and this justifies the
present of extra-factors including $\vec{\hat{l}}$. When the
particle also has a deflection $\vec{\hat{\eta}}$, the laser-like
beam is tilted, but the wavefront remains spherical. Since the
observation plane remains orthogonal to the $z$ axis, the phase
factor before $\Psi_f$ does not include $\eta$ and thus it does
not depend on the combination
$\vec{\hat{\theta}}-{\hat{l}}/{\hat{z}}-\vec{\hat{\eta}}$.}:

\begin{eqnarray}
\hat{E}_{f}\left(\hat{z},\vec{\hat{\eta}}, \vec{\hat{l}},
\vec{\hat{\theta}}\right)&=& \frac{1}{\hat{z}} \exp\left[i\left(
\frac{\hat{z} \hat{\theta}^2}{2}-
\vec{\hat{\theta}}\cdot\vec{\hat{l}}
+\frac{\hat{l}^2}{2\hat{z}^2}~\right)\right]
\Psi_f\left(\left|\vec{\hat{\theta}}-\frac{\hat{l}}{\hat{z}}-\vec{\hat{\eta}}\right|\right)~
, \label{undurad4}
\end{eqnarray}
where

\begin{eqnarray}
\Psi_f(\alpha) \equiv
\mathrm{sinc}\left[\frac{\alpha^2}{4}\right]~ , \label{bisgg}
\end{eqnarray}
subscript $f$ indicating the "far-zone". The field distribution of
the virtual source positioned at $z=0$, corresponding to the waist
of our laser-like beam was found to be:

\begin{eqnarray}
\hat{{E}}_{0}\left(0,\vec{\hat{\eta}}, \vec{\hat{l}},
\vec{\hat{r}}_{{}}\right) &=& - i \pi \exp\left[i \vec{\hat{\eta}}
\cdot \left(\vec{\hat{r}}_{}-\vec{\hat{l}}~\right) \right]
\Psi_0\left({\left|~\vec{\hat{r}}_{
{}}-\vec{\hat{l}}~~\right|}\right)~, \label{undurad5gg}
\end{eqnarray}
where

\begin{eqnarray}
\Psi_0(\alpha) \equiv \frac{1}{\pi}\left[\pi - 2\mathrm{Si}
\left(\alpha^2\right)\right]~ . \label{undurad4bisgg0}
\end{eqnarray}
where $\mathrm{Si}(z)=\int_0^z dt \sin(t)/t$ indicates the sin
integral function and subscript "0" is indicative of the source
position. Plots of $\Psi_f$ and $\Psi_0$ are given in Fig.
\ref{psif} and Fig. \ref{psio}. It should be noted here that the
independent variable in both plots is the dummy variable $\alpha$.
The characteristic transverse range of the field in the far zone
is in units of the radiation diffraction angle
$\sqrt{\lambdabar/L_w}$, while the characteristic transverse range
of the field at the source is in units of the radiation
diffraction size $\sqrt{\lambdabar L_w}$.

Finally, with the help of the Fresnel propagation formula Eq.
(\ref{fieldpropback}), we found the following expression for the
field distribution at any distance $\hat{z} > 1/2$ from the
virtual source:

\begin{eqnarray}
\hat{E}\left(\hat{z},\vec{\hat{\eta}},
\vec{\hat{l}},\vec{\hat{\theta}}_{}\right) &=& \exp\left[\frac{i
\hat{z} }{2}\left|\vec{\hat{\theta}}_{}
-\frac{\vec{\hat{l}}}{\hat{z}}~\right|^2\right]
\Psi\left(\hat{z},\left|\vec{\hat{\theta}}_{}-
\frac{\vec{\hat{l}}}{\hat{z}}-\vec{\hat{\eta}}\right|\right)~,
\label{Esumg}
\end{eqnarray}
where we defined

\begin{eqnarray}
\Psi(\hat{z},\alpha) \equiv \exp\left[-i \frac{\hat{z}\alpha^2}{2}
\right] \left\{\mathrm{Ei} \left[\frac{i \hat{z}^2\alpha^2
}{2{\hat{z}} - 1}\right]- \mathrm{Ei} \left[\frac{i
\hat{z}^2\alpha^2 }{2{\hat{z}}  + 1}\right]\right\}~,
\label{undurad4bisggz}
\end{eqnarray}
and $\mathrm{Ei}(z)=-\int_{-z}^{\infty} dt \exp(-t)/t$ indicates
the exponential integral function. Eq. (\ref{Esumg}) is a
particular case of Eq. (\ref{undunormfin}) at perfect resonance.
Note that free space basically acts as a spatial Fourier
transformation. This means that the field in the far zone is,
aside for a phase factor, the Fourier transform of the field at
any position $z$ down the beamline. It is also, aside for a phase
factor, the spatial Fourier transform of the virtual source:

\begin{eqnarray}
{\hat{E}}_0\left(0,\vec{\hat{\eta}},
\vec{\hat{l}},\vec{\hat{r}}_{}\right)&=& -\frac{i  \hat{z}}{2 \pi
} \int d\vec{\hat{\theta}}\exp{\left[-\frac{i  |\vec{
\hat{\theta}}|^2}{2} \hat{z}\right]}{
\hat{E}}_f\left(\hat{z},\vec{\hat{\eta}}, \vec{\hat{l}},
\vec{\hat{\theta}}\right)\exp\left[{i } \vec{\hat{r}}_{{}}\cdot
\vec{\hat{\theta}}\right] ~ . \label{virfiemody}
\end{eqnarray}
It follows that

\begin{eqnarray}
\int d \vec{{\theta}}~ \Psi_f(\theta) \exp{\left[{i  }
{\vec{\hat{r}}}_{}\cdot\vec{{\theta}} \right]} &=& 2 \pi
\int_0^{\infty} d \theta~ \theta J_o\left( {|\vec{\hat{r}}}_{}|
\theta \right) \Psi_f(\theta)   =  2\pi^2
\Psi_0\left(|\vec{\hat{r}}_{}|\right) ~.\label{psi0psif}
\end{eqnarray}
We conclude verifying that Eq. (\ref{Esumg}) is in agreement with
Eq. (\ref{undurad4}) and Eq. (\ref{undurad5gg}) $\hat{z} \gg 1$
and for $\hat{z}=0$, respectively. Consider first $\hat{z}=0$. For
positive numbers $\alpha^2 \equiv
\left|\vec{\hat{r}}-\vec{\hat{l}}-\hat{z}\vec{\hat{\eta}}\right|^2
> 0$, we have

\begin{eqnarray}
-i [\pi +2 \mathrm{Si}(\alpha^2)] = \mathrm{Ei}(- i \alpha^2)
-\mathrm{Ei}( i \alpha^2)~, \label{siei}
\end{eqnarray}
and Eq. (\ref{Esumg})  yields back Eq. (\ref{undurad5gg}).

Consider now the limit for $\hat{z} \gg 1$. For positive numbers
$\xi^2 \equiv \left|\vec{\hat{\theta}} -\vec{\hat{\eta}}\right|^2
> 0$ one has

\begin{eqnarray}
&&\exp\left[-\frac{i\hat{z}\xi^2}{2}\right]\left[
\mathrm{Ei}\left(\frac{i \xi^2 \hat{z}^2}{2 \hat{z} -1}\right) -
\mathrm{Ei}\left(\frac{i \xi^2 \hat{z}^2}{2 \hat{z}
+1}\right)\right] \longrightarrow
\frac{1}{\hat{z}}\mathrm{sinc}\left(\frac{\xi^2}{4}\right)~,\label{sincei}
\end{eqnarray}
as it can be directly seen comparing Eq. (\ref{Esumg}) with Eq.
(\ref{undunormfin})), where integration in Eq. (\ref{psig}) is
performed directly at $\hat{C} = 0$ and in the limit for $\hat{z}
\gg 1$. Thus, Eq. (\ref{Esumg}) yields back Eq. (\ref{undurad4}),
as it must be.

Finally, it should be noted that expressions in the present
Section \ref{sub:fiel} have been derived for $\omega > 0$.
Expressions for the field at negative values of $\omega$ can be
obtained based on the property $\vec{\bar{E}}(-\omega) =
\vec{\bar{E}}^*(\omega)$ starting from explicit expressions at
$\omega>0$.

\subsection{\label{sub:evol} Cross-spectral density of an undulator source
and its free-space propagation.}

From now on we consider a normalized expression of the
cross-spectral density, $\hat{G}$, that is linked with $G$ in Eq.
(\ref{coore}) by a proportionality factor

\begin{equation}
\hat{G} =  \left(\frac{2 c^2  \gamma}{K \omega e A_{JJ}}\right)^2
G ~.\label{inetransf}
\end{equation}
Moreover, we introduce variables

\begin{equation}
\Delta \vec{{r}} = {\vec{\hat{r}}_{1}-\vec{\hat{r}}_{2}}~,~~
\vec{\bar{r}} = \frac{\vec{\hat{r}}_{1}+\vec{\hat{r}}_{2}}{2}~.
\label{deltabarx}
\end{equation}
and

\begin{equation}
\Delta \vec{\theta} =
{\vec{\hat{\theta}}_{1}-\vec{\hat{\theta}}_{2}}~,~~
\vec{\bar{\theta}} =
\frac{\vec{\hat{\theta}}_{1}+\vec{\hat{\theta}}_{2}}{2}~,
\label{deltabarth}
\end{equation}
where, as before, $\vec{\hat{\theta}} = \vec{\hat{r}}/z$. Thus,
Eq. (\ref{coore}) can now be written as

\begin{equation}
\hat{G}(\hat{z},\vec{\bar{r}},\Delta \vec{r}, \hat{C}) \equiv
\left\langle \hat{E} \left(\vec{\hat{\eta}},\vec{\hat{l}},\hat{z},
\vec{\bar{r}}+ \frac{\Delta\vec{r}}{2}, \hat{C}\right)
\hat{E}^*\left(\vec{\hat{\eta}},\vec{\hat{l}},\hat{z},
\vec{\bar{r}}- \frac{\Delta\vec{r}}{2}, \hat{C}\right)
\right\rangle~.\label{coore2}
\end{equation}
On the one hand, the cross-spectral density as is defined in Eq.
(\ref{coore2}) includes the product of fields which obey the free
space propagation relation Eq. (\ref{fieldpropback}). On the other
hand, the averaging over random variables commutes with all
operations involved in the calculation of the field propagation.
More explicitly, introducing the notation
$\hat{E}(z)=\mathcal{O}[\hat{E}(z_s)]$ as a shortcut for Eq.
(\ref{fieldpropback}) one can write

\begin{eqnarray}
\hat{G}(z)  = \left\langle
\mathcal{O}\left[\hat{E}(\hat{z}_s)\right]\mathcal{O}^*\left[\hat{E}^*(\hat{z}_s)\right]\right\rangle
=
\mathcal{O}\cdot\mathcal{O}^*\left[\left\langle\hat{E}(\hat{z}_s)\hat{E}^*(\hat{z}_s)\right\rangle\right]
= \mathcal{O}\cdot\mathcal{O}^*\left[G(\hat{z}_s)\right] ~,
\label{comm}
\end{eqnarray}
where $\mathcal{O}$ may also represent, more in general, any
linear operator. Once the cross-spectral density at the source is
known, Eq. (\ref{comm}) provides an algorithm to calculate the
cross-spectral density at any position $\hat{z}$ down the beamline
(in the free-space case). Similarly, propagation through a complex
optical system can be performed starting from the knowledge of
$\hat{G}(\hat{z}_s)$. As a result, the main problem to solve in
order to characterize the cross-spectral density at the specimen
position is to calculate the cross-spectral density at the virtual
source. For the undulator case, we fix the position of the source
in the center of the undulator $\hat{z}_s=0$. This is the main
issue this paper is devoted to. However, free-space propagation is
also treated, and may be considered an illustration of how our
main result can be used in a specific case.

Based on Eq. (\ref{coore2}) and on results in Section
\ref{sub:fiel} we are now ready to present an expression for the
cross-spectral density at any position down the beamline, always
keeping in mind that the main result we are looking for is the
cross-spectral density at the virtual source position.

We begin giving a closed expression for $\hat{G}$ valid at any
value of the detuning parameter $\hat{C}$ by substituting Eq.
(\ref{undunormfin}) in Eq. (\ref{coore2}), and replacing the
ensemble average with integration over the transverse beam
distribution function. We thus obtain

\begin{eqnarray}
&&\hat{G}\left(\hat{z},\hat{C},\vec{\bar{\theta}},\Delta
\vec{\theta}\right) = \exp\left[{i \hat{z} \vec{\bar{\theta}}\cdot
\Delta\vec{\theta}} \right]\int d\vec{\hat{l}}
\exp\left[-{i}\vec{\hat{l}}\cdot \Delta \vec{\theta}\right]\cr
&&\times  \int d\vec{\hat{\eta}}
f_{\bot}\left(\vec{\hat{l}},\vec{\hat{\eta}}\right)
{\Psi}_{C}\left(\hat{z},\hat{C},\left|\vec{\bar{\theta}}+\frac{\Delta
\vec{\theta}}{2}-\frac{\vec{\hat{l}}}{\hat{z}}-\vec{\hat{\eta}}\right|\right)
{\Psi}^*_{C}\left(\hat{z},\hat{C},\left|\vec{\bar{\theta}}-\frac{\Delta
\vec{\theta}}{2}-\frac{\vec{\hat{l}}}{\hat{z}}-\vec{\hat{\eta}}\right|\right)~.\cr
&& \label{Cany}
\end{eqnarray}
It is often useful to substitute the integration variables
$\vec{\hat{l}}$ with $\vec{\phi} \equiv
-\vec{\bar{\theta}}+\vec{\hat{l}}/\hat{z}+\vec{\hat{\eta}}$ In
fact, in this way, Eq. (\ref{Cany}) becomes

\begin{eqnarray}
&&\hat{G}\left(\hat{z},\hat{C},\vec{\bar{\theta}},\Delta
\vec{\theta}\right) = \hat{z}^2 \exp\left[{i
\hat{z}\vec{\bar{\theta}}\cdot \Delta\vec{\theta}} \right]\int
d\vec{\phi} \int d\vec{\hat{\eta}} \exp\left[-{i \hat{z}}
\left(\vec{\phi}+\vec{\bar{\theta}}-\vec{\hat{\eta}}\right)\cdot
\Delta \vec{\theta}\right]\cr &&\times
f_\bot\left(\vec{\phi}+\vec{\bar{\theta}}-\vec{\hat{\eta}},\vec{\hat{\eta}}\right)
{\Psi}_{C}\left(\hat{z},\hat{C},\left|\vec{\phi}-\frac{\Delta
\vec{\theta}}{2}\right|\right)
{\Psi}^*_{C}\left(\hat{z},\hat{C},\left|\vec{\phi}+\frac{\Delta
\vec{\theta}}{2}\right|\right)~, \label{Cany2}
\end{eqnarray}
For choices of $f$ of particular interest (e.g. product of
Gaussian functions for both transverse and angle distributions),
integrals in $d\vec{\eta}$ can be performed analytically, leaving
an expression involving two integrations only, and still quite
generic.

Eq. (\ref{Cany2}) is as far as we can get with this level of
generality, and can be exploited with the help of numerical
integration techniques. However, it still depends on six
parameters at least: four parameters\footnote{At least. This
depends on the number of parameters needed to specify $f_\bot$.
For a Gaussian distribution in phase space, four parameters are
needed, specifying rms transverse size and angular divergence in
the horizontal and vertical direction.} are needed to specify
$f_\bot$, plus the detuning parameter $\hat{C}$ and the distance
$\hat{z}$.

In the following we will assume $\hat{C} \ll 1$, that allows us to
take advantage of analytical presentations for the single-particle
field obtained in \cite{OURF} and reported before. This means that
monochromatization is good enough to neglect finite bandwidth of
the radiation around the fundamental frequency. By this, we
automatically assume that monochromatization is performed around
the fundamental frequency. It should be noted, however, that our
theory can be applied to the case monochromatization is performed
at other frequencies too. Analytical presentation of the
single-particle field cannot be used in full generality, but for
any fixed value $\hat{C}$ of interest one may tabulate the special
function $\Psi_C$ once and for all, and use it in place of $\Psi$
throughout the paper\footnote{Of course, selection of a particular
value $\hat{C}$ still implies a narrow monochromator bandwidth
around that value.}. From this viewpoint, although the case of
prefect resonance studied here is of practical importance in many
situations, it should be considered as a particular illustration
of our theory only.

Also note that in Eq. (\ref{Cany}) the electron beam energy spread
is assumed to be negligible. Contrarily to the monochromator
bandwidth, the energy spread is fixed for a given facility: its
presence constitutes a fundamental effect. In order to
quantitatively account for it, one should sum the dimensionless
energy-spread parameter $\hat{\Delta}_E = 4 \pi N_w
\delta\gamma/\gamma$ to $\hat{C}$ in Eq. (\ref{Cany}) and,
subsequently, integration should be extended over the
energy-spread distribution. Typical energy spread $\delta
\gamma/\gamma$ for third generation light sources is of order $0.1
\%$. For ERL sources this figure is about an order of magnitude
smaller, $\delta \gamma/\gamma \sim 0.01 \%$.

\begin{figure}
\begin{center}
\includegraphics*[width=140mm]{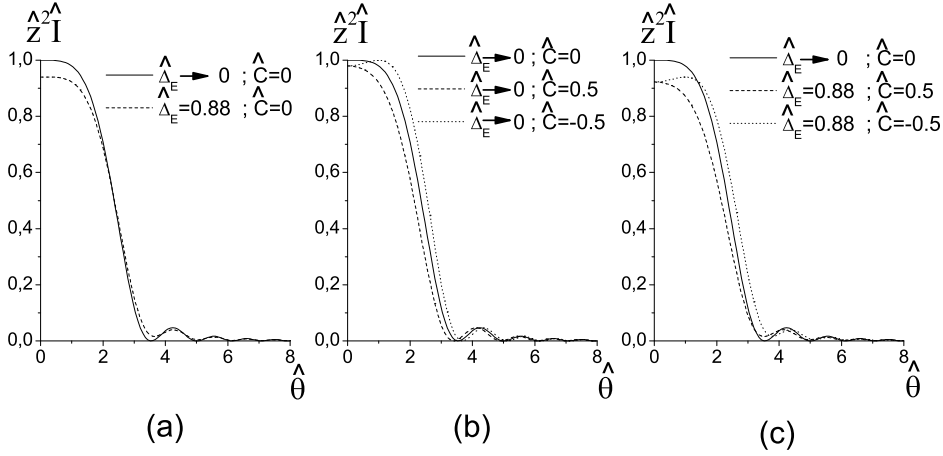}% Here is how to import EPS art
\caption{Study of $\hat{I}$ given in Eq. (\ref{enspread}) for
different valued of $\hat{C}$ and $\hat{\Delta}_E$. Here $N_w
=70$. Plot (a) : comparison, at $\hat{C}=0$, of the case for
negligible energy spread with the case $\hat{\Delta}_E = 0.88$.
Plot (b) : comparison, at negligible energy spread, of the case
$\hat{C}=0$ with the case $\hat{C} = \pm 0.5$. Plot (c) :
comparison of the case $\hat{C}=0$ at negligible energy spread
with the case $\hat{C} = \pm 0.5$ at $\Delta \omega/\omega \simeq
0.1 \%$. \label{espread} }
\end{center}
\end{figure}

In order to study the impact of a finite energy spread parameter
$\hat{\Delta}_{E}$, of a finite radiation bandwidth and of a
relatively small detuning from the fundamental we consider an
expression for the intensity of a diffraction-limited beam
$\hat{z}^2  \hat{I} = \hat{z}^2 \langle |\hat{E}|^2 \rangle$
including both $\hat{\Delta}_{E}$ and $\hat{C}$:

\begin{eqnarray}
\hat{z}^2 \hat{I}(\hat{\theta},\hat{C},\hat{\Delta}_E) &&=
\frac{1}{\sqrt{2\pi}\hat{\Delta}_E} \int_{-\infty}^{\infty} d
\hat{\xi}_E \exp\left[-\frac{\hat{\xi}_E^2}{2 \hat{\Delta}_E^2}
\right] \mathrm{sinc}^2
\left(\frac{\hat{C}+\hat{\xi}_E}{2}+\frac{\hat{\theta}^2}{4}\right)~.\cr
&& \label{enspread}
\end{eqnarray}
We plotted $\hat{z}^2 \hat{I}$ for different values of $\hat{C}$
and $\hat{\Delta}_E$ in Fig. \ref{espread}. First, in Fig.
\ref{espread} (a), we compared, at $\hat{C}=0$, the case for
negligible energy spread with the case $\hat{\Delta}_E = 0.88$,
corresponding to $\delta \gamma/\gamma \simeq 0.1 \%$ at $N_w
=70$, typical of third generation sources. As one can see, maximal
intensity differences are within $10 \%$. Second, in Fig.
\ref{espread} (b) we compared, at negligible energy spread, the
case $\hat{C}=0$ with $\hat{C} = \pm 0.5$ at $N_w = 70$,
corresponding to a shift $\Delta \omega/\omega \simeq 0.1 \%$. One
can see that also in this case maximal intensities differ of about
$10 \%$. Analysis of Fig. \ref{espread} (c), where the case
$\hat{C}=0$ at negligible energy spread is compared with cases
$\hat{C} = \pm 0.5$ at $\Delta \omega/\omega \simeq 0.1 \%$ and
$N_w = 70$ leads to a similar result. This reasoning allows to
conclude that our simplest analytical illustrations can be applied
to practical cases of interest involving third generation sources
and undulators with up to $70$ periods with good accuracy. It
should be remarked that such illustration holds for the first
harmonic only. In fact, while the shape of $\hat{z}^2 \hat{I}$ is
still given by Eq. (\ref{enspread}) in the case of odd harmonic of
order $h$, parameters $\hat{C}$ and $\hat{\Delta}_E$ are modified
according to $\hat{C}_h = h \hat{C}$ and $\hat{\Delta}_{E~h}=h
\hat{\Delta}_{E}$, decreasing the applicability of our analytical
results.

%When $\delta\gamma/\gamma \sim 0.01 \%$ the asymptotic case for a
%monoenergetic beam, Eq. (\ref{Cany}), can be used with high
%accuracy. For $\delta\gamma/\gamma \sim 0.1 \%$ Eq. (\ref{Cany})
%can be used as a first estimation for the cross-spectral density.

With this in mind, we can present an expression for the
cross-spectral density at $\hat{C} \ll 1$ based on Eq.
(\ref{coore2}) and Eq. (\ref{Esumg}). Substituting the latter in
the former we obtain an equation for
$\hat{G}(\hat{z},\vec{\bar{\theta}},\Delta \vec{\theta})$ that can
be formally derived from Eq. (\ref{Cany}) by substitution of
$\Psi_C$ with $\Psi$. Similarly as before, one may give
alternative presentation of
$\hat{G}(\hat{z},\vec{\bar{\theta}},\Delta \vec{\theta})$
replacing the integration variables $\vec{\hat{l}}$ with
$\vec{\phi} \equiv
-\vec{\bar{\theta}}+\vec{\hat{l}}/\hat{z}+\vec{\hat{\eta}}$. This
results in another expression for
$\hat{G}(\hat{z},\vec{\bar{\theta}},\Delta \vec{\theta})$ that can
be formally derived from Eq. (\ref{Cany2}) by substituting
$\Psi_C$ with $\Psi$. This last expression presents the
cross-spectral density in terms of a convolution of the transverse
electron beam phase space distribution with an analytical
function, followed by Fourier transformation\footnote{Aside for an
inessential multiplicative constant. This remark also applies in
what follows.}.

One may obtain an expression for $\hat{G}$ at $\hat{z}=0$ as a
limiting case of Eq. (\ref{Cany}) or Eq. (\ref{Cany2}) at
$\hat{C}=0$. It is however simpler to do so by substituting Eq.
(\ref{undurad5gg}) in Eq. (\ref{coore2}) that gives

\begin{eqnarray}
\hat{G}\left(0,\vec{\bar{r}},\Delta \vec{r}\right) &=& \int
d\vec{\eta} \exp\left[i \vec{{\eta}} \cdot \Delta \vec{r}\right]
\cr && \times \int d\vec{l} f_\bot\left(\vec{l},\vec{\eta}\right)
\pi^2 \Psi_0\left(\left|\vec{\bar{r}}-\frac{\Delta \vec{r}}{2} -
\vec{l}\right|\right) \Psi_0\left(\left|\vec{\bar{r}}+\frac{\Delta
\vec{r}}{2} - \vec{l}\right|\right)~,\label{Gnor}
\end{eqnarray}
where the function $\Psi_0$ has already been defined in Eq.
(\ref{undurad4bisgg0}). The product
$\Psi_0\left(\left|\vec{\bar{r}}-{\Delta \vec{r}}/{2}
\right|\right)
 \Psi_0\left(\left|\vec{\bar{r}}+{\Delta \vec{r}}/{2}\right| \right)$ is a
four-dimensional , analytical function in $\vec{\bar{r}}$ and
$\Delta \vec{r}$. Eq. (\ref{Gnor}) tells that the cross-spectral
density at the virtual source position can be obtained convolving
$\Psi_0\left(\left|\vec{\bar{r}}-{\Delta \vec{r}}/{2}
\right|\right) \Psi_0\left(\left|\vec{\bar{r}}+{\Delta
\vec{r}}/{2} \right|\right)$ with the transverse beam distribution
at $z=0$, $f_\bot\left(\vec{\hat{l}},\vec{\hat{\eta}}\right)$,
considered as a function of $\vec{\hat{l}}$, and taking Fourier
transform with respect to $\vec{\hat{\eta}}$. When the betatron
functions have minima in the center of the undulator we have

\begin{eqnarray}
f_\bot\left(\vec{\hat{l}},\vec{\hat{\eta}}\right) =
f_l\left(\vec{\hat{l}}\right)
f_\eta\left(\vec{\hat{\eta}}\right)~. \label{ff}
\end{eqnarray}
Then, Eq. (\ref{Gnor}) becomes

\begin{eqnarray}
\hat{G}\left(0,\vec{\bar{r}},\Delta \vec{r}\right) &=& \int
d\vec{\hat{\eta}} f_\eta\left(\vec{\hat{\eta}}\right) \exp\left[i
\vec{\hat{\eta}} \cdot \Delta \vec{r}\right] \cr && \times \int
d\vec{\hat{l}} f_l\left(\vec{\hat{l}}\right) \pi^2
\Psi_0\left(\left|\vec{\bar{r}}-\frac{\Delta \vec{r}}{2} -
\vec{\hat{l}}\right|\right)
\Psi_0\left(\left|\vec{\bar{r}}+\frac{\Delta \vec{r}}{2} -
\vec{\hat{l}}\right|\right)~,\label{Gnor2}
\end{eqnarray}
that will be useful later on. In this case, the cross-spectral
density is the product of two separate factors. First, the Fourier
transform of the distribution of angular divergence of electrons.
Second, the convolution of the transverse electron beam
distribution with the four-dimensional function
$\Psi_0\left(\left|\vec{\bar{r}}-{\Delta \vec{r}}/{2}
\right|\right) \Psi_0\left(\left|\vec{\bar{r}}+{\Delta
\vec{r}}/{2} \right|\right)$.

Eq. (\ref{Cany}) (or Eq. (\ref{Cany2})) constitutes the most
general result in in the calculation of the cross-spectral density
for undulator sources. Its applicability is not restricted to
third generation light sources. In particular, it can be used for
arbitrary undulator sources like ERLs \cite{EDGA} or XFEL
spontaneous undulators \cite{XFEL}. Eq. (\ref{Cany2}) has been
derived, in fact, under the only constraints $\gamma^2 \gg 1$,
$N_w \gg 1$ and $\sigma_T \omega_r \gg N_w$. Note that $\sigma_T
\sim 30 $ ps for a typical SR source, whereas $\sigma_T \sim 100$
fs for an XFEL spontaneous undulator source or an ERL. Yet, for
all practical cases of interest, $\sigma_T \omega \gg N_w$. As we
have seen before, Eq. (\ref{Cany2}) further simplifies  in the
particular but practical case of perfect resonance, i.e. in the
limit for $\hat{C} \ll 1$. A particularly important asymptote of
Eq. (\ref{Cany}) at perfect resonance is at the virtual source
position, described by Eq. (\ref{Gnor}), which express the
cross-spectral density in the undulator center. While Eq.
(\ref{Cany}) (or Eq. (\ref{Cany2})) solves all problems concerning
characterization of transverse coherence properties of light in
free-space, the knowledge of Eq. (\ref{Gnor}) constitutes, in the
presence of optical elements, the first (and main) step towards
the characterization of SR light properties at the specimen
position. In fact, the tracking of the cross-spectral density can
be performed with the help of standard statistical optics
formalism developed for the solution of problems dealing with
partially coherent sources. Finally, it should be noted that in
the case of XFELs and ERLs, there is no further simplification
that we may apply to previously found equations. In particular,
the transverse electron beam phase space should be considered as
the result of a start-to-end simulation or, better, of
experimental diagnostics measurements in a operating machine. On
the contrary, as we will see, extra-simplifications can be
exploited in the case of third-generation light sources, allowing
for the development of a more specialized theory.

Inspection of  Eq. (\ref{Cany}) or Eq. (\ref{Gnor}) results in the
conclusion that a Gaussian-Schell model cannot be applied to
describe partially coherent SR light. In fact, functions $\Psi_C$,
$\Psi$, $\Psi_0$ and $\Psi_f$ are of non-Gaussian nature, as the
laser-like beam they can be ascribed to is non-Gaussian. This
explains our words in the Introduction, where we stated that
\cite{COI1,COI2,COI3} are of general theoretical interest, but
they do not provide a satisfactory approximation to third
generation SR sources.

As a final remark to this Section, we should discuss the relation
of our approach with that given, in terms of Wigner distribution,
in \cite{KIM2,KIM3}. As said in Section \ref{sec:intro}, treatment
based on Wigner distribution is equivalent to treatment based on
cross-spectral density. We chose to use cross-spectral density
because such quantity is straightforwardly physically measurable,
being related to the outcome of a Young's experiment. Essentially,
one can obtain a Wigner distribution $\hat{W}$ from $\hat{G}$ by
means of an inverse Fourier transformation:

\begin{eqnarray}
\hat{W} \left(\hat{z},\vec{\bar{r}}, \vec{\bar{u}}\right) =
\frac{1}{4\pi^2} \int d(\Delta\vec{r}~) \exp\left[-i\Delta
\vec{r}\cdot \vec{\bar{u}}
\right]\hat{G}\left(\hat{z},\vec{\bar{r}},\Delta \vec{r}\right) ~.
\label{W1}
\end{eqnarray}
Thus, Eq. (\ref{Gnor2}) gives

\begin{eqnarray}
\hat{W}\left(0,\vec{\bar{r}},\vec{\bar{u}}\right) &=& \int
d\vec{\hat{l}} \int d\vec{\hat{\eta}}
f\left(\vec{\hat{l}},\vec{\hat{\eta}}\right)
\hat{W}_o\left(0,\vec{\bar{r}}-\vec{\hat{l}},\vec{\bar{u}}-\vec{\hat{\eta}}\right)~.
\label{W2}
\end{eqnarray}
where

\begin{eqnarray}
\hat{W}_o\left(0,\vec{\alpha},\vec{\delta}\right) = \frac{1}{4}
\int d(\Delta \vec{r}~) \exp\left[-i\Delta \vec{r}\cdot
\vec{\delta} \right] \Psi_0\left(\left|\vec{\alpha}-\frac{\Delta
\vec{r}}{2}\right|\right)
\Psi_0\left(\left|\vec{\alpha}+\frac{\Delta
\vec{r}}{2}\right|\right)~,\label{WWW}
\end{eqnarray}
The Wigner distribution at $\hat{z}=0$ is presented as a
convolution product between the electron phase-space and a
universal function $\hat{W}_o$. This result may be directly
compared (aside for different notation) with \cite{KIM2,KIM3},
where the Wigner distribution is presented as a convolution
between the electrons phase space and a universal function as
well. The study in \cite{KIM2,KIM3} ends at this point, presenting
expressions for arbitrary detuning parameter. On the contrary, in
the following Sections we will take advantage of expressions at
perfect resonance, of small and large parameters related to third
generation light sources and of specific characteristics of the
electron beam distribution. This will allow us to develop a
comprehensive theory of third generation SR sources.

\section{\label{sec:main} Theory of transverse coherence for
third-generation light sources}

\subsection{\label{sub:cross} Cross-Spectral Density}

We now specialize our discussion to third-generation light
sources.  We assume that the motion of particles in the horizontal
and vertical directions are completely uncoupled. Additionally, we
assume a Gaussian distribution of the electron beam in the phase
space. These two assumptions are practically realized, with good
accuracy, in storage rings. For simplicity, we also assume that
the minimal values of the beta-functions in horizontal and
vertical directions are located at the virtual source position
$\hat{z} = 0$, that is often (but not
always\footnote{Generalization to the case when this assumption
fails is straightforward.}) the case in practice. $f_\bot =
f_{\eta_x}(\hat{\eta}_x) f_{\eta_x}(\hat{\eta}_y)
f_{l_x}(\hat{l}_x) f_{l_x}(\hat{l}_y)$ with

\begin{eqnarray}
&& f_{\eta_x}(\hat{\eta}_x) = \frac{1}{\sqrt{2\pi D_x}}
\exp{\left(-\frac{\hat{\eta}_x^2}{2 D_x}\right)}~,~~~~
f_{\eta_y}(\hat{\eta}_y)  = \frac{1}{\sqrt{2\pi D_y}}
\exp{\left(-\frac{\hat{\eta}_y^2}{2 D_y}\right)}~,\cr &&
f_{l_x}(~\hat{l}_x) =\frac{1}{\sqrt{2\pi N_x} }
\exp{\left(-\frac{\hat{l}_x^2}{2 N_x}\right)}~,~~~~~
f_{l_y}(~\hat{l}_y)=\frac{1}{\sqrt{2\pi N_y} }
\exp{\left(-\frac{\hat{l}_y^2}{2 N_y}\right)}~.\label{distr}
\end{eqnarray}
Here

\begin{equation}
D_{x,y} = \frac{\sigma_{x',y'}^2} {{\lambdabar}/L_w}~,~~~~~
N_{x,y} = \frac{\sigma^2_{x,y}} {\lambdabar L_w}~,\label{enne}
\end{equation}
$\sigma_{x,y}$ and $\sigma_{x',y'}$ being rms transverse bunch
dimensions and angular spreads. Parameters $N_{x,y}$ will be
indicated as the beam diffraction parameters, are analogous to
Fresnel numbers and correspond to the normalized square of the
electron beam sizes, whereas $D_{x,y}$ represent the normalized
square of the electron beam divergences.  Consider the reduced
emittances $\hat{\epsilon}_{x,y} = \epsilon_{x,y}/\lambdabar$,
where $\epsilon_{x,y}$ indicate the geometrical emittance of the
electron beam in the horizontal and vertical directions. Since we
restricted our model to third generation light sources, we can
consider $\hat{\epsilon}_x\gg 1$. Moreover, since betatron
functions are of order of the undulator length, we can also
separately accept

\begin{eqnarray}
N_x \gg 1~,~~~~~D_x \gg 1, \label{major}
\end{eqnarray}
still retaining full generality concerning values of $N_y$ and
$D_y$, due to the small coupling coefficient between horizontal
and vertical emittance.

Exploitation of the extra-parameter $\hat{\epsilon}_x\gg 1$ (or
equivalently $N_x\gg 1$ and $D_x \gg 1$) specializes our theory to
the case of third-generation sources.

With this in mind we start to specialize our theory beginning with
the expression for the cross-spectral density at the virtual
source, i.e. Eq. (\ref{Gnor2}). After the change of variables
$\vec{\phi} \longrightarrow -\vec{\bar{r}}+\vec{\hat{l}}$, and
making use of Eq. (\ref{distr}), Eq. (\ref{Gnor2}) becomes

\begin{eqnarray}
\hat{G}\left(0,\vec{\bar{r}},\Delta \vec{r}\right) &=&
\frac{\pi}{2 \sqrt{N_x N_y}} \exp \left[-\frac{(\Delta x)^2
D_x}{2}\right] \exp \left[-\frac{(\Delta y)^2 D_y}{2}\right]\cr &&
\times \int_{-\infty}^{\infty} d \phi_x \int_{-\infty}^{\infty} d
\phi_y \exp\left[-\frac{\left(\phi_x+\bar{x}\right)^2}{2
N_x}\right] \exp\left[-\frac{\left(\phi_y+\bar{y}\right)^2}{2
N_y}\right] \cr && \times
\Psi_0\left\{\left[\left({\phi_{x}}+\frac{\Delta
x}{2}\right)^2+\left({\phi_{y}}+\frac{\Delta y}{2} \right)^2
\right]^{1/2}\right\} \cr && \times
\Psi_0\left\{\left[\left({\phi_{x}}-\frac{\Delta x}{2}
\right)^2+\left({\phi_{y}}-\frac{\Delta y}{2} \right)^2
\right]^{1/2}\right\}~,\label{Gnor3}
\end{eqnarray}
where $\Delta x$ and $\Delta y$ indicate components of $\Delta
\vec{r}$. In Eq. (\ref{Gnor3}) the range of variable $\phi_x$  is
effectively limited up to values $|\phi_x| \sim 1$. In fact,
$\phi_x$ enters the expression for $\Psi_0$. It follows that at
values larger than unity the integrand in Eq. (\ref{Gnor3}) is
suppressed. Then, since $N_x \gg 1$, we can neglect $\phi_x$ in
the exponential function. Moreover $D_x \gg 1$ and from the
exponential function in $D_x$ follows that $\Delta x \ll 1$ can be
neglected in $\Psi_0$. As a result, Eq. (\ref{Gnor3}) is
factorized in the product of a horizontal cross-spectral density
$G_x(0,\bar{x}, \Delta x)$ and a vertical cross-spectral density
$G_y(0,\bar{x}, \Delta x)$:

\begin{eqnarray}
\hat{G}\left(0,\vec{\bar{r}},\Delta \vec{r}\right) =
\hat{G}_x\left(0,{\bar{x}},\Delta {x}\right)
\hat{G}_y\left(0,{\bar{y}},\Delta {y}\right)~, \label{factorizeee}
\end{eqnarray}
where

\begin{eqnarray}
\hat{G}_x\left(0,{\bar{x}},\Delta {x}\right) &=&
\sqrt{\frac{\pi}{{N_x}}} \exp \left[-\frac{(\Delta x)^2D_x
}{2}\right] \exp\left[-\frac{\bar{x}^2}{2 N_x}\right]
~,\label{Gnor3x}
\end{eqnarray}
\begin{eqnarray}
\hat{G}_y\left(0,{\bar{y}},\Delta {y}\right) &=&
\frac{1}{2}\sqrt{\frac{\pi}{N_y}}  \exp \left[-\frac{(\Delta
y)^2D_y}{2}\right] \int_{-\infty}^{\infty} d \phi_y
\int_{-\infty}^{\infty} d \phi_x
\exp\left[-\frac{\left(\phi_y+\bar{y}\right)^2}{2 N_y}\right] \cr
&& \times
\Psi_0\left\{\left[{\phi_{x}}^2+\left({\phi_{y}}+\frac{\Delta
y}{2} \right)^2 \right]^{1/2}\right\}
\Psi_0\left\{\left[{\phi_{x}}^2+\left({\phi_{y}}-\frac{\Delta
y}{2} \right)^2 \right]^{1/2}\right\}~.\cr &&\label{Gnor3y}
\end{eqnarray}
Note that, in virtue of Eq. (\ref{fieldpropback}), factorization
holds in general, at any position $\hat{z}$. This allows us to
separately study $\hat{G}_x$ and $\hat{G}_y$. $\hat{G}_x$
describes a quasi-homogeneous Gaussian source, which will be
treated in Section \ref{sub:gaus}. Here we will focus our
attention on $\hat{G}_y$ only. It should be remarked that the
quasi-homogenous Gaussian source asymptote is obtained from Eq.
(\ref{Gnor3y}) in the limit $N_y \gg 1$ and $D_y \gg 1$. In other
words, normalization constants in Eq. (\ref{Gnor3x}) and Eq.
(\ref{Gnor3y}) are chosen in such a way that Eq. (\ref{Gnor3y})
reduces to Eq. (\ref{Gnor3x}) in the limit $N_y \gg 1$ and $D_y
\gg 1$ (with the obvious substitution $x \longrightarrow y$). It
should be clear that this normalization is most natural, but not
unique. The only physical constraint that normalization of Eq.
(\ref{Gnor3x}) and Eq. (\ref{Gnor3y}) should obey is that the
product $\hat{G}_x \hat{G}_y$ should not change.

Let us define the two-dimensional universal function
$\mathcal{S}(\alpha,\delta)$ as\footnote{$\mathcal{S}$ stands for
"Source".}

\begin{eqnarray}
\mathcal{S}(\alpha,\delta) = \mathcal{K}_S \int_{-\infty}^{\infty}
d \phi_x
\Psi_0\left\{\left[{\phi_{x}}^2+\left(\alpha+\frac{\delta}{2}\right)^2
\right]^{1/2}\right\}
\Psi_0\left\{\left[{\phi_{x}}^2+\left(\alpha-
\frac{\delta}{2}\right)^2 \right]^{1/2}\right\}~. \cr &&
\label{mtcm}
\end{eqnarray}
The normalization constant $\mathcal{K}_S \simeq 0.714$ is chosen
in such a way that $\mathcal{S}(0,0)=1$. We can present
$\hat{G}_y$ at the virtual source with the help of $\mathcal{S}$
as

\begin{figure}
\begin{center}
\includegraphics*[width=140mm]{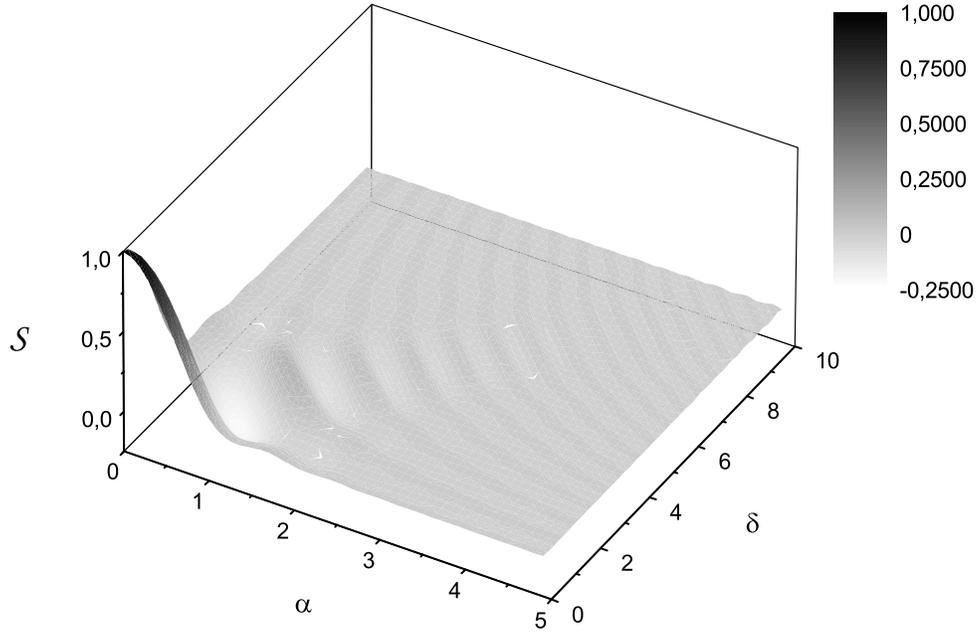}% Here is how to import EPS art
\caption{\label{FTM} Plot of the universal function
${\mathcal{S}}$, used to calculate the cross-spectral density at
the virtual source when $N_x \gg1$, $D_x \gg 1$, $N_y$ and $D_y$
are arbitrary. }
\end{center}
\end{figure}

\begin{eqnarray}
\hat{G}_y\left(0,{\bar{y}},\Delta {y}\right) &=&
\frac{1}{2}\sqrt{\frac{\pi}{N_y}}  \exp \left[-\frac{(\Delta y)^2
D_y }{2}\right] \int_{-\infty}^{\infty} d \phi_y
\exp\left[-\frac{\left(\phi_y+\bar{y}\right)^2}{2 N_y}\right] \cr
&& \times~\frac{1}{\mathcal{K}_S} \mathcal{S}(\phi_y,\Delta y) ~.
\label{Gnor4y}
\end{eqnarray}
Therefore, $\hat{G}_y$ at the virtual source is found by
convolving an universal function, $\mathcal{S}$, with a Gaussian
function and multiplying the result by another Gaussian function.

Note that in the limit $N_x \gg 1$ and $D_x \gg 1$ there is no
influence of the electron beam distribution along the vertical
direction on $\hat{G}_x$. In spite of this, in the same limit, Eq.
(\ref{Gnor4y}) shows that there is an influence of the horizontal
electron beam distribution on $\hat{G}_y$, due to the
non-separability of the function $\Psi_0$ in $\mathcal{S}$. In
fact, contrarily to the case of a Gaussian laser beam,
$\Psi_0[{\phi_{x}}^2+(\alpha- {\delta}/{2})^2 ] \ne
\Psi_0[{\phi_{x}}^2 ] \Psi_0[(\alpha- {\delta}/{2})^2 ]$, and the
integral in $d \phi_x$, that is a remainder of the integration
along the x-direction, is still present in the definition of
$\mathcal{S}$. However, such influence of the horizontal electron
beam distribution is independent of $N_x$ and $D_x$. As a
consequence, $\mathcal{S}$ is a universal function.

A plot of ${\mathcal{S}}$ is presented in Fig. \ref{FTM}.
${\mathcal{S}}$ is a real function. Then, $\hat{G}$ at the virtual
plane is also real. Moreover, ${\mathcal{S}}$ is invariant for
exchange of $\alpha$ with $\delta/2$.

\begin{figure}
\begin{center}
\includegraphics*[width=140mm]{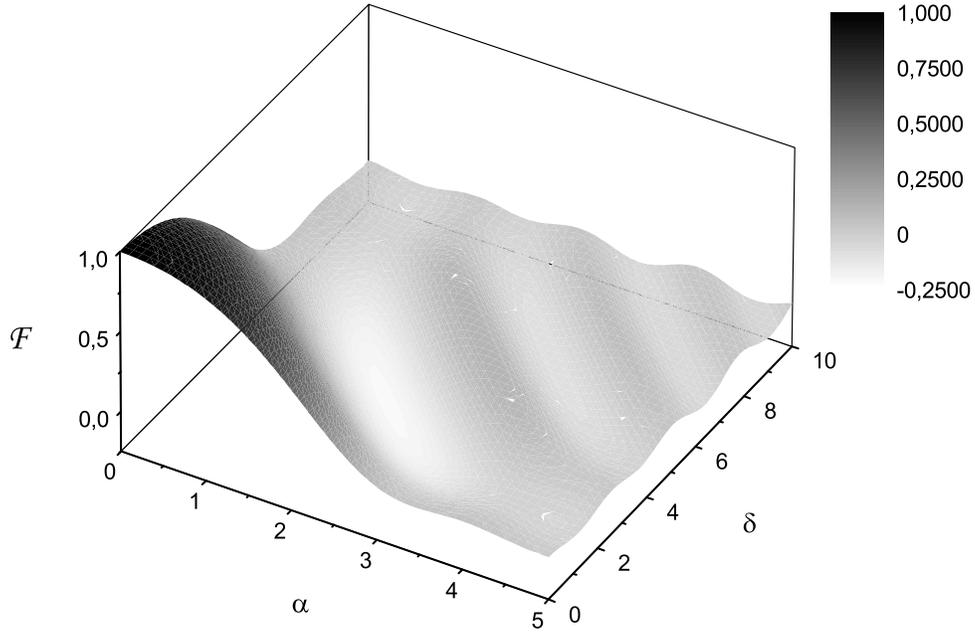}% Here is how to import EPS art
\caption{\label{M} Plot of the universal function $\mathcal{F}$,
used to calculate the cross-spectral density in the far zone when
$N_x \gg1$, $D_x \gg 1$, $N_y$ and $D_y$ are arbitrary. }
\end{center}
\end{figure}

Let us now deal with the evolution of the cross-spectral density
$\hat{G}_y$ in free-space. In principle, one may use Eq.
(\ref{Gnor4y}) and apply Eq. (\ref{comm}), remembering Eq.
(\ref{fieldpropback}). It is however straightforward to use
directly Eq. (\ref{Cany2}) at $\hat{C} \ll 1$. Under the
assumption $N_x \gg 1$ and $D_x \gg 1$, as has been already
remarked, factorization of $\hat{G}$ as a product of $\hat{G}_x$
and $\hat{G}_y$ holds for any value of $\hat{z}$.  Isolating these
factors in Eq. (\ref{Cany2}) at $\hat{C} \ll 1$ and using Eq.
(\ref{distr}) one obtains

\begin{eqnarray}
&&\hat{G}_y\left(\hat{z},{\bar{\theta}_y},\Delta {\theta}_y\right)
= \frac{\hat{z}^2}{2 \pi^2 \sqrt{2 N_y D_y}} \exp\left[{i
\hat{z}\bar{\theta}_y \Delta{\theta}_y } \right]\int d{\phi}_y
\exp\left[-{i \hat{z}}\left({\phi}_y+{\bar{\theta}_y}\right)
\Delta {\theta}_y\right] \cr && \times \int d\hat{\eta}_y
\exp\left[i \hat{z}{\hat{\eta}_y}\Delta {\theta}_y\right]
\exp\left[-\frac{\hat{\eta}_y^2}{2 D_y}\right]
\exp\left[-\frac{\left({\phi}_y
+\bar{\theta}_y-\hat{\eta}_y\right)^2}{2 N_y/\hat{z}^2} \right]\cr
&& \times \int d \phi_x
{\Psi}\left\{\hat{z},\left[{\phi}_x^2+\left(\phi_y-\frac{\Delta
\theta_y }{2}\right)^2\right]^{1/2}\right\}
{\Psi}^*\left\{\hat{z},\left[{\phi}_x^2+\left(\phi_y+\frac{\Delta
\theta_y}{2}\right)^2\right]^{1/2}\right\}~. \label{Gnoryany}
\end{eqnarray}
The integral in $d \hat{\eta}_y$ (second line) can be calculated
analytically yielding

\begin{eqnarray}
&&\hat{G}_y\left(\hat{z},{\bar{\theta}}_y,\Delta \theta_y\right) =
\frac{\hat{z}}{2\pi\sqrt{ \pi} \sqrt{N_y/\hat{z}^2 + D_y}}
\exp\left[{\hat{z}}{i \bar{\theta}_y \Delta\theta_{y} } \right]
\exp\left[-\frac{D_y N_y \Delta \theta_y^2 }{2\left(N_y/\hat{z}^2
+ D_y \right)}\right] \cr && \times \int d{\phi}_y
\exp\left[-\frac{i \left(\phi_y+\bar{\theta}_y\right) \Delta
\theta_y N_y/\hat{z}}{N_y/\hat{z}^2 + D_y }\right]
\exp\left[-\frac{\left({\phi}_y+{\bar{\theta}_y}\right)^2}{2\left(N_y/\hat{z}^2
+ D_y \right)}\right] \cr && \times \int d \phi_x
{\Psi}\left\{\hat{z},\left[{\phi}_x^2+\left(\phi_y-\frac{\Delta
\theta_y }{2}\right)^2\right]^{1/2}\right\}
{\Psi}^*\left\{\hat{z},\left[{\phi}_x^2+\left(\phi_y+\frac{\Delta
\theta_y}{2}\right)^2\right]^{1/2}\right\}~. \label{Gnoryany2}
\end{eqnarray}
We are now interested in discussing the far-zone limit of Eq.
(\ref{Gnoryany2}). Up to now we dealt with the far-zone region of
the field from a single particle, Eq. (\ref{undurad4}). In this
case, the field exhibits a spherical wavefront. Such wavefront
corresponds to the quadratic phase factor in Eq. (\ref{undurad4}).
Note that when the electron is moving on-axis, Eq.
(\ref{undurad4}) consists of the product of $\hat{z}^{-1} \exp[i
\hat{z} \hat{\theta}^2/2]$ by a real function independent of
$\hat{z}$. Such field structure can be taken as a definition of
far-zone. A similar definition can be used for the far-zone
pertaining the cross-spectral density. We regard the quadratic
phase factor $\exp[{\hat{z}}{i \bar{\theta}_y \Delta\theta_{y} }
]$ in Eq. (\ref{Gnoryany2}) as the equivalent, in terms of
cross-spectral density, of the quadratic phase factor for the
single-particle field. We therefore take as definition of far-zone
the region of parameters where $\hat{G}_y(\hat{z},
\bar{\theta}_y,\Delta \theta_y) = \hat{z}^{-1}
h(\bar{\theta}_y,\Delta \theta_y) \exp[{\hat{z}}{i \bar{\theta}_y
\Delta\theta_{y} }]$, $h$ being a real function, and remains like
that for larger values of $\hat{z}$.

Let us discuss the definition of the far-zone region in terms of
problem parameters. In the single-particle situation, the only
parameter of the problem was $\hat{z}$ and, as is intuitively
sound, the far-zone region was shown to coincide with the limit
$\hat{z} \gg 1$. In the case of Eq. (\ref{Gnoryany2}), we deal
with three parameters $\hat{z}$, ${N}_y$ and ${D}_y$. Therefore we
should find that the far-zone is defined in terms of conditions
involving all three parameters.

When $N_y\lesssim 1$ and $D_y \lesssim 1$, analysis of Eq.
(\ref{Gnoryany2}) shows that the far-zone region is for $\hat{z}
\gg 1$. In this case the definition of far-zone for $\hat{G}_y$
coincides with that of far-zone for the field of a single
particle.

However, when either or both $N_y \gg 1$ or $D_y \gg 1$ the
situation is different, and one finds that the far-zone condition
is a combination of $\hat{z}$, $N_y$, and $D_y$. In all these
cases, detailed mathematical analysis of Eq. (\ref{Gnoryany2})
shows that far-zone is reached when

\begin{eqnarray}
\max[N_y,1] \ll \hat{z}^2 \max[D_y,1] ~.\label{farzo}
\end{eqnarray}
As it will be clearer after reading Section \ref{sub:geop}, but
can also be seen considering the definition of our dimensionless
units, the physical meaning of comparisons of $N_y$ and $D_y$ with
unity in condition (\ref{farzo}) is that of a comparison between
diffraction-related parameters (diffraction angle and diffraction
size) and beam-related parameters (divergence and size of the
electron beam).

When ${N}_y \gg 1$, but $D_y \lesssim 1$ condition (\ref{farzo})
reads $\hat{z}^2 \gg N_y \gg 1$. This result is in agreement with
intuition. The far-zone condition for $\hat{G}_y$ does not
coincide with that for the field of a single electron, but it is
anyway reached far away from the source, at $\hat{z} \gg 1$. As we
will see in Section \ref{sub:nong}, the case ${N}_y \gg 1$ with
$D_y \lesssim 1$ corresponds to a quasi-homogeneous non-Gaussian
source. In Section \ref{sub:nong} we will see that in this case
the VCZ theorem is applicable, and its region of applicability is
in agreement with our definition of far-zone $\hat{z}^2 \gg N_y
\gg 1$.

When $D_y \gg 1$ (with arbitrary $N_y$) and condition
(\ref{farzo}) holds, analysis shows that the phase factor under
the integral sign in Eq. (\ref{Gnoryany2}) can be neglected.
Moreover, in this case, the Gaussian function in $\phi_y +
\bar{\theta}_y$ has a width in $\phi_y$ much larger than unity,
while the integral in $d \phi_x$ in second line of Eq.
(\ref{Gnoryany2}) has a width in $\phi_y$ of order unity, because
$\Psi$ does not depend on parameters. Therefore, the dependence in
$\phi_y$ in the Gaussian function can
always be omitted, and the Gaussian function factors out of the integral sign in $d \phi_y$. %It follows that Eq. (\ref{Gnoryany2})
%simplifies to
%
%
%\begin{eqnarray}
%&&\hat{G}_y\left(\hat{z},{\bar{y}},\Delta {y}\right) = \frac{
%\hat{z}\exp\left[{i \hat{z}\bar{\theta}_y \Delta{\theta}_y }
%\right]}{2 \pi\sqrt{ \pi D_y}} \exp\left[-\frac{N_y \Delta
%\theta_y^2 }{2}\right] \exp\left[-\frac{{\bar{\theta}_y}^2}{2 D_y
%}\right]  \int d{\phi}_y \cr && \times\int d \phi_x
%{\Psi}\left\{\hat{z},\left[{\phi}_x^2+\left(\phi_y-\frac{\Delta
%\theta_y }{2}\right)^2\right]^{1/2}\right\}
%{\Psi}^*\left\{\hat{z},\left[{\phi}_x^2+\left(\phi_y+\frac{\Delta
%\theta_y }{2}\right)^2\right]^{1/2}\right\}~. \cr &&
%\label{Gnoryany4}
%\end{eqnarray}
%%
One is left with the product of exponential functions and the
double integral

\begin{eqnarray}
\tilde{f}(\Delta \theta_y) &\equiv&  \int d{\phi}_y\int d \phi_x
\cr && \times {\Psi}\left\{\hat{z},\left[{\phi}_x^2+
\left(\phi_y-\frac{\Delta
\theta_y}{2}\right)^2\right]^{1/2}\right\}
{\Psi}^*\left\{\hat{z},\left[{\phi}_x^2+\left(\phi_y+ \frac{\Delta
\theta_y}{2}\right)^2\right]^{1/2}\right\}~, \cr &&
\label{doubleint}
\end{eqnarray}
which has a very peculiar property. In fact, it does not depend on
$\hat{z}$. The proof is based on the autocorrelation theorem in
two dimensions, and is given in detail in Appendix C of reference
\cite{OURU}. This quite remarkable property of $\tilde{f}$ carries
the consequence that substitution of $\Psi$ with $\Psi_f$ can be
performed in Eq. (\ref{doubleint}) without altering the final
result.

When both $D_y \gg 1$ and $N_y \gg 1$, one may neglect the
dependence in $\Delta \theta_y$ in functions $\Psi$ and $\Psi^*$
in Eq. (\ref{doubleint}), because the exponential function in
$\Delta \theta_y$ before the integral sign limits the range of
$\Delta \theta_y$ to values of order $1/\sqrt{N_y} \ll 1$. As a
result, the double integration in $d \phi_x $ and $d \phi_y$
yields a constant, and the description of the photon beam is
independent of $\Psi$, i.e. does not include diffraction effects.
This result is intuitive: when the electron beam size and
divergence is large compared to the diffraction size and
divergence, the photon beam can be described in terms of the
phase-space distribution of the electron beam. This approach will
be treated in more detail in Section \ref{sub:geop}.

Finally, when $D_y \gg 1$ and $N_y \lesssim 1$, diffraction
effects cannot be neglected, nor can be the dependence in $\Delta
\theta_y$ in $\Psi$ and $\Psi^*$. In this case, from condition
(\ref{farzo}) we obtain that the far-zone coincides with $D_y
\hat{z}^2 \gg 1$. This result is completely counterintuitive. In
fact, since $D_y \gg 1$, the far-zone is reached for values
$\hat{z} \sim 1$, i.e. at the very end of the undulator. Yet,
diffraction effects cannot be neglected, and the field from a
single electron is far from exhibiting a spherical wavefront at
$\hat{z} \sim 1$. This paradox is solved by the special property
of the double integral in Eq. (\ref{doubleint}), that allows one
to substitute $\Psi$ with $\Psi_f$ independently of the value of
$\hat{z}$.

As it will be discussed in Section \ref{sub:gaus} and Section
\ref{sub:nong}, the case $D_y \gg 1$ corresponds to a
quasi-homogeneous Gaussian source when $N_y \gg 1$ and to a
quasi-homogeneous non-Gaussian source when $N_y \lesssim 1$. It
will be shown that the VCZ theorem is applicable to these
situations. In particular, the applicability region of the VCZ
theorem will be seen to be in agreement with our definition of
far-zone.

Our discussion can be summarized in a single statement. The far
zone is defined, in terms of problem parameters, by condition
(\ref{farzo}). Remembering this condition one can derive the
far-zone expression for $\hat{G}_y$ simplifying Eq.
(\ref{Gnoryany2}):

\begin{eqnarray}
\hat{G}_y(\hat{z},\bar{\theta}_y,\Delta {\theta}_y) &=&
\frac{1}{\hat{z}} \frac{1}{2\pi\sqrt{\pi D_y}} {\exp{\left[i
\hat{z} \bar{\theta}_y \Delta {\theta}_y \right]}} \exp{\left[-
\frac{N_y \Delta {\theta}_y^2}{2} \right]} \cr &&\times
\int_{-\infty}^{\infty} d {\phi}_y
\exp{\left[-\frac{(\bar{\theta}_y+{\phi}_y)^2}{2 D_y}\right]}
\frac{1}{\mathcal{K}_F}\mathcal{F}\left({\phi}_y,\Delta
{\theta}_y\right) ~,\label{resu1}
\end{eqnarray}
where the universal two-dimensional
function\footnote{$\mathcal{F}$ stands for "Far".}
$\mathcal{F}(\alpha,\delta)$ is normalized in such a way that
$\mathcal{F}(0,0)=1$ and reads :

\begin{eqnarray}
\mathcal{F}(\alpha,\delta)&=& \mathcal{K}_F
{\int_{-\infty}^{\infty} d {\phi}_x \Psi_f\left\{\left[{{\phi}_x^2
+\left(\alpha-\frac{\delta}{2}\right)^2}\right]^{1/2}\right\}}
\Psi_f\left\{\left[{{\phi}_x^2
+\left(\alpha+\frac{\delta}{2}\right)^2}\right]^{1/2}\right\}~,\cr
&& \label{Motherofall}
\end{eqnarray}
with $\mathcal{K}_F={3}/({8 \sqrt{\pi}})$.

Note that, similarly to the source case, the right hand side of
Eq. (\ref{resu1}) is found by convolving an universal function,
$\mathcal{F}$, with a Gaussian function and multiplying the result
by another Gaussian function.

A plot of $\mathcal{F}$ function defined is given in Fig. \ref{M}.
$\mathcal{F}$ is a real function.  Thus,  only the geometrical
phase factor in Eq. (\ref{resu1}) prevents $\hat{G}_y$ in the
far-zone from being real. Another remarkable property of
$\mathcal{F}$ is its invariance for exchange of $\alpha$ with
$\delta/2$. Also, $\mathcal{F}$ is invariant for exchange of
$\alpha$ with $-\alpha$ (or $\delta$ with $-\delta$).

Finally, from Eq. (\ref{psi0psif}) and Eq. (\ref{mtcm}) we have

\begin{eqnarray}
{\mathcal{S}}(x,y) = \frac{\mathcal{K}_S}{4\pi^3 \mathcal{K}_F}
\int_{-\infty}^{\infty} d\alpha \int_{-\infty}^{\infty} d\delta
\exp\left[ 2i x \alpha +\frac{i y \delta}{2}\right]
\mathcal{F}(\alpha,\delta)~. \label{calM}
\end{eqnarray}

\subsection{\label{sub:cross2} Intensity distribution}

With expressions for the cross-spectral density at hand, it is now
possible to investigate the intensity distribution\footnote{Words
"intensity distribution" include some abuse of language here and
in the following. What we really calculate is the ensemble average
of the square modulus of the normalized field, $\hat{I}_x
\hat{I}_y = \langle |\hat{E}|^2\rangle$. Conversion to dimensional
units, followed by multiplication by $c/(4\pi^2)$ yields the
spectral density normalized to the electron number $N_e$.} along
the beamline, letting $\Delta \vec{r} = 0$ in the expression for
$\hat{G}$. Since ${N}_x \gg 1$ and $D_x \gg 1$, factorization of
the cross-spectral density still holds. Therefore we can
investigate the intensity profile along the vertical direction
without loss of generality.

Posing $\Delta y = 0$ in Eq. (\ref{Gnor4y}) we obtain the
intensity profile at the virtual source,
$\hat{I}_{y}\left(0,{\bar{y}}\right)$, as a function of $\bar{y}$:

\begin{eqnarray}
\hat{I}_y\left(0,{\bar{y}}\right) &=& \frac{1}{2 \mathcal{K}_S
}\sqrt{\frac{\pi}{N_y}} \int_{-\infty}^{\infty} d \phi_y
\exp\left[-\frac{\left(\phi_y+\bar{y}\right)^2}{2 N_y}\right]~
\mathcal{I}_S(\phi_y) ~, \label{Inty0}
\end{eqnarray}
where we introduced the universal function

\begin{eqnarray}
\mathcal{I}_S(\alpha)\equiv\mathcal{S}(\alpha,0) = \mathcal{K}_S
\int_{-\infty}^{\infty} d \phi_x
\Psi_0^2\left(\sqrt{\phi_x^2+\alpha^2}\right)~. \label{Bdefine}
\end{eqnarray}
A change of the integration variable: $\phi_x \longrightarrow x
\equiv (\phi_x^2+\alpha^2)^{1/2}$ allows the alternative
representation:

\begin{eqnarray}
\mathcal{I}_S(\alpha)= 2\mathcal{K}_S \int_{0}^{\infty} d x
\frac{\mathrm{rect}\left[\alpha/(2 x)
\right]}{\sqrt{1-\alpha^2/x^2}}\Psi_0^2\left(x\right)~, \label{B2}
\end{eqnarray}
where the function $\mathrm{rect}(\xi)$ is defined following
\cite{GOOD}: it is equal to unity for $|\xi|\leqslant 1/2$ and
zero otherwise.

\begin{figure}
\begin{center}
\includegraphics*[width=140mm]{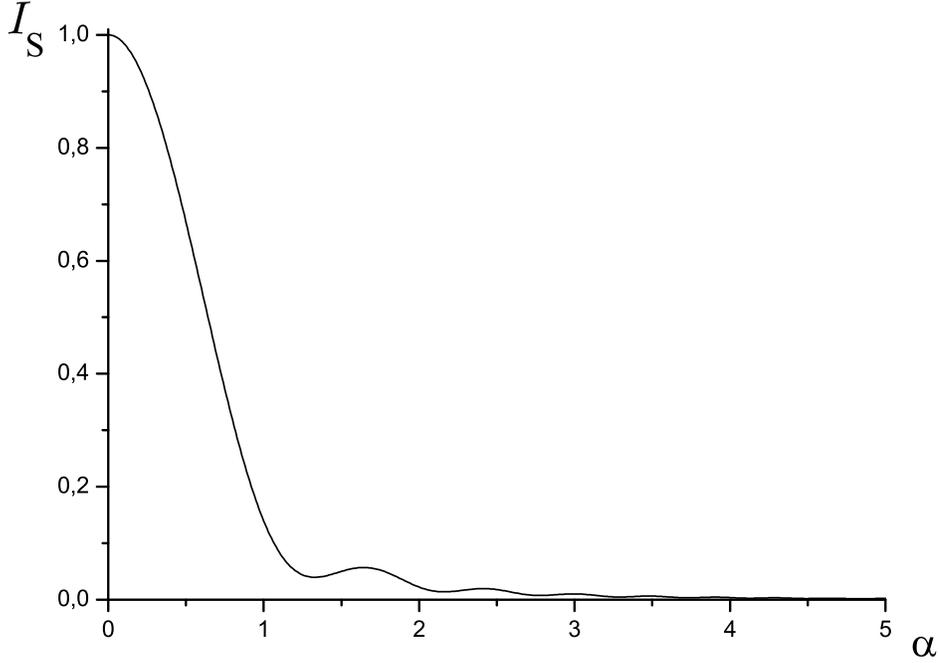}% Here is how to import EPS art
\caption{\label{ftbetaplot} The universal function
${\mathcal{I}_S}$, used to calculate intensity at the
virtual-source position.}
\end{center}
\end{figure}

The intensity at the virtual-source position is given in terms of
a convolution of a Gaussian function with the universal function
$\mathcal{I}_S$. A plot of $\mathcal{I}_S$ is given in Fig.
\ref{ftbetaplot} .

Similar derivations can be performed in the far zone. Posing
$\Delta \theta_y = 0$ in Eq. (\ref{resu1}) we obtain the
directivity diagram of the radiation as a function of
$\bar{\theta}_y$:

\begin{eqnarray}
\hat{I}_y(\hat{z},\bar{\theta}_y) &=& \frac{1}{\hat{z}}
\frac{1}{\sqrt{2 \pi D_y} \mathcal{K}_F } \int_{-\infty}^{\infty}
d {\phi}_y \exp{\left[-\frac{(\bar{\theta}_y+{\phi}_y)^2}{2
D_y}\right]} \mathcal{I}_F\left({\phi}_y\right) ~,\label{Intyf}
\end{eqnarray}
where we defined

\begin{eqnarray}
\mathcal{I}_F(\alpha) \equiv  \mathcal{F}\left(\alpha,0\right)
&=&\mathcal{K}_F {\int_{-\infty}^{\infty} d {\phi}_x
\Psi_f^2\left[\left({{\phi}_x^2 +\alpha^2}\right)^{1/2}\right]}
\cr &=&\mathcal{K}_F {\int_{-\infty}^{\infty} d {\phi}_x
\mathrm{sinc}^2\left[\frac{{{\phi}_x^2 +\alpha^2}}{4}\right]}~.
\label{Isintro}
\end{eqnarray}

\begin{figure}
\begin{center}
\includegraphics*[width=140mm]{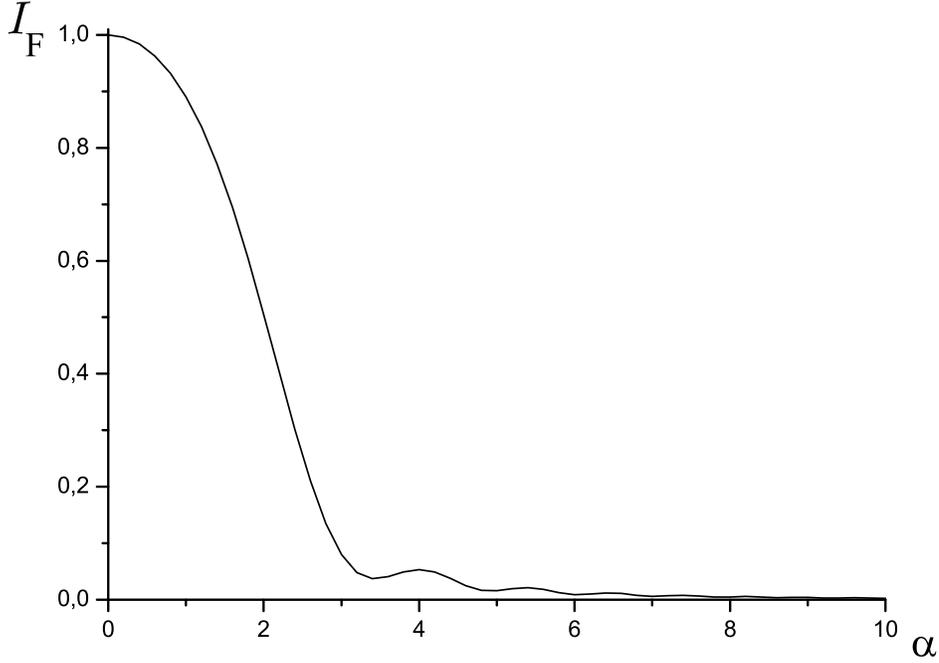}% Here is how to import EPS art
\caption{\label{ISnorm} The universal function $\mathcal{I}_F$,
used to calculate the intensity in the far zone.}
\end{center}
\end{figure}

The intensity in the far-zone is given in terms of a convolution
of a Gaussian function with the universal function
$\mathcal{I}_F$. A plot of $\mathcal{I}_F$ is given in Fig.
\ref{ISnorm}.

\subsection{\label{sub:degree} Spectral degree of coherence}

We can now present expressions for the spectral degree of
coherence at the virtual source (that is a real quantity) and in
the far-zone (that is not real). Eq. (\ref{normfine}) can be
written for the vertical direction and in normalized units as:

\begin{equation}
\hat{g}_y\left(\hat{z},{\bar{y}},\Delta y\right) =
\frac{\hat{G}_y\left(\hat{z},{\bar{y}},\Delta {y}\right)}{
\left[\hat{I}_y(\hat{z},\bar{y}+\Delta y/2)\right]^{1/2}\left[
\hat{I}_y(\hat{z},\bar{y}-\Delta y/2)\right]^{1/2}} ~.
\label{normfine2}
\end{equation}
Substitution of Eq. (\ref{Gnor4y}) and Eq. (\ref{Inty0}) in Eq.
(\ref{normfine2}) gives the spectral degree of coherence at the
virtual source:

\begin{eqnarray}
\hat{g}_y\left(0,{\bar{y}},\Delta {y}\right) &=& \exp
\left[-\frac{(\Delta y)^2 D_y }{2}\right] \int_{-\infty}^{\infty}
d \phi_y \exp\left[-\frac{\left(\phi_y+\bar{y}\right)^2}{2
N_y}\right] ~ \mathcal{S}(\phi_y,\Delta y)\cr && \times \left\{
\int_{-\infty}^{\infty} d \phi_y
\exp\left[-\frac{\left(\phi_y+\bar{y}+\Delta y/2\right)^2}{2
N_y}\right]~ \mathcal{I}_S(\phi_y)\right\}^{-1/2} \cr &&\times
\left\{ \int_{-\infty}^{\infty} d \phi_y
\exp\left[-\frac{\left(\phi_y+\bar{y}-\Delta y/2\right)^2}{2
N_y}\right]~ \mathcal{I}_S(\phi_y)\right\}^{-1/2} ~. \label{g0}
\end{eqnarray}
Similarly, substitution of Eq. (\ref{resu1}) and Eq. (\ref{Intyf})
in Eq. (\ref{normfine2}) gives the spectral degree of coherence in
the far zone:

\begin{eqnarray}
\hat{g}_y(\hat{z},\bar{\theta}_y,\Delta {\theta}_y) &=&
{\exp{\left[i  \hat{z} \bar{\theta}_y \Delta {\theta}_y \right]}}
\exp{\left[-\frac{  (\Delta {\theta}_y)^2 N_y}{2} \right]}\cr &&
\times \int_{-\infty}^{\infty} d {\phi}_y
\exp{\left[-\frac{(\bar{\theta}_y+{\phi}_y)^2}{2 D_y}\right]}
\mathcal{F}\left({\phi}_y,\Delta {\theta}_y\right)\cr && \times
\left\{\int_{-\infty}^{\infty} d {\phi}_y
\exp{\left[-\frac{(\bar{\theta}_y+\Delta \theta_y/2+
{\phi}_y)^2}{2 D_y}\right]}
\mathcal{I}_F\left({\phi}_y\right)\right\}^{-1/2}\cr && \times
\left\{\int_{-\infty}^{\infty} d {\phi}_y
\exp{\left[-\frac{(\bar{\theta}_y-\Delta \theta_y/2+
{\phi}_y)^2}{2 D_y}\right]}
\mathcal{I}_F\left({\phi}_y\right)\right\}^{-1/2}~.\label{gfar}
\end{eqnarray}

\subsection{\label{sub:limit} Influence of horizontal emittance on
vertical coherence for $D_y \ll 1$ and $N_y \ll 1$}

The theory developed up to now is valid for arbitrary values of
$N_y$ and $D_y$. In the present Section \ref{sub:limit} we discuss
an application in the limiting case for ${D}_y \ll 1$ and ${N}_y
\ll 1$ corresponding to third generation light sources operating
in  the soft X-ray range.

\begin{figure}
\begin{center}
\includegraphics*[width=140mm]{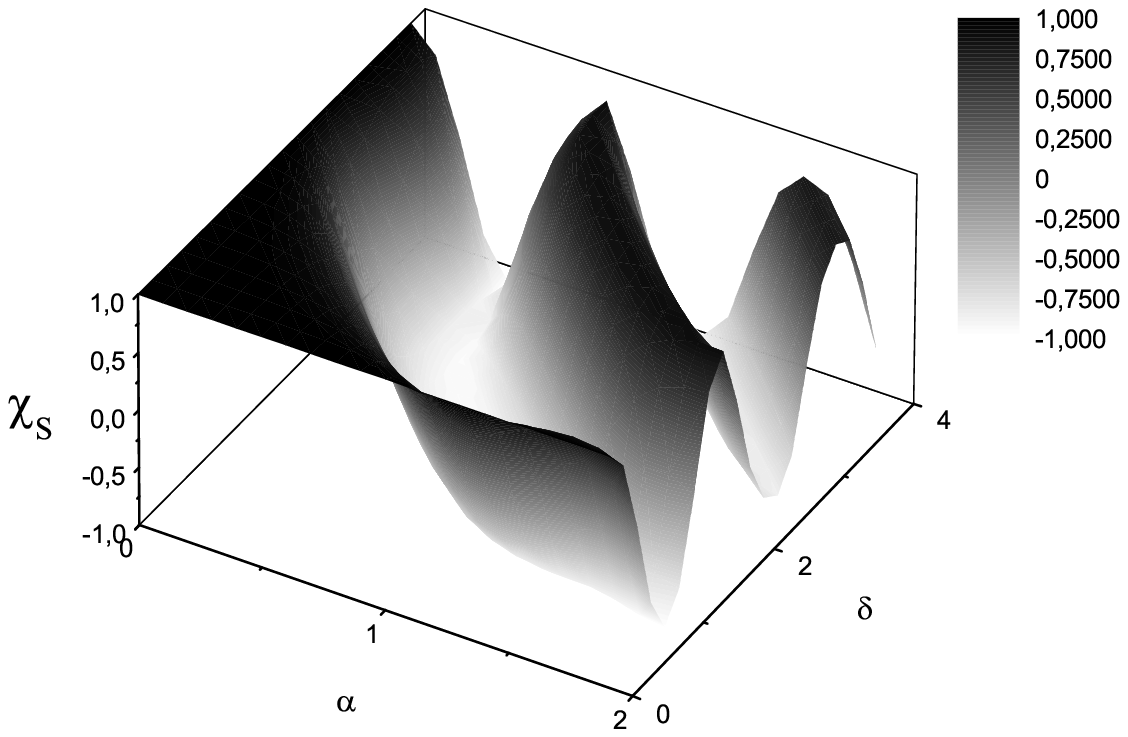}% Here is how to import EPS art
\caption{\label{ftchi} Plot of the universal function
$\mathcal{X}_S$, used to calculate the modulus of the spectral
degree of coherence of the source when $\hat{N}_x \gg1$,
$\hat{D}_x \gg 1$, $\hat{N}_y \ll 1$ and $\hat{D}_y \ll 1$.}
\end{center}
\end{figure}

At the virtual source position, the following simplified
expression for the spectral degree of coherence in the vertical
direction is derived from Eq. (\ref{g0}):

\begin{eqnarray}
\hat{g}_y\left(0,{\bar{y}},\Delta {y}\right) =
\mathcal{X}_S({\bar{y}},\Delta {y}) ~, \label{g0soft}
\end{eqnarray}
where

\begin{equation}
\mathcal{X}_S(\alpha,\delta) = \frac{{\mathcal{S}}(\alpha,\delta)}
{\left[{\mathcal{I}_S}(\alpha-\delta/2)\right]^{1/2}
\left[{\mathcal{I}_S}(\alpha+\delta/2)\right]^{1/2}}~.
\label{mathcalchi}
\end{equation}
Thus, $\hat{g}_y$ is given by the universal function
$\mathcal{X}_S$. A plot of $\mathcal{X}_S$ is given in Fig.
\ref{ftchi}.

\begin{figure}
\begin{center}
\includegraphics*[width=140mm]{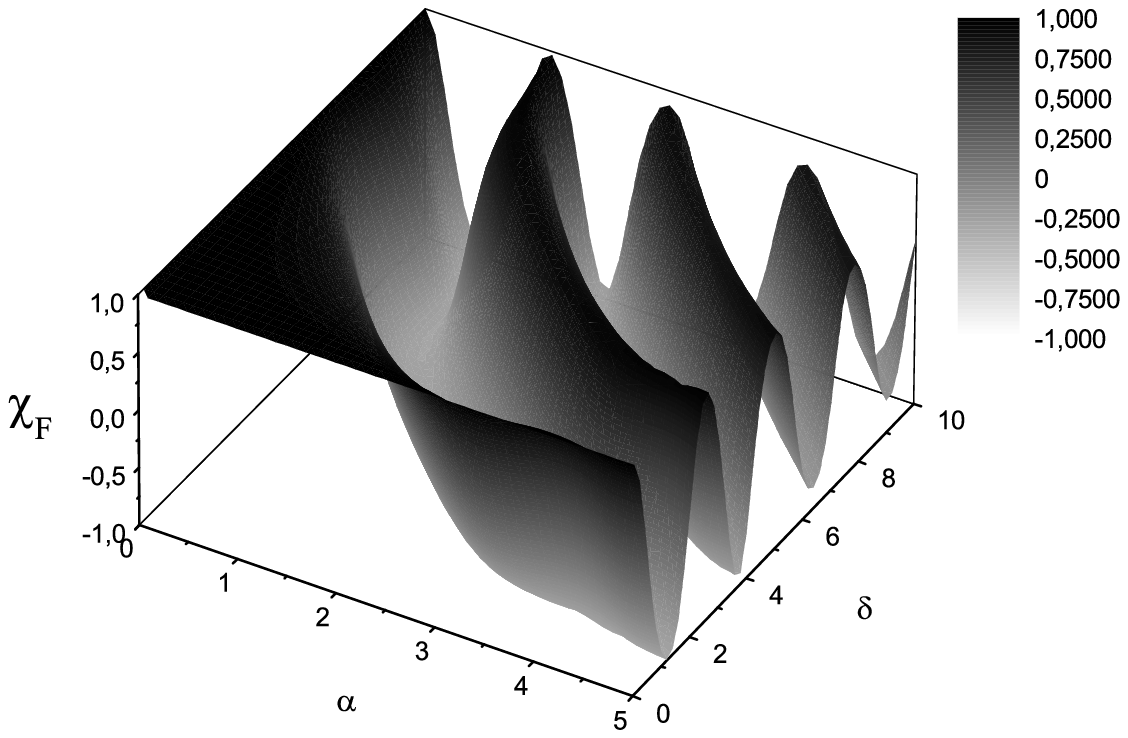}% Here is how to import EPS art
\caption{\label{chi} Plot of the universal function
$\mathcal{X}_F$, used to calculate the modulus of the spectral
degree of coherence in the far zone when $\hat{N}_x \gg1$,
$\hat{D}_x \gg 1$, $\hat{N}_y \ll 1$ and $\hat{D}_y \ll 1$.}
\end{center}
\end{figure}

Similarly, in  the far zone, one obtains from Eq. (\ref{gfar}):

\begin{eqnarray}
\hat{g}_y(\hat{z},\bar{\theta}_y,\Delta {\theta}_y) &=&
{\exp{\left[i \hat{z} \bar{\theta}_y \Delta {\theta}_y \right]}}
\mathcal{X}_F({\bar{\theta}_y},\Delta \theta_{y}) ~.
\label{gfarsoft}
\end{eqnarray}
Here the universal function $\mathcal{X}_F$ is given by

\begin{equation}
\mathcal{X}_F{\left(\alpha,\delta\right)} =
\frac{\mathcal{F}(\alpha,\delta)}{\left[\mathcal{I}_F(\alpha-\delta/2)\right]^{1/2}
\left[\mathcal{I}_F(\alpha+\delta/2)\right]^{1/2}}~.
\label{chidef}
\end{equation}
Thus, $|\hat{g}_y|$ is given by the universal function
$\mathcal{X}_F$. A plot of $\mathcal{X}_F$ is presented in Fig.
\ref{chi} as a function of dummy variable $\alpha$ and $\delta$.
It is straightforward to see that
$\mathcal{X}_F({\bar{\theta}_y},\Delta \theta_{y})$ is symmetric
with respect to $\Delta {\theta}_y$ and with respect to the
exchange of $\Delta {\theta}_y/2$  with $\bar{\theta}_y$. When
$\bar{\theta}_y =0$, i.e. $\hat{\theta}_{y1}= -\hat{\theta}_{y2}$,
we obviously obtain $\mathcal{X}_F(0,\Delta {\theta}_y)=1$ that
corresponds to complete coherence at this particular value of
$\bar{\theta}_y$. However, since $|\hat{g}_y| = \mathcal{X}_F$
oscillates from positive to negative values, in general one never
has full coherence in the vertical direction, even in the case of
zero vertical emittance. Note that this effect does not depend, in
the limit for ${N}_x \gg 1$ and ${D}_x \gg 1$, on the actual
values of ${N}_x$ and ${D}_x$.

\begin{figure}
\begin{center}
\includegraphics*[width=140mm]{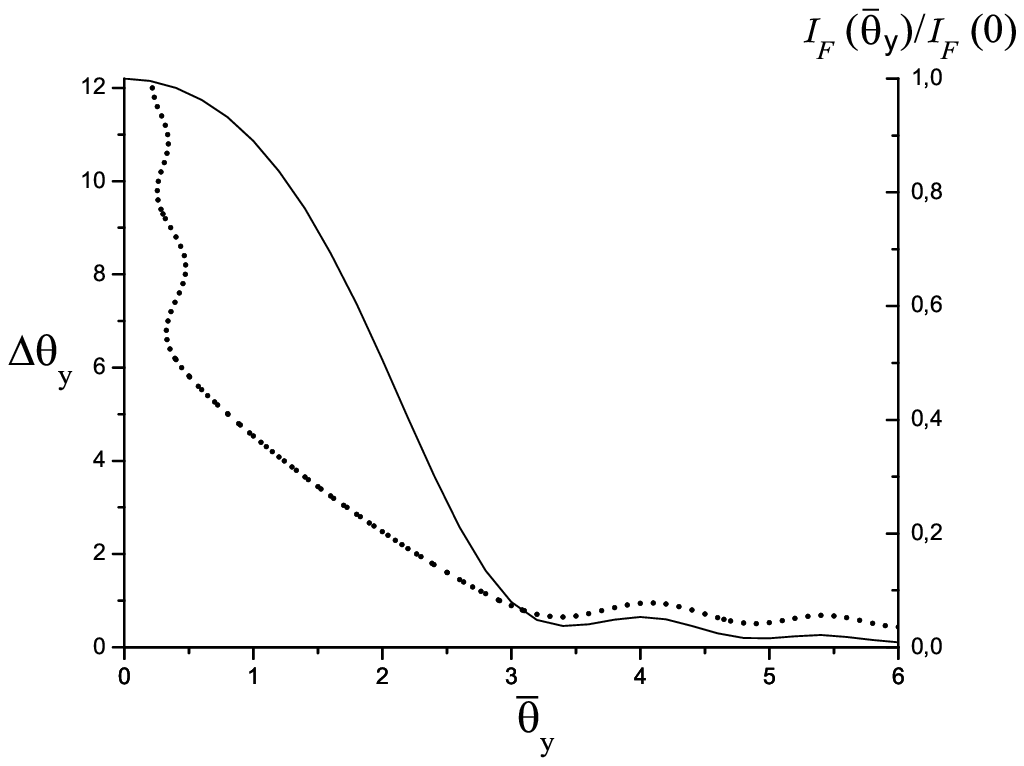}% Here is how to import EPS art
\caption{\label{plot2Dcomp} Comparison between some zeros of
$\mathcal{X}_F$, at coordinates  $(\bar{\theta}_y,\Delta
{\theta}_{y}$ (black circles), and the directivity diagram of
undulator radiation in the vertical direction at very large
horizontal electron beam divergence $\hat{D}_x \gg 1$ and
negligible vertical divergence $\hat{D}_y \ll 1$ (solid line).}
\end{center}
\end{figure}
In other words, as it is evident from Fig. \ref{chi},
$\mathcal{X}_F({\bar{\theta}_y},\Delta \theta_{y})$ exhibits many
different zeros in $\bar{\theta}_y$ for any fixed value of $\Delta
{\theta}_y$. In Fig. \ref{plot2Dcomp} some of these zeros are
illustrated with black circles on the plane
$(\bar{\theta}_y,\Delta {\theta}_{y})$. Consider a two-pinhole
experiment as in Fig. \ref{ed12c}. Once a certain distance
$\hat{z} \Delta {\theta}_y$ between the two pinholes is fixed,
Fig. \ref{plot2Dcomp} illustrates at what position of the pinhole
system, $\bar{\theta}_{y}$, the spectral degree of coherence in
the vertical direction drops from unity to zero for the first
time.

To estimate the importance of this effect, it is crucial to
consider the position of $\bar{\theta}_{y}$ in the directivity
diagram of the radiant intensity, that coincides in this
case\footnote{In other words, it can be shown that $\mathcal{I}_F$
is the directivity diagram corresponding to the case $N_x\gg1$,
$D_x \gg1$, $N_y \ll 1$, $D_y \ll 1$.} with
$\mathcal{I}_F(\bar{\theta}_y)$  (solid line in Fig.
\ref{plot2Dcomp}).
From Fig. \ref{plot2Dcomp} one can see that
$\mathcal{X}_F({\bar{\theta}_y},\Delta \theta_{y})$ drops to zero
for the first time at $\Delta {\theta}_y \sim 2$ $\bar{\theta}_y
\sim 2$, where the X-ray flux is still intense. This behavior of
the degree of coherence may influence particular kind of
experiments. To give an example we go back to the two-pinhole
setup in Fig. \ref{ed12c}. After spatial filtering in the
horizontal direction, one will find that for some vertical
position $\bar{\theta}_y$ of the pinholes (at fixed $\Delta
{\theta}_y$), well within the radiation pattern diagram, there
will be no fringes, while for some other vertical position there
will be perfect visibility. Without the knowledge of the function
$\mathcal{X}_F$ it would not be possible to fully predict the
outcomes of a two-pinhole experiment. Results described here
should be considered as an illustration of our general theory that
may or may not, depending on the case under study, have practical
influence. It may, however, be the subject of experimental
verification.

\subsubsection{Discussion}

Although the $\mathcal{X}_F$ is independent of $N_x$ and $D_x$,
its actual shape is determined by the presence of a large
horizontal emittance. Let us show this fact. If both $N_{x,y} \ll
1$ and $D_{x,y}\ll 1$ (filament beam limit), one would have had
$|\hat{G}| = \Psi_f\left(\left|\vec{\theta}_1\right|\right)
\Psi_f\left(\left|\vec{\theta_2}\right|\right)$, so that
$|\hat{g}|=1$, strictly. Note that in this case $\hat{G}$ could
not be factorized. When instead $N_x \gg 1$ and $D_x \gg 1$ the
cross-spectral density can be factorized according to Eq.
(\ref{factorizeee}), but the integration in $d\phi_x$, that
follows from an integration over the horizontal electron beam
distribution, is still included in the vertical cross-spectral
density $\hat{G}_x$ (and, therefore, in $\mathcal{X}_F$). This
results in the outcome described above and can be traced back to
the non-Gaussian nature of $\Psi_f$. Note that if one adopted a
Gaussian-Schell model, the cross-spectral density could have been
split in the product of Gaussian intensity and Gaussian spectral
degree of coherence and $\Psi_f$, being a product of Gaussians,
would have been separable. Then, $|\hat{g}_y|$ would have been
constant for $N_x\gg 1$ and $D_x \gg 1$. As a result, the effect
described here would not have been recognized. This fact
constitutes a particular realization of our general remarks about
Gaussian-Schell models at the end of Section \ref{sub:evol}.

As a final note to the entire Section, we stress the fact that our
theory of partial coherence in third generation light sources is
valid under several non-restrictive assumptions. Alongside with
previously discussed conditions $\gamma^2 \gg 1$, $N_w \gg 1$,
$\sigma_T \omega \gg N_w$ and the assumption of perfect resonance
(i.e. the limit $\hat{C} \ll 1$), we assumed separability and
particular shape of the electron beam phase space (see Eq.
(\ref{distr})). Moreover, for third generation light sources we
assumed $\epsilon_x \gg \lambdabar$ (up to the VUV range).
Together with $\beta_{x,y} \simeq L_w$ this implies $N_x \gg 1$
and $D_x \gg 1$, allowing for separability of the cross-spectral
density in horizontal and vertical factors. Moreover, due to $N_x
\gg 1$ and $D_x \gg 1$, we are dealing with quasi-homogenous
Gaussian sources in the horizontal direction. This particular kind
of sources will be treated in detail as an asymptote of our
general theory in the next Section. In the vertical direction
instead, we still have fully generic sources. We showed how the
vertical cross-spectral density can be expressed in terms of
convolutions between two-dimensional universal functions and
Gaussian functions. A particularly interesting case is that of
quasi-homogeneous non-Gaussian sources that will also be treated
as an asymptotic case in the following Section dedicated to
quasi-homogeneous sources.

\section{\label{sec:quas} Quasi-homogeneous asymptotes for
undulator sources}

In Section \ref{sec:main} we developed a general theory of
transverse coherence properties of third generation light sources.
In this Section we consider the class of quasi-homogeneous sources
for undulator devices as an asymptotic limit for that theory.

Quasi-homogeneous sources are defined by the fact that the
cross-spectral density of the virtual source (positioned at $z=0$)
can be written as:

\begin{eqnarray}
\hat{G}\left(0,\vec{\bar{r}},\Delta \vec{r}\right) =
\hat{I}\left(0,\vec{\bar{r}}\right) \hat{g}\left(\Delta
\vec{r}\right) ~.\label{introhdopo}
\end{eqnarray}
The definition of quasi-homogeneity amounts to a factorization of
the cross-spectral density as the product of the field intensity
distribution and the spectral degree of coherence. A set of
necessary and sufficient conditions for such factorization to be
possible follows: (i) the radiation intensity at the virtual
source varies very slowly with the position across the source on
the scale of the field correlation length and (ii) the spectral
degree of coherence depends on the positions across the source
only through the difference $\Delta \vec{r}$.

Factorization of Eq. (\ref{Gnor3}) as in Eq. (\ref{introhdopo}),
for third generation light sources, is equivalent to a particular
choice of the region of parameters for the electron beam: ${N}_x
\gg 1$, ${D}_x \gg 1$ and either (or both) ${N}_y \gg 1$ and
${D}_y \gg 1$\footnote{These conditions describe the totality of
third generation quasi-homogeneous sources. In fact, while a
purely mathematical analysis indicates that factorization of Eq.
(\ref{Gnor3}) is equivalent to more generic conditions (${N}_x
\gg1$ and ${N}_y \gg 1$, or ${D}_x \gg1$ and ${D}_y \gg 1$),
comparison with third generation source parameters ($N_x \gg 1$
and $D_x \gg 1$) reduces such conditions to the already mentioned
ones.}. In this case, the reader may verify that conditions (i)
and (ii) are satisfied.

In the horizontal direction, we have both $N_x \gg 1$ and $D_x \gg
1$ for wavelengths up to the VUV range, so that factorization of
the cross-spectral density in horizontal and vertical
contributions $\hat{G}_x$ and $\hat{G}_y$ always holds.

Let us first consider $\hat{G}_y$. Depending on the values of
$N_y$ and $D_y$ we may have Gaussian quasi-homogeneous sources
characterized by a Gaussian transverse distribution of intensity
($N_y \gg 1$ and $D_y \gg 1$), as well as non-Gaussian
quasi-homogeneous sources ($N_y \gg 1$ and $D_y \lesssim 1$ or
$N_y \lesssim 1$ and $D_y \gg 1$). Gaussian quasi-homogenous
sources are to be expected in the vertical direction in the hard
X-ray limit, where diffraction effects play no role. On the
contrary, diffraction effects must be accounted for when dealing
with non-Gaussian quasi-homogeneous sources.

Let us now consider $\hat{G}_x$. In the horizontal direction both
$N_x \gg 1$ and $D_x \gg 1$. It follows that Gaussian
quasi-homogeneous sources find a very natural application in the
description of the cross-spectral density in the horizontal
direction, from the hard X-ray to the VUV range.

We will see that the VCZ theorem applies to all quasi-homogeneous
cases. Actually, the concept of far-zone for quasi-homogeneous
sources can be introduced as the region in the parameter space
$\hat{z}, N_{x,y}, D_{x,y}$ such that the VCZ theorem holds. We
will see that these condition coincides with condition
(\ref{farzo}) given before.

\subsection{\label{sub:gaus} Gaussian undulator sources}

When $N_x \gg 1$ and $D_x \gg 1$ Eq. (\ref{Gnor3x}) applies. When
$N_y \gg 1$ and $D_y \gg 1$, Eq. (\ref{Gnor3y}) reduces to

\begin{eqnarray}
\hat{G}_y\left(0,{\bar{y}},\Delta {y}\right) &=&
\sqrt{\frac{\pi}{{N_y}}} \exp \left[-\frac{(\Delta y)^2D_y
}{2}\right] \exp\left[-\frac{\bar{y}^2}{2 N_y}\right]
~,\label{Gnorgauss0}
\end{eqnarray}
that is equivalent to Eq. (\ref{Gnor3x}): thus, identical
treatments hold separately in the horizontal and vertical
directions. For this reasons, and for simplicity of notation, we
drop all subscripts "x" or "y" in the present Section
\ref{sub:gaus}, and we substitute letters "x" and "y" in variables
with the more generic "r". However, as stated before, the Gaussian
quasi-homogeneous model primarily describes third generation light
sources in the horizontal direction.

Eq. (\ref{introhdopo}) is satisfied. Moreover,

\begin{eqnarray}
\hat{I}\left(0,{\bar{r}}\right)=\sqrt{\frac{\pi}{{N}}}
\exp\left[-\frac{\bar{r}^2}{2 N}\right] \label{integauss}
\end{eqnarray}
and

\begin{eqnarray}
\hat{g}(\Delta r) = \exp \left[-\frac{(\Delta r)^2D }{2}\right]
~.\label{degregauss0}
\end{eqnarray}
Propagation of Eq. (\ref{Gnorgauss0}) can be found taking the
limit of Eq. (\ref{Gnoryany2}) for $N\gg 1$ and $D\gg 1$, which
yields an analytical expression for the evolution of the
cross-spectral density based on Eq. (\ref{Cany2}) at $\hat{C} \ll
1$ :

\begin{eqnarray}
\hat{G}(\hat{z},\bar{r},\Delta r) &=& \frac{\sqrt{\pi}}{{\hat{z}
\sqrt{A+D}}} \exp\left[-\frac{\bar{r}^{2}}{2({A}+{D})
\hat{z}^2}\right] \exp\left[ i \frac{\bar{r}\Delta
{r}}{\hat{z}}\right]\cr &&\times \exp\left[-  i
\frac{{A}\bar{r}\Delta {r}}{\hat{z} ({A}+{D})}\right]
\exp\left[-\frac{{A} {D} (\Delta r)^{2}}{2({A}+{D}) }\right]
~,\label{Ggaussz}
\end{eqnarray}
where

\begin{equation}
{A} = \frac{{N}}{\hat{z}^2} ~.\label{a2set}
\end{equation}
We have

\begin{eqnarray}
\hat{I}(\hat{z},\bar{r}) &=& \frac{\sqrt{\pi}}{{\hat{z}
\sqrt{A+D}}} \exp\left[-\frac{\bar{r}^{2}}{2({A}+{D})
\hat{z}^2}\right] ~,\label{Igaussz}
\end{eqnarray}
and

\begin{eqnarray}
\hat{g}(\hat{z},\Delta r) &=& \exp\left[i \frac{\bar{r}\Delta
{r}}{\hat{z}}\right] \exp\left[-  i \frac{{A}\bar{r}\Delta
{r}}{\hat{z} ({A}+{D})}\right] \exp\left[-\frac{{A} {D} (\Delta
r)^{2}}{2({A}+{D}) }\right] ~.\label{degregaussz}
\end{eqnarray}
Note that, due to the phase factors in Eq. (\ref{Ggaussz}), only
the virtual source at $\hat{z}=0$ constitutes a quasi-homogeneous
source.

Geometrical interpretation of ${A}$ is the dimensionless square of
the apparent angular size of the electron bunch at the observer
point position, calculated as if the beam was positioned at
$\hat{z}=0$.

The far-zone for quasi-homogeneous Gaussian sources is given by
condition $A \ll D$, which can be retrieved by condition
(\ref{farzo}) or directly by Eq. (\ref{Ggaussz}). In this case,
simplification of Eq. (\ref{Ggaussz}) or use of Eq. (\ref{resu1})
in the limit for $N\gg 1$ and $D \gg 1$ yields the far-zone
cross-spectral density:

\begin{eqnarray}
\hat{G}\left(\hat{z},{\bar{\theta}},\Delta {\theta}\right) &=&
\frac{1}{\hat{z}} \sqrt{\frac{\pi}{D}} \exp\left[i \hat{z}
{\bar{\theta}} \Delta {\theta}\right] \exp\left[-\frac{N (\Delta
\theta)^2}{2}\right] \exp\left[-\frac{\bar{\theta}^2}{2 D}\right]
~,\cr &&\label{Gnorgaussf}
\end{eqnarray}
so that

\begin{eqnarray}
\hat{I}\left(\hat{z},{\bar{\theta}}\right) &=& \frac{1}{\hat{z}}
\sqrt{\frac{\pi}{D}} \exp\left[-\frac{\bar{\theta}^2}{2 D}\right]
~,\label{Integaussf}
\end{eqnarray}
and

\begin{eqnarray}
\hat{g}\left(\hat{z},\Delta {\theta}\right) &=&  \exp\left[i
\hat{z} {\bar{\theta}}  \Delta {\theta}\right] \exp\left[-\frac{N
(\Delta \theta)^2}{2}\right] ~.\label{degregaussf}
\end{eqnarray}
Similarly as before we suppressed subscripts "x" or "y" in the
symbol $\theta$.

Analysis of Eq. (\ref{Gnorgauss0}) and Eq. (\ref{Gnorgaussf})
allows to conclude that

$(a_2)$ the spectral degree of coherence of the field at the
source plane $g({\Delta {r}})$ and the angular distribution of the
radiant intensity $I({\bar{\theta}})$ are a Fourier pair.

$(b_2)$ the spectral degree of coherence of the far field
$g(\Delta {\theta})$ and the source-intensity distribution
$I({\bar{r}})$ are, apart for a simple geometrical phase factor, a
Fourier pair.

The statement $(b_2)$ is a version of the VCZ theorem valid for
quasi-homogeneous sources.  Statement $(a_2)$ instead, regards the
symmetry between space and angle domains, and can be seen as an
inverse VCZ theorem.

This discussion underlines the link between the VCZ theorem and
the Wiener-Khinchin theorem. Exactly as the space domain has a
reciprocal description in terms of transverse (two-dimensional)
wave vectors, the time domain has a reciprocal description in
terms of frequency. The reader will recognize the analogy between
statements $(a_2)$ and $(b_2)$, with statements $(a_1)$ and
$(b_1)$ discussed in Section \ref{sub:temp}. In particular, the
VCZ is analogous to the inverse Wiener-Khincin theorem. Similarly,
separability of $G$ in Eq. (\ref{introhdopo}) in the product of
spectral degree of coherence and intensity is analogous to
separability of $\Gamma_\omega$, in Eq. (\ref{gamma6prima}) in the
product of spectral correlation function and spectral density
distribution of the source.

Let us calculate the transverse coherence length $\hat{\xi}_c$ as
a function of the observation distance $\hat{z}$. We introduce the
coherence length following the definition by Mandel \cite{WOLF}.
The coherence length, naturally normalized to the diffraction
length $\sqrt{L_w c/\omega}$ is defined as

\begin{equation}
\hat{\xi}_c(\hat{z}) =  \int_{-\infty}^{\infty} |g(\hat{z},\Delta
r )|^2 d(\Delta r)~, \label{cohlen}
\end{equation}
Performing the integration in Eq. (\ref{cohlen}) with the help of
Eq. (\ref{degregaussz}) yields:

\begin{equation}
\hat{\xi}_c(\hat{z}) =  {\sqrt{\pi}}
\left(\frac{1}{{A}}+\frac{1}{{D}}\right)^{1/2}~.\label{cohlen2}
\end{equation}
\begin{figure}
\begin{center}
\includegraphics*[width=140mm]{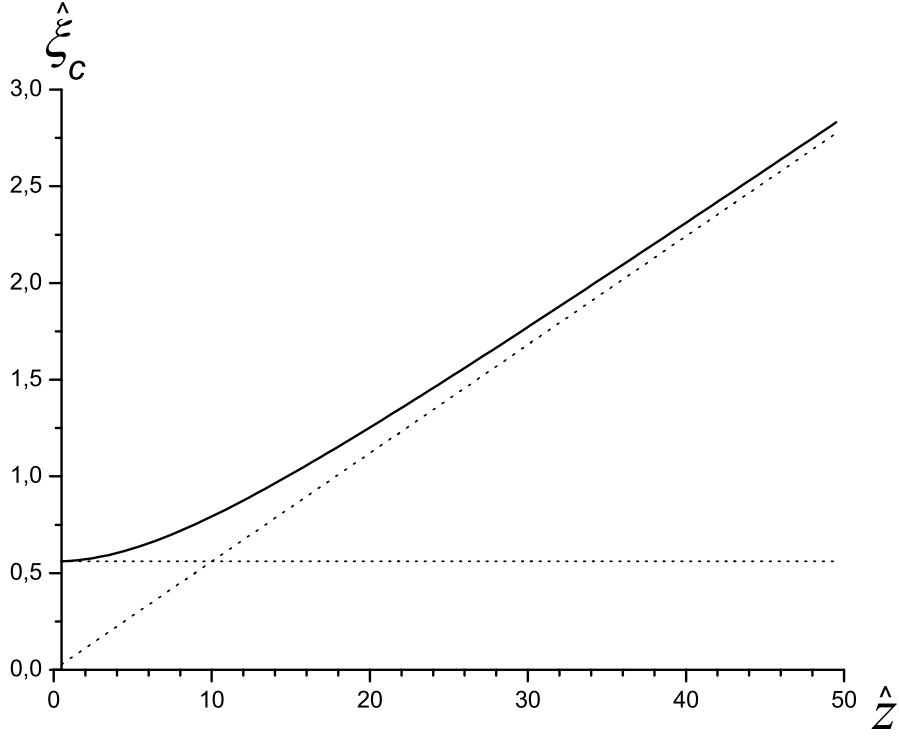}% Here is how to import EPS art
\caption{\label{uno} Coherence length $\hat{\xi}_c$ as a function
of $\hat{z}$ and asymptotic behaviors for $\hat{z} \longrightarrow
1/2$ and $\hat{z} \gg 1$. Here ${N} = 10^3$ and ${D} = 10$.}
\end{center}
\end{figure}

\begin{figure}
\begin{center}
\includegraphics*[width=140mm]{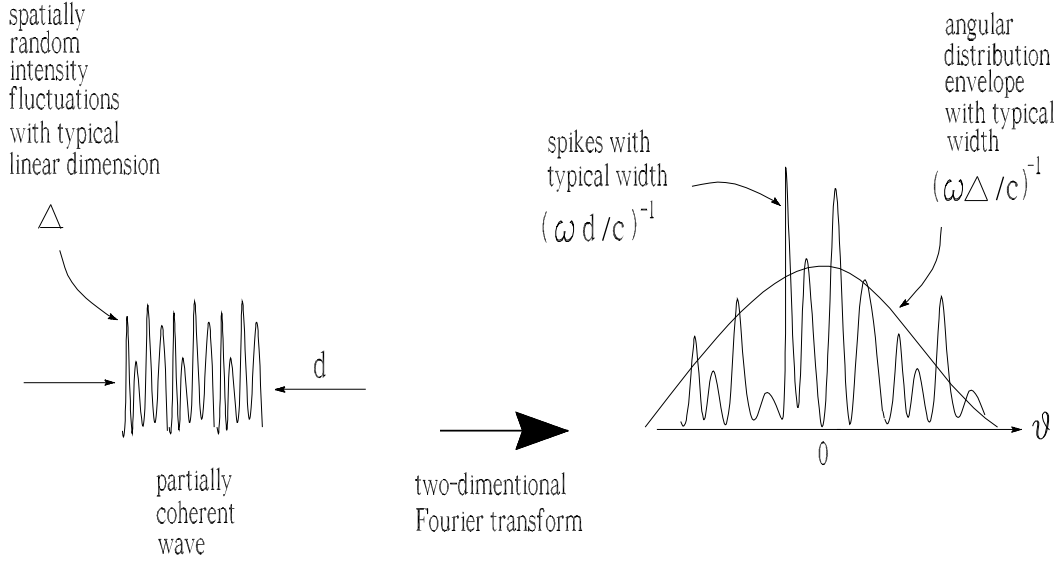}% Here is how to import EPS art
\caption{\label{spikesp} Physical interpretation of the
generalized VCZ theorem. If the radiation beyond the source plane
is partially coherent, a spiky angular distribution of intensity
is expected. The nature of these spikes is easily described in
Fourier transform notations. We can expect that typical width of
the angular distribution of intensity should be of order $(\omega
\Delta/c)^{-1}$, where $\Delta$ is the typical linear dimension of
spatially random intensity fluctuations. If the source has
transverse size $d$, the angular distribution of intensity should
contain spikes with typical width of about $(\omega d/c)^{-1}$, a
consequence of the reciprocal width relations of Fourier transform
pairs.}
\end{center}
\end{figure}

The coherence length in Eq. (\ref{cohlen2}) exhibits linear
dependence on $\hat{z}$, that is $\hat{\xi}_c \longrightarrow
\sqrt{\pi/ {N}}~{\hat{z}}$ while for $\hat{z} \longrightarrow 1/2$
that is at the end of the undulator, it converges to a constant
$\hat{\xi}_c \longrightarrow [\pi/(4{N})+\pi/{D}]^{1/2}$. Eq.
(\ref{cohlen2}) and its asymptotes are presented in Fig. \ref{uno}
for the case ${N} = 10^3$, ${D} = 10$. At the exit of the
undulator, $\hat{\xi}_c \sim 1/\sqrt{{D}}$, because ${N} \gg {D}$.
On the other hand, horizontal dimension of the light spot is
simply proportional to $\sqrt{{N}}$. This means that the
horizontal dimension of the light spot is determined by the
electron beam size, as is intuitive, while the beam angular
distribution is printed in the fine structures of the intensity
function that are of the dimension of the coherence length.  In
the limit for $A \ll D$ the situation is reversed. The radiation
field at the source can be presented as a superposition of plane
waves, all at the same frequency $\omega$, but with different
propagation angles with respect to the $z$-direction. Since the
radiation at the exit of the undulator is partially coherent, a
spiky angular distribution of intensity is to be expected. The
nature of the spikes is easily described in terms of Fourier
transform theory. From Fourier transform theorem  we can expect an
angular spectrum with Gaussian envelope and rms width
$\sqrt{{D}}$. Also, the angular distribution of intensity should
contain spikes with characteristic width $1/\sqrt{N}$, as a
consequence of the reciprocal width relations of Fourier transform
pairs (see Fig. \ref{spikesp}). This is realized in mathematics by
the expression for the cross-spectral density, Eq. (\ref{Ggaussz})
and by the equation for the coherence length, Eq. (\ref{cohlen2}).

It is also important to remark that the asymptotic behavior for
${A} \ll 1$ of $\hat{g}$, that is Eq. (\ref{degregaussf}) and
$\hat{\xi}_c$, that is

\begin{equation}
\hat{\xi}_c \longrightarrow \sqrt{\frac{\pi}{
N}}{\hat{z}}\label{cohlen2vcz}
\end{equation}
are direct application of VCZ theorem.  In fact, the last
exponential factor on the right hand side of Eq. (\ref{Ggaussz})
is simply linked with the Fourier transform of
$f_{{l}}\left(\hat{l}\right)$. We derived Eq. (\ref{Ggaussz}) for
${N} \gg 1$ and ${D} \gg 1$, with $ A \ll D$: in non-normalized
units these conditions mean that the VCZ theorem is applicable
when the electron beam divergence is much larger than the
diffraction angle, i.e. $\sigma'^2 \gg \lambda /(2\pi L_w)$, the
electron beam dimensions are much larger than the diffraction
size\footnote{We do not agree with statement in \cite{TAKA}: "the
electron-beam divergence must be much smaller than the photon
divergence" for the VCZ theorem to apply. This would imply
$\sigma' \ll \sqrt{\lambda /(2\pi L_w)}$ (reference \cite{TAKA},
page 571, Eq. (57)).}, i.e. $\sigma^2 \gg \lambda L_w/2\pi $, and
$({\sigma'} {z})^2 \gg {\sigma}^2$.

In \cite{GOOD} (paragraph 5.6.4) a rule of thumb is given for the
applicability region of the generalization of the VCZ theorem to
quasi-homogeneous sources. The rule of thumb requires $z > 2 d
\Delta/\lambda$ where $d$ is "the maximum linear dimension of the
source", that is the diameter of a source with uniform intensity
and $\Delta$ "represents the maximum linear dimension of a
coherence area of the source". In our case, since $\sigma$ is the
rms source dimension, $d\simeq 2\sigma$. Moreover, from Eq.
(\ref{cohlen2}) we have $\Delta = \xi_c \simeq \lambda/(2
\sqrt{\pi} \sigma')$. The rule of thumb then requires $z >2
\sigma/( \sqrt{\pi} \sigma')$: in dimensionless form this reads $
\hat{z} \gtrsim \sqrt{{N}/{D}}$. This is parametrically in
agreement with our limiting condition $A \ll D$, even though these
two conditions are different when it come to actual estimations:
our condition $A \ll D$ is, in fact, only an asymptotic one. To
see how well it works in reality we might consider the plot in
Fig. \ref{uno}. There ${N} = 10^3$ and ${D} = 10$. Following
\cite{GOOD} we may conclude that a good condition for the
applicability of the VCZ theorem should be $\hat{z} \gtrsim 10$.
However as it is seen from the figure, the linear asymptotic
behavior is not yet a good approximation at $\hat{z} \simeq 10$.
This may be ascribed to the fact that the derivation in
\cite{GOOD} is not generally valid, but has been carried out for
sources which drop to zero very rapidly outside the maximum linear
dimension $d$ and whose correlation function also drops rapidly to
zero very rapidly outside maximum linear dimension $\Delta$.
However, at least parametrically, the applicability of the VCZ
theorem in the asymptotic limit $A \ll D$ can be also expected
from the condition $z > 2 d \Delta/\lambda$ in \cite{GOOD}.

We conclude that the far field limit ${A} \ll {D}$ corresponds
with the applicability region of the VCZ theorem. When this is the
case, the VCZ theorem applies and the modulus of the spectral
degree of coherence in the far field, forms a Fourier pair with
the intensity distribution of the virtual source.  In particular
one concludes that the rms width of the virtual source is
$\sqrt{N}$. In our study case for $N \gg 1$ and $D \gg 1$, such a
relation between the rms width of the spectral degree of coherence
in the far field and the rms dimension of the virtual source is
also a relation between the rms width of the cross-spectral
density function in the far field and the rms dimension of the
electron beam at the plane of minimal beta function in the center
of the undulator. In dimensional units one can write the value
$\sigma_c$ of the rms width of the spectral degree of coherence
$\hat{g}(\Delta \vec{r})$ in the far field as

\begin{equation}
\sigma_c = \frac{\lambda z}{2\pi \sigma} ~.\label{eqdimboh}
\end{equation}
Here $\sigma$ is, as usual, the rms dimension of the electron
beam. These few last remarks help to clarify what is the size of
the source in the VCZ theorem, that is far from being a trivial
question. In several papers \cite{YAB1,PFEI} the  rms electron
beam size is recovered from the measurement of the transverse
coherence length and subsequent application of the VCZ theorem,
under the assumptions that the VCZ theorem can indeed be applied.
In this regard, in \cite{YAB1} Section V, one may find a statement
according to which the rms electron beam size "is only the average
value along the undulator" because "the beta function has a large
variation along the undulator". Another example dealing with the
same issue is given in reference \cite{PFEI}. This paper (as well
as reference \cite{YAB1}) reports experimental results. However,
authors of \cite{PFEI} observe a disagreement between the electron
beam rms size reconstructed from the VCZ theorem and beam
diagnostics result of about a factor $2$. They ascribe this
variation to the variation of the electron beam size along the
undulator. In footnote 25 of reference \cite{PFEI}, one may read:
"The precise shape and width of the x-ray intensity distribution
in the source plane are directly connected to the properties of
the electron beam. It would not be surprising if the limited depth
of focus of the parabolically shaped electron beta function in the
undulator translates into a virtually enlarged x-ray source
size.". At first glance it looks like if the SR source had a
finite longitudinal dimension, and the virtual source size
depended on variations of the beta function along the undulator.
However, as we have seen before, the concept of virtual source
involves a single transverse plane, and in the most general case
any variation of the beta function does not affect the virtual
source size. In our case of quasi-homogeneous Gaussian source, the
virtual source is located where the beta function is minimal, and
its size coincides with the transverse size of the electron beam
at that location.

Finally, it should be remarked that the VCZ theorem can only be
applied to quasi-homogeneous sources. Consider for example
\cite{YAB1,YAB2}, where a characterization of the vertical
emittance in Spring-8 is reported, based on the measurement of the
X-Ray beam coherence length in the far zone. The experiment was
performed at the beamline BL29XU. Based on the assumption of
validity of the VCZ theorem, it was found that the rms electron
beam size at the undulator center (corresponding to the minimal
value of the beta function) was $s_y \simeq 4.5 ~\mu$m, and that
the coupling factor between horizontal and vertical emittance was
down to the value $\zeta \simeq 0.12 \%$, which corresponds to an
extremely small vertical emittance $\epsilon_y = 3.6$
pm$\cdot$rad. A resolution limit of this method was also
discussed, based on numerical calculations of the radiation size
from a single electron performed at the exit of the undulator,
$s_p \simeq 1.6 ~\mu$m at $E_p = 14.41 $ keV for the $4.5$ m long
undulator used in the experiment. The resolution limit of the
measurement of $s_y$ was estimated to be about $1~ \mu$m.
Considering propagation laws for Gaussian beams it seems
reasonable that the radiation spot size at the virtual source,
located in the center of the undulator, be smaller than $s_p$.
However, undulator radiation from a single electron cannot be
considered a Gaussian beam. In particular, under the resonance
approximation the field at the exit of the undulator exhibits a
singularity and is not suitable for evaluation of the radiation
spot size.

Use of $s_p$ led to an underestimation of the virtual source size
and, thus, an overestimation of the resolution. Let us show this
fact. Based on Eq. (\ref{undurad5gg}), we can determine the
correct virtual source size of undulator radiation from a single
electron. Let us consider the case when the single electron is
emitting photons at the fundamental harmonic with energy $E =
14.41$ keV. The angular frequency of light oscillations is given,
in this case, by $\omega = 2.2 \cdot 10^{19}$ Hz. For an undulator
length $L_w = 4.5$ m, the normalization factor for the transverse
size, $(L_w c/\omega)^{1/2}$, is about $8~\mu$m. From Fig.
\ref{psio} obtain the dimensionless Half Width Half Maximum (HWHM)
radiation size from a single electron (i.e. the HWHM width of the
intensity distribution at the virtual source, located at the
center of the undulator). This HWHM dimensionless value is about
$0.7$. It follows that the HWHM value of the radiation spot size
from a single electron is about $0.7 \cdot (c L_w/\omega)^{1/2}
\simeq 6~ \mu$m. Therefore, the resolution in \cite{YAB1,YAB2}, is
estimated to be better than the correct value.

Note that the HWHM radiation spot size from a \textit{single
electron} is larger than the \textit{rms electron beam size} $s_y
\simeq 4.5 ~ \mu$m found by means of coherence measurements. One
concludes that the uncertainty due to finite resolution is larger
than the measured electron beam size. This suggests that the
analysis of experimental results \cite{YAB1,YAB2} might include a
logical flaw. Authors of that reference assume the validity of the
VCZ theorem in the vertical direction. If one assumes their result
of a vertical emittance ${\epsilon}_y \simeq 0.3 \lambda/(2 \pi)$,
it follows \textit{a posteriori} that the VCZ theorem could not
have been applied in first instance (in this experiment the value
of the beta function was ${\beta} \simeq L_w$). We suggest that
analysis of experimental results in \cite{YAB1,YAB2} should be
based, instead, on the study of transverse coherence for
non-homogeneous undulator sources in free space made in the
previous Section \ref{sec:main}.

\subsection{\label{sub:nong} Non-Gaussian undulator sources}

Let us now turn to the analysis of non-Gaussian quasi-homogeneous
sources on the basis of Eq. (\ref{Gnor3y}) and Eq. (\ref{resu1}).
We still assume that $N_x \gg 1$ and $D_x \gg 1$. Moreover, as
before, we assume that the minimal beta function is located at the
undulator center.

\subsubsection{Case of a large electron-beam size $N_y \gg 1$ and $D_y
\lesssim 1$.}

Eq. (\ref{Gnor3y}) simplifies to

\begin{eqnarray}
\hat{G}_y\left(0,{\bar{y}},\Delta {y}\right) &=&
\sqrt{\frac{\pi}{N_y}}  \exp \left[-\frac{(\Delta
y)^2D_y}{2}\right] \exp\left[-\frac{\bar{y}^2}{2 N_y}\right]
\gamma_S(\Delta y)~,\cr &&\label{nong10}
\end{eqnarray}
where we defined the universal function $\gamma_S(\alpha)$ as

\begin{eqnarray}
\gamma_S(\alpha)&=&\frac{1}{2}\int_{-\infty}^{\infty} d \phi_y
\int_{-\infty}^{\infty} d \phi_x
\Psi_0\left\{\left[{\phi_{x}}^2+\left({\phi_{y}}+\frac{\alpha}{2}
\right)^2 \right]^{1/2}\right\}
\Psi_0\left\{\left[{\phi_{x}}^2+\left({\phi_{y}}-\frac{\alpha}{2}
\right)^2 \right]^{1/2}\right\}\cr && = \frac{1}{2 \mathcal{K}_S}
\int_{-\infty}^{\infty} d\phi_y \mathcal{S}(\phi_y,\alpha) ~.
\label{gammadef}
\end{eqnarray}

\begin{figure}
\begin{center}
\includegraphics*[width=140mm]{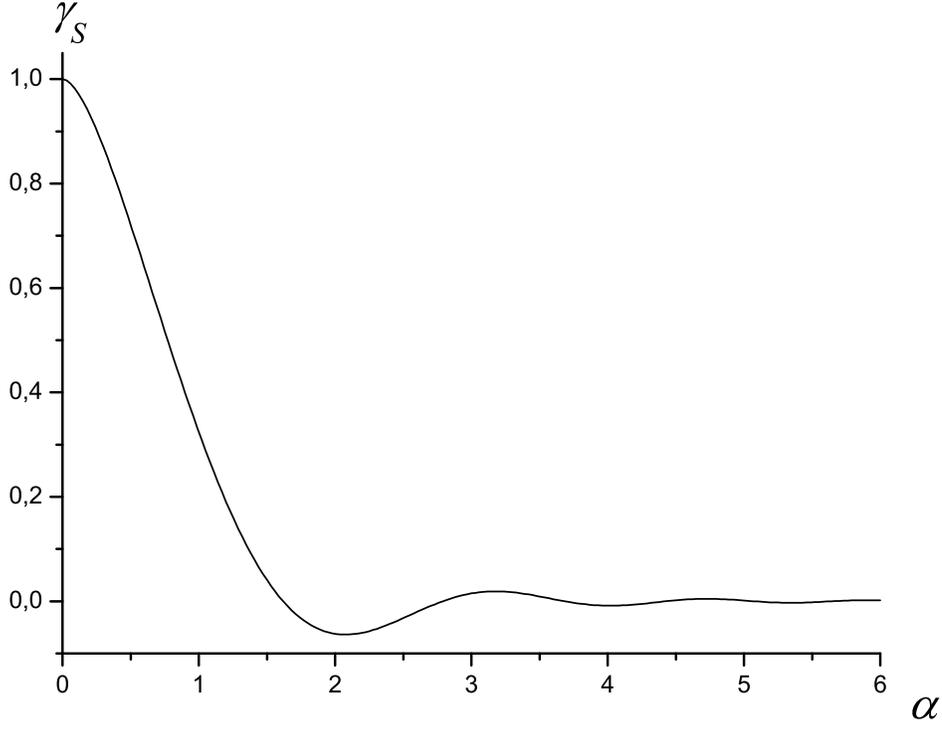}% Here is how to import EPS art
\caption{\label{gammas} Plot of the universal function
$\gamma_S(\alpha)$, used to calculate the cross-spectral density
of a quasi-homogeneous source when ${N}_x \gg1$, ${D}_x \gg 1$,
${N}_y \gg 1$ and ${D}_y \lesssim 1$.}
\end{center}
\end{figure}

A plot of $\gamma_S(\alpha)$ is given in Fig. \ref{gammas}. Main
features of $\gamma_S$ are a strong non-Gaussian shape, and the
fact that also negative values are assumed. Eq. (\ref{introhdopo})
is satisfied. Moreover,

\begin{eqnarray}
\hat{I}_y\left(0,{\bar{y}}\right) &=& \sqrt{\frac{\pi}{ N_y}}
\exp\left[-\frac{\bar{y}^2}{2 N_y}\right] ~,\cr &&\label{nong10In}
\end{eqnarray}
and

\begin{eqnarray}
\hat{g}_y\left(0,\Delta {y}\right) &=& \exp \left[-\frac{(\Delta
y)^2D_y}{2}\right] \gamma_S(\Delta y)~,\cr &&\label{nong10de}
\end{eqnarray}
In the far-zone, Eq. (\ref{resu1}) simplifies to

\begin{eqnarray}
\hat{G}_y(\hat{z},\bar{\theta}_y,\Delta {\theta}_y) &=&
\frac{1}{2\pi \hat{z} \sqrt{\pi D_y} \mathcal{K}_F} {\exp{\left[i
\hat{z} \bar{\theta}_y \Delta {\theta}_y \right]}} \exp{\left[-
\frac{N_y \Delta {\theta}_y^2}{2} \right]}\cr &&\times
\int_{-\infty}^{\infty} d \hat{\phi}_y
\exp{\left[-\frac{(\bar{\theta}_y+\hat{\phi}_y)^2}{2 D_y}\right]}
\mathcal{I}_F(\phi_y) ~,\label{nong1f}
\end{eqnarray}
where $\mathcal{I}_F$ has already been defined in Eq.
(\ref{ISnorm}).  Moreover,

\begin{eqnarray}
\hat{I}_y(\hat{z},\bar{\theta}_y) &=& \frac{1}{2\pi \hat{z}
\sqrt{\pi D_y} \mathcal{K}_F} \int_{-\infty}^{\infty} d
\hat{\phi}_y \exp{\left[-\frac{(\bar{\theta}_y+\hat{\phi}_y)^2}{2
D_y}\right]} \mathcal{I}_F(\phi_y) ~,\label{nong1fIn}
\end{eqnarray}
and

\begin{eqnarray}
\hat{g}_y(\hat{z},\Delta {\theta}_y) &=& {\exp{\left[i \hat{z}
\bar{\theta}_y \Delta {\theta}_y \right]}} \exp{\left[- \frac{N_y
\Delta {\theta}_y^2}{2} \right]} ~\label{nong1fde}
\end{eqnarray}
Note that, due to geometrical phase factor in Eq. (\ref{nong1f}),
only the virtual source at $\hat{z}=0$ constitutes a
quasi-homogeneous source.

Also, $\gamma_S$ and $\mathcal{I}_F$ basically form a Fourier
pair:

\begin{eqnarray}
\gamma_S(\Delta {y}) = \frac{1}{{2}\pi^2 \mathcal{K}_F}
\int_{-\infty}^{\infty} d {\phi}_y \exp{\left[i \Delta {y}
{\phi}_y\right]} \mathcal{I}_F({\phi}_y)~,\label{G2D3lastlastno2}
\end{eqnarray}
meaning that the inverse VCZ theorem is satisfied for $N_{x,y} \ll
\hat{z}^2$.

Finally, it is possible to calculate $\gamma_S$ analytically. To
this purpose, it is sufficient to note that the Fourier transform

\begin{eqnarray}
{\widetilde{\gamma}}_S(\xi,\eta) = \int_{-\infty}^{\infty} d
\hat{\phi}_x \int_{-\infty}^{\infty} d \hat{\phi}_y
 \exp{\left[i(\xi \hat{\phi}_x+\eta \hat{\phi}_y)\right]}
\mathrm{sinc}^2\left(\frac{\hat{\phi}_x^2+\hat{\phi}_y^2}{4}\right)
~\label{G2D3lastlastno3}
\end{eqnarray}
can be evaluated with the help of the Bessel-Fourier formula as

\begin{eqnarray}
{\widetilde{\gamma}}_S(\lambda) &=& 2\pi \int_{0}^{\infty} d
{\phi} ~{\phi} J_0\left({\phi} \lambda\right)
\mathrm{sinc}^2\left(\frac{{\phi}^2}{4}\right)  \cr &=& 2\pi
\left[\pi+ \lambda^2 \mathrm{Ci}\left(\frac{\lambda^2}{2}\right)-
2 \sin\left(\frac{\lambda^2}{2}\right)- 2
\mathrm{Si}\left(\frac{\lambda^2}{2}\right)
\right]~,\label{G2D3lastlastno4}\end{eqnarray}
where $\lambda^2 = \xi^2+\eta^2$, $\phi^2 = \phi_x^2+\phi_y^2$,
$\mathrm{Si}(\cdot)$ is the sine integral function and
$\mathrm{Ci}(\cdot)$ is the cosine integral function. Thus,
letting $\xi=0$ and $\eta = \Delta y$ one has

\begin{eqnarray}
\gamma_S(\Delta {y}) =\frac{1}{\pi} \left[\pi+ (\Delta y)^2
\mathrm{Ci}\left(\frac{(\Delta y)^2}{2}\right)- 2
\sin\left(\frac{(\Delta y)^2}{2}\right)- 2
\mathrm{Si}\left(\frac{(\Delta y)^2}{2}\right)
\right]~.\label{G2D3lastlastno5}
\end{eqnarray}

\subsubsection{Case of a large electron-beam divergence $D_y \gg 1$ and $N_y \lesssim 1$.}

A similar analysis can be given in the case of a large
electron-beam divergence. In this case Eq. (\ref{Gnor3y})
simplifies to give

\begin{eqnarray}
\hat{G}_y\left(0,{\bar{y}},\Delta {y}\right) &=&
\frac{1}{2\mathcal{K}_S} \sqrt{\frac{\pi}{N_y}}  \exp
\left[-\frac{(\Delta y)^2D_y}{2}\right] \int_{-\infty}^{\infty} d
\phi_y \exp\left[-\frac{\left(\phi_y+\bar{y}\right)^2}{2
N_y}\right] \mathcal{I}_S(\phi_y)~,\cr &&\label{nong20}
\end{eqnarray}
where $\mathcal{I}_S$ has already been defined in Eq. (\ref{B2}).
Eq. (\ref{introhdopo}) is satisfied. Moreover,

\begin{eqnarray}
\hat{I}_y\left(0,{\bar{y}}\right) &=& \frac{1}{2 \mathcal{K}_S
}\sqrt{\frac{\pi}{N_y}} \int_{-\infty}^{\infty} d \phi_y
\exp\left[-\frac{\left(\phi_y+\bar{y}\right)^2}{2 N_y}\right]
\mathcal{I}_S(\phi_y)~,\cr &&\label{nong20In}
\end{eqnarray}

\begin{eqnarray}
\hat{g}_y\left(0,\Delta {y}\right) &=& \exp \left[-\frac{(\Delta
y)^2D_y}{2}\right]~.\cr &&\label{nong20de}
\end{eqnarray}
In the far zone, Eq. (\ref{resu1}) simplifies to

\begin{eqnarray}
\hat{G}(\hat{z},\bar{\theta}_y,\Delta {\theta}_y) &=&
\frac{1}{\hat{z}} \sqrt{\frac{\pi}{D_y}} {\exp{\left[i \hat{z}
\bar{\theta}_y \Delta {\theta}_y \right]}} \exp{\left[- \frac{N_y
\Delta {\theta}_y^2}{2} \right]}
\exp{\left[-\frac{\bar{\theta}_y^2}{2 D_y}\right]}\gamma_F(\Delta
\theta_y) ~,\cr &&\label{nong2f}
\end{eqnarray}
where

\begin{eqnarray}
\gamma_F(\alpha) &=& \frac{1}{2\pi^2} \int_{-\infty}^{\infty} d
{\phi}_y \int_{-\infty}^{\infty} d {\phi}_x \cr && \times
\Psi_f\left\{\left[{{\phi}_x^2
+\left(\phi_y-\frac{\alpha}{2}\right)^2}\right]^{1/2}\right\}
\Psi_f\left\{\left[{{\phi}_x^2
+\left(\phi_y+\frac{\alpha}{2}\right)^2}\right]^{1/2}\right\} \cr
&& = \frac{1}{2\pi^2 \mathcal{K}_F}\int_{-\infty}^{\infty} d
\phi_y \mathcal{F}(\phi_y,\alpha) ~.\label{betadef}
\end{eqnarray}
Moreover

\begin{eqnarray}
\hat{I}_y(\hat{z},\bar{\theta}_y) &=& \frac{1}{\hat{z}}
\sqrt{\frac{\pi}{D_y}} \exp{\left[-\frac{\bar{\theta}_y^2}{2
D_y}\right]} ~ ,\cr &&\label{nong2fIn}
\end{eqnarray}
and

\begin{eqnarray}
\hat{g}(\hat{z},\Delta {\theta}_y) &=& {\exp{\left[i \hat{z}
\bar{\theta}_y \Delta {\theta}_y \right]}} \exp{\left[- \frac{N_y
\Delta {\theta}_y^2}{2} \right]} \gamma_F(\Delta \theta_y) ,\cr
&&\label{nong2fde}
\end{eqnarray}

\begin{figure}
\begin{center}
\includegraphics*[width=140mm]{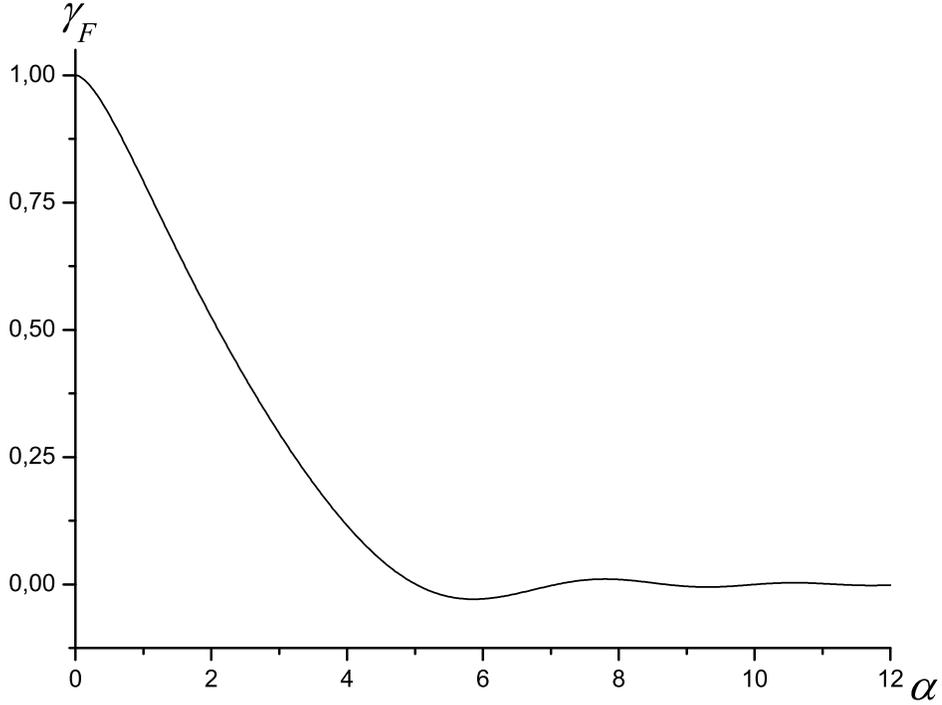}% Here is how to import EPS art
\caption{\label{gammaf} Plot of the universal function $\gamma_F$,
used to calculate the cross-spectral density in the far zone when
$\hat{N}_x \gg1$, $\hat{D}_x \gg 1$, $\hat{N}_y \lesssim 1$ and
$\hat{D}_y \gg 1$.}
\end{center}
\end{figure}

A plot of $\gamma_F(\alpha)$ is given in Fig. \ref{gammaf}. As for
$\gamma_S$, main features of $\gamma_F$ are a strong non-Gaussian
shape, and the fact that also negative values are assumed. Due to
the phase factors in Eq. (\ref{nong2f}), only the virtual source
at $\hat{z}=0$ constitutes a quasi-homogeneous source.

Also, $\mathcal{I}_S$ and $\gamma_F$ form a Fourier pair:

\begin{eqnarray}
{\mathcal{I}_S}(\bar{y}) = \frac{\mathcal{K}_S}{\pi}
\int_{-\infty}^{\infty} d {\phi}_y \exp\left[-i \bar{y}
{\phi}_y\right] \gamma_F({\phi}_y) ~,\label{BB}
\end{eqnarray}
meaning that the VCZ theorem is satisfied for $\hat{z}^2 D_{x,y}
\gg 1$. Since $D_{x,y} \gg 1$ this means that, in this case, the
far zone begins at the very exit of the undulator, at $\hat{z}
\sim 1$.

Finally, starting with the representation of $\mathcal{I}_S$ in
Eq. (\ref{B2}), we can write (see \cite{GOOD} Appendix A.3.)

\begin{eqnarray}
{\mathcal{I}_S}(\bar{y}) = \mathcal{K}_S \int_{-\infty}^{\infty} d
{\phi}_y \exp\left[-i \bar{y} {\phi}_y\right] \int_{0}^{\infty} d
\alpha ~ \alpha J_0(\alpha \phi_y) \Psi_0^2(\alpha) ~.\label{BBQ1}
\end{eqnarray}
Comparison with Eq. (\ref{BB}) yields the following alternative
representation of $\gamma_F$ in terms of a one-dimensional
integral involving special functions:

\begin{eqnarray}
\gamma_F(\phi_y) = \pi \int_{0}^{\infty} d \alpha ~ \alpha
J_0(\alpha \phi_y) \Psi_0^2(\alpha) ~.\label{BBQ2}
\end{eqnarray}

\subsection{\label{sub:accu} Accuracy of the quasi-homogeneous approximation}

As we have discussed before, the quasi-homogeneous approximation
can be applied when either or both $N_y \gg 1$ or $D_y \gg 1$. In
this Section we will see that the accuracy of the
quasi-homogeneous approximation scales as $(\sqrt{\max[N_y,1]
\max[D_y,1]}~)^{-1}$. As we have previously seen, in the
particular case when both $N_y \gg 1$ and $D_y \gg 1$, a Gaussian
model may be used. We will see that the accuracy of such model is
worse than that of the quasi-homogeneous model, and scales as
$\max(1/\sqrt{D_y},1/\sqrt{N_y})$.

We begin demonstrating that, when the source is quasi-homogeneous,
we may take the approximation $\hat{\mathcal{S}}(\Delta {y},
\hat{\phi}) \simeq \gamma_S(\Delta {y}) \mathcal{I}_S(\hat{\phi})$
in Eq. (\ref{Gnor4y}) with an accuracy scaling as
$(\sqrt{\max[N_y,1] \max[D_y,1]}~)^{-1}$. First, let us introduce
a normalized version of the one-dimensional inverse Fourier
transform of the function $\mathcal{F}$, that is

\begin{eqnarray}
\bar{\mathcal{F}}(u,v) &=& \frac{1}{2\pi^2
\mathcal{K}_F}\int_{-\infty}^{\infty}
\mathcal{F}\left(\alpha,v\right) \exp\left[2 i u \alpha \right] d
\alpha = \frac{1}{2\mathcal{K}_S}\int_{-\infty}^{\infty}
{\mathcal{S}}\left(u,\delta\right) \exp\left[-i v\frac{\delta}{2}
\right] d \delta,\cr &&\label{WIG2}
\end{eqnarray}
where the normalization factor is chosen in such as way that
$\bar{\mathcal{F}}(0,0)=1$. The cross-spectral density  in Eq.
(\ref{Gnor4y}) can therefore be written as

\begin{eqnarray}
\hat{G}_y(0,\bar{y},\Delta {y}) &=&
\frac{1}{4\sqrt{2}\pi^2}\exp\left[- \frac{{D}_y (\Delta {y}^2)}{2}
\right] \cr && \times \int_{-\infty}^{\infty} d {u} \exp\left[-i
\frac{{u}\bar{y}}{2}\right] \exp\left[-\frac{{N}_y
{u}^2}{2}\right] \bar{\mathcal{F}} (\Delta {y}, {u})~,\cr
&&\label{gen2sepa}
\end{eqnarray}
having used the convolution theorem. Under the quasi-homogeneous
assumption, we can approximate $\bar{\mathcal{F}} (\Delta {y},
{u}) \simeq \bar{\mathcal{F}} (\Delta {y}, 0)\bar{\mathcal{F}} (0,
{u})$. To show this, let us represent $\bar{\mathcal{F}}(u,v)$
using a Taylor expansion around the point $(0,0)$. One obtains

\begin{eqnarray}
\bar{\mathcal{F}}(u,v) &=& 1 + \sum_{k=1}^{\infty} \frac{1}{k!}
\left[u^k \frac{\partial^k \bar{\mathcal{F}}(u,0)}{\partial
u^k}\Bigg|_{u=0}+v^k \frac{\partial^k
\bar{\mathcal{F}}(0,v)}{\partial v^k}\Bigg|_{v=0}\right] \cr && +
O(uv) ~, \label{Mpexp}
\end{eqnarray}
where the normalization relation $\bar{\mathcal{F}}(0,0)=1$ has
been taken advantage of. Similarly, one may consider the following
representation of the product $\bar{\mathcal{F}}(u,0)$
$\bar{\mathcal{F}}(0,v)$ also obtained by means of a Taylor
expansion:

\begin{eqnarray}
\bar{\mathcal{F}}(u,0)\bar{\mathcal{F}}(0,v) &=&
\left[\bar{\mathcal{F}}(0,0)+ \sum_{k=1}^{\infty} \frac{u^k}{k!}
\frac{d^k \bar{\mathcal{F}}(u,0)}{du^k}\Bigg|_{u=0}\right]\cr
&\times& \left[\bar{\mathcal{F}}(0,0)+ \sum_{j=1}^{\infty}
\frac{v^j}{j!} \frac{d^j
\bar{\mathcal{F}}(0,v)}{dv^j}\Bigg|_{v=0}\right] \cr &=& 1 +
\sum_{n=1}^{\infty} \frac{1}{n!} \left[u^n \frac{d^n
\bar{\mathcal{F}}(u,0)}{du^n}\Bigg|_{u=0} \right.\cr && \left.
+v^n \frac{d^n \bar{\mathcal{F}}(0,v)}{dv^n}\Bigg|_{v=0}\right] +
O(uv) ~,\label{Mpexp2}
\end{eqnarray}
having used $\bar{\mathcal{F}}(0,0)=1$. Comparison of the last
equality in (\ref{Mpexp2}) with the right hand side of Eq.
(\ref{Mpexp}) shows that $\bar{\mathcal{F}}(u,v)\simeq
\bar{\mathcal{F}}(u,0)\bar{\mathcal{F}}(0,v)$ up to corrections of
order $uv \sim 1/\sqrt{\max[{N}_y,1]\max[{D}_y,1]}$, that is the
quasi-homogeneous accuracy. Using this approximation in Eq.
(\ref{gen2sepa}) yields

\begin{eqnarray}
\hat{G}(0,\bar{y},\Delta {y}) &=& \frac{1}{4\sqrt{2}\pi^2}
\exp\left[-\frac{{D}_y \Delta {y}^2}{2} \right] \bar{\mathcal{F}}
(\Delta {y}, 0) \cr &&\times \int_{-\infty}^{\infty} d {u}
\exp\left[-i \frac{{u}\bar{y}}{2}\right] \exp\left[-\frac{ {N}_y
{u}^2}{2}\right] \bar{\mathcal{F}} (0,{u})~.\label{gen2sepabis}
\end{eqnarray}
Finally, recalling the definitions of $\gamma_S$ and
$\mathcal{I}_S$ we can write Eq. (\ref{gen2sepabis}) as

\begin{eqnarray}
\hat{G}(0,\bar{y},\Delta {y}) &=& \sqrt{\frac{\pi}{N_y
}}\frac{1}{2\mathcal{K}_S }\exp\left[-\frac{{D}_y \Delta {y}^2}{2}
\right] \gamma_S(\Delta {y}) \cr && \times \int_{-\infty}^{\infty}
d {\phi} \exp\left[-\frac{(\bar{y}+{\phi})^2}
{2{N}_y}\right]\mathcal{I}_S({\phi})~.\label{gen2qhapp}
\end{eqnarray}
Eq. (\ref{gen2qhapp}) is valid in any quasi-homogeneous case.

Note that Eq. (\ref{gen2qhapp}) accounts for diffraction effects
through the universal functions $\gamma_S$ and $\mathcal{I}_S$.
This may be traced back to the use of the inhomogeneous wave
equation to calculate the cross-spectral density for the virtual
source, from which Eq. (\ref{gen2qhapp}) follows. Deriving Eq.
(\ref{gen2qhapp}), we assume a large number of modes, and this
justifies the use of phase space representation as an alternative
characterization of the source, in place of the cross-spectral
density (i.e. Eq. (\ref{gen2qhapp}) itself).

Setting $\Delta {y}=0$, Eq. (\ref{gen2qhapp}) gives the exact
intensity distribution at the virtual source, i.e. Eq.
(\ref{Inty0}). The spectral degree of coherence on the virtual
source is then recovered using the definition of quasi-homogeneous
source $\hat{G} = \hat{I}(\bar{y}) g(\Delta {y})$. Since the
source is quasi-homogeneous, the Fourier transform of the spectral
degree of coherence $g(\Delta {y})$ yields the intensity in the
far zone. Remembering that $\gamma_S$ and $\mathcal{I}_F$ form a
Fourier pair, we conclude that, starting from Eq.
(\ref{gen2qhapp}) it is possible to reproduce the exact result for
the intensity in the far zone, Eq. (\ref{Intyf}). Quite
remarkably, Eq. (\ref{gen2qhapp}), which is derived under the
quasi-homogeneous approximation and is related to an accuracy
$1/\sqrt{\max[{N}_y,1]\max[{D}_y,1]}$, yields back two results,
Eq. (\ref{Inty0}) and Eq. (\ref{Intyf}) which are valid regardless
the fact that the source is quasi-homogeneous or not.

Let us now consider the quasi-homogeneous case when both
$\hat{N}_y \gg 1$ and $\hat{D}_y \gg 1$. Worsening the accuracy in
the calculation of the cross-spectral density of the source, we
may reduce Eq. (\ref{gen2qhapp}) to

\begin{eqnarray}
\hat{G}_y\left(0,{\bar{y}},\Delta {y}\right) &=&
\sqrt{\frac{\pi}{{N_y}}} \exp \left[-\frac{(\Delta y)^2D_y
}{2}\right] \exp\left[-\frac{\bar{y}^2}{2 N_y}\right]
~,\label{gen2qhappworse}
\end{eqnarray}
that is Eq. (\ref{Gnorgauss0}). Note that neglecting the product
with the $\gamma_S$ function can be done with an accuracy
$1/\sqrt{{D}_y}$, while extraction of the exponential function in
$\bar{y}$ from the convolution product with the $\mathcal{I}_S$
function can be done with an accuracy $1/\sqrt{{N}_y}$. In our
study case when ${D}_y \gg 1$ and ${N}_y \gg 1$, the overall
accuracy of Eq. (\ref{gen2qhappworse})  can be estimated as
$\max(1/\sqrt{{D}_y},1/\sqrt{{N}_y})$, that is the accuracy of the
Gaussian approximation. Such accuracy is much worse than that of
the quasi-homogeneous assumption in Eq. (\ref{gen2qhapp}), that is
$1/\sqrt{{N}_y {D}_y}$.

When ${N}_y \gg 1$ and ${D}_y \simeq 1$ the accuracy of the
quasi-homogeneous approximation becomes $1/\sqrt{{N}_y
\max[1,{D}_y]}$. When ${N}_y \simeq 1$ and ${D}_y \gg 1$ it
becomes, instead, $1/\sqrt{\max[1,{N}_y]\hat{D}_y }$. In these
cases, the accuracy of the quasi-homogeneous approximation is
comparable to the accuracy of the Gaussian approximation. To be
specific, when ${N}_y \gg 1$ and ${D}_y \simeq 1$ the accuracy of
the quasi-homogeneous approximation is of order $1/\sqrt{{N}_y}$
and Eq. (\ref{gen2qhapp}) can be substituted with

\begin{eqnarray}
\hat{G}(0,\bar{y},\Delta {y}) &=& \sqrt{\frac{\pi}{N_y
}}\exp\left[-\frac{\bar{y}^2}{2{N}_y}\right]\exp\left[-\frac{{D}_y
\Delta {y}^2}{2} \right] \gamma_S(\Delta {y})~.\label{gen2qhappN}
\end{eqnarray}
without loss of accuracy, because the relative accuracy of the
convolution is of order $1/\sqrt{{N}_y}$ as the accuracy of the
quasi-homogenous approximation. Eq. (\ref{gen2qhappN}) is just Eq.
(\ref{nong10}). A similar reasoning can be done when ${D}_y \gg 1$
and ${N}_y \simeq 1$. In this case the accuracy of the
quasi-homogeneous approximation is of order $1/\sqrt{{D}_y}$, and
Eq. (\ref{gen2qhapp}) can be substituted with

\begin{eqnarray}
\hat{G}(0,\bar{y},\Delta {y}) &=& \sqrt{\frac{\pi}{N_y
}}\frac{1}{2\mathcal{K}_S }\exp\left[-\frac{{D}_y \Delta {y}^2}{2}
\right]\int_{-\infty}^{\infty} d {\phi}
\exp\left[-\frac{(\bar{y}+{\phi})^2}
{2{N}_y}\right]\mathcal{I}_S({\phi})~.\label{gen2qhappD}
\end{eqnarray}
without loss of accuracy. Eq. (\ref{gen2qhappD}) is just Eq.
(\ref{nong20}).

\subsection{\label{sub:geop} Quasi-homogeneous sources in terms of phase space}

The cross-spectral density at the virtual source can be written as
in Eq. (\ref{introhdopo}) that we rewrite here for convenience in
terms of coordinates $\bar{r}_{x,y}$ and $\Delta {{r}_{x,y}}$:

\begin{eqnarray}
\hat{G}_o({\bar{r}_x},\bar{r}_y,\Delta {{r}_x},\Delta {{r}_y}) =
\hat{I}\left(\bar{r}_x,\bar{r}_y\right) {g}(\Delta {{r}_x},\Delta
{{r}_y}) ~.\label{introhdopob}
\end{eqnarray}
For notational simplicity we use notation $\hat{G}_o$ to indicate
$\hat{G}$ at $\hat{z}=0$. The Fourier transform of Eq.
(\ref{introhdopob}) with respect to all variables will be
indicated with

\begin{eqnarray}
\hat{\mathcal{G}}_o( {\bar{\theta}_x}, {\bar{\theta}_y},{\Delta
{\theta}_x},\Delta {\theta}_y) &=& \int_{-\infty}^{\infty} d
\Delta { {r}_x'}\int_{-\infty}^{\infty}d \Delta { {r}_y'}
\int_{-\infty}^{\infty} d { \bar{r}_x'}\int_{-\infty}^{\infty}d
 { \bar{r}_y'} ~ \hat{G}_o({ \bar{r}_x'},{ \bar{r}_y'},
\Delta{{r}'_x},\Delta{{r}'_y}) \cr &&\times\exp [2i (
\bar{\theta}_x \Delta{{r}'_x}+\bar{\theta}_y \Delta{{r}'_y} )]\exp
[i ( \Delta {\theta}_x { \bar{r}'_x}+\Delta {\theta}_y {
\bar{r}'_y} )]~.\cr && \label{ftgdeffor}
\end{eqnarray}
The two quantities
$\hat{I}(\bar{r}_x,\bar{r}_y)=\hat{G}_o({\bar{r}_x},\bar{r}_y,0,0)$
and $\hat{\Gamma}({\bar{\theta}_x}, {\bar{\theta}_y})
=\hat{\mathcal{G}}_o( {\bar{\theta}_x}, {\bar{\theta}_y},0,0)$ are
always positive because, by definition of $\hat{G}_o$, they are
ensemble averages of quantities under square modulus.

Let us now introduce the Fourier transform of Eq.
(\ref{introhdopo}) with respect to $\Delta {r_{x,y}}$:

\begin{eqnarray}
\hat{\Phi}_o({\bar{r}_x},\bar{r}_y, {\bar{\theta}_x},
{\bar{\theta}_y}) &=& \int_{-\infty}^{\infty} d \Delta {
{r}_x'}\int_{-\infty}^{\infty}d \Delta {r_y'}~ \hat{G}_o({
\bar{r}_x},{ \bar{r}_y}, \Delta{{r}'_x},\Delta{{r}'_y}) \cr
&&\times\exp [i ( \bar{\theta}_x \Delta{ {r}'_x}+\bar{\theta}_y
\Delta{{r}'_y} )]~. \label{wigdef}
\end{eqnarray}
Accounting for Eq. (\ref{introhdopob}), i.e. in the particular
case of a (virtual) quasi-homogeneous source, Eq. (\ref{wigdef})
can be written as

\begin{eqnarray}
\hat{\Phi}_o({\bar{r}_x},\bar{r}_y, {\bar{\theta}_x},
{\bar{\theta}_y}) &=& \hat{I}\left(\bar{r}_x,\bar{r}_y\right)
\hat{\Gamma}({\bar{\theta}_x}, {\bar{\theta}_y})~,
\label{wigdefdef}
\end{eqnarray}
having recognized that $\hat{\Gamma}({\bar{\theta}_x},
{\bar{\theta}_y}) =\hat{\mathcal{G}}_o( {\bar{\theta}_x},
{\bar{\theta}_y},0,0)$ is the Fourier transform of the spectral
degree of coherence ${g}$. The distribution $\hat{\Phi}_o$, being
the product of two positive quantities, never assumes negative
values. Therefore it may always be interpreted as a phase space
distribution\footnote{Physically, in the quasi-homogeneous case,
$\hat{\Gamma}$ can be identified with the radiant intensity of the
virtual source. This follows from a statement similar to the van
Cittert-Zernike theorem for quasi-homogeneous sources (see
\cite{MAND}). Note that the intensity and the Fourier transform of
the spectral degree of coherence are obtained back from the phase
space distribution, Eq. (\ref{wigdefdef}), by integration over
coordinates $\bar{\theta}_{x,y}$ and $\bar{r}_{x,y}$
respectively.}. This analysis shows that quasi-homogeneous sources
can always be characterized in terms of Geometrical Optics. It
also shows that, in this particular case, the coordinates in the
phase space, $\bar{r}_{x,y}$ and $\bar{\theta}_{x,y}$, are
separable.

Eq. (\ref{wigdef}) is the definition of a Wigner distribution. In
the case of quasi-homogenous sources, as we have just seen, the
Wigner distribution is never negative and, therefore, can always
be interpreted as a phase space distribution.  In the case of non
quasi-homogeneous sources one may still define a Wigner
distribution using Eq. (\ref{wigdef}). However the Wigner function
itself is not always a positive function. As a consequence it
cannot always be interpreted as a phase space distribution. On the
one hand, quasi-homogeneity is a sufficient condition for the
Geometrical Optics approach to be possibly used in the
representation of the source. On the other hand though, necessary
and sufficient conditions for $\hat{\Phi}_o$ to be a positive
function are more difficult to find.

Note that in the case of quasi-homogeneous non-Gaussian sources
one should account to diffraction effects when calculating source
properties. However, using a Wigner function approach, one can
still use a phase-space representation. In this case diffraction
effects have the effect of complicating the structure of the
phase-space describing the source. Also note that Gaussian-Schell
model cannot be applied in all generality even to
quasi-homogeneous sources, as it does not properly describe the
case of non-Gaussian sources, which is most natural for third
generation facilities in the vertical direction.

As we have seen before, third generation light sources are
characterized, up to the VUV range, by ${N}_x \gg 1$ and ${D}_x
\gg 1$, which implies factorization of the cross-spectral density.
Thus, considering the vertical direction separately, equivalent
condition for quasi-homogeneity requires that either (or both)
${N}_y \gg 1$ and ${D}_y \gg 1$.

An intuitive picture in the real space is given by a (virtual)
quasi-homogeneous source with characteristic (normalized) square
sizes of order $\max[{N}_{y},1]$ and $N_x$, and characteristic
(normalized) correlation length square sizes of order
$\min[1/{D}_{y},1]$ and $1/D_x$, in the vertical and horizontal
directions. Since horizontal and the vertical directions can be
treated separately, we have a large number of independently
radiating sources given by the product

\begin{equation}
M_{y}= \max[{N}_{y},1]\max[{D}_{y},1]~ \label{modesappr}
\end{equation}
in the vertical direction, and

\begin{equation}
M_{x}= {N}_{x}{D}_{x}~ \label{modesappr2}
\end{equation}
in the horizontal direction. The number $M_{x,y}$ is, in other
words, an estimation of the number of coherent modes in the
horizontal and in the vertical direction\footnote{This is in
agreement with an intuitive picture where the photon-beam phase
space reproduces the electron-beam phase space up to the limit
imposed by the intrinsic diffraction of undulator radiation.
Imagine to start from a situation with ${N}_{x,y}\gg 1$ and
${D}_{x,y} \gg 1$ and to "squeeze" the electron-beam phase space
in the vertical direction by diminishing ${N}_{y}$ and ${D}_{y}$.
On the one hand the characteristic sizes of the phase space of the
electron beam are always of order ${N}_{y}$ and ${D}_{y}$. On the
other hand the characteristic sizes of the phase space of the
photon beam are of order $\max[{N}_{y},1]$ and $\max[{D}_{y},1]$:
diffraction effects limit the "squeezing" of the phase space of
the photon beam. }. The number $M_{x,y}^{-1}$ is the accuracy of
Geometrical Optics results compared with Statistical Optics
results or, better, the accuracy of the quasi-homogeneous
assumption. It should be noted that, as $M_{x,y}$ approaches
unity, the accuracy of the quasi-homogeneous assumption becomes
worse and worse and $M_{x,y}$ cannot be taken anymore as a
meaningful estimation of the number of modes: it should be
replaced by a more accurate concept based on Statistical Optics.
To complete the previous statement we should add that $M_{x,y}$
completely loses the meaning of "number of modes" when Geometrical
Optics cannot be applied. For instance when both ${N}_y$ and
${D}_y$ are of order unity (or smaller), one can state that the
Geometrical Optics approach fails in the vertical direction
because the phase space area is getting near to the uncertainty
limit. In this case it is not possible to ascribe the meaning of
"number of modes" to the number $M_y$ simply because the
Geometrical Optics approach in the vertical direction fails.
However, when ${N}_y$ and ${D}_y$ are of order unity (or smaller),
but both ${N}_x \gg 1$ and ${D}_x \gg 1$, the cross-spectral
density admits factorization in the horizontal and in the vertical
direction and the source in the horizontal direction can be still
described, independently, with the help of Geometrical Optics.

We should remark that Statistical Optics is the only mean to deal,
in general, with the stochastic nature of SR. Only in particular
cases SR can be treated in terms of Geometrical Optics, where
planning of experiments can take advantage of ray-tracing code
techniques. One of these cases is constituted by second generation
light sources, because ${N}_{x,y} \gg 1$ and ${D}_{x,y} \gg 1$.

Description in terms of quasi-homogeneous sources is also
important for third generation facilities in the case of bending
magnet beamlines (surprisingly, in both horizontal and vertical
plane), horizontal plane in undulator beamlines, but it cannot be
applied with accuracy to future sources (e.g. ERL-based sources).

Quasi-homogenous sources can be described in terms of geometrical
optics, the outcome being equivalent to description in terms of
statistical optics. In order to decide whether Geometrical Optics
or Wave Optics is applicable, in all generality one should
separately compare the \textit{photon beam} size and divergence
with the radiation diffraction size and diffraction angle, which
are quantities pertaining the single electron radiation. Let us
fix a given direction $x$ or $y$. The square of the diffraction
angle is defined by $(\sigma_d')^2 \sim \lambda/(2 \pi L_f)$,
$L_f$ being the formation length of the radiation at wavelength
$\lambda$. The diffraction size of the source is given by
$\sigma_d \sim \sigma'_d L_f$. In calculating the photon beam size
and divergence one should always include diffraction effects. As a
result, if $\sigma^2$ and $(\sigma')^{2}$ indicate the square of
the electron beam size and divergence, the corresponding square of
the photon beam size and divergence will be respectively of order
$\max[\sigma^2, \sigma_d^2]$ and $\max[(\sigma')^{2},
(\sigma_d')^2]$. These quantities can be rewritten in terms of the
electron beam emittance  as $\max[\epsilon \beta, \sigma_d^2]$ and
$\max[\epsilon/\beta, (\sigma'_d)^2]$, $\beta$ being the beta
function value at the virtual source position for the radiator
(undulator, bending magnet, or other). Dividing these two
quantities respectively by $\sigma_d^2$ and $(\sigma'_d)^2$  give
natural values, normalized to unity, for the photon beam size
$\max[2 \pi \epsilon \beta/(L_f \lambda), 1]$ and divergence
$\max[2 \pi \epsilon L_f/ (\beta \lambda), 1]$. When the product
between these two quantities is much larger than unity one can use
a Geometrical Optics approach. In this case, this product
represents the normalized \textit{photon beam} emittance. When
$\beta \sim L_f$, as in many undulator cases, one may compare, for
rough estimations, the electron beam emittance and the radiation
wavelength as we have done before. However, in the case of a
bending magnet one may typically have $\beta$ of order $10$ m and
$L_f\simeq (\rho^2 \lambdabar)^{1/3}$ ($\rho$ being the bending
radius) of order $10^{-3} \div 10^{-2} $ m. The ratio $\beta/L_f
\gg 1$ now constitutes an extra large parameter of the problem. In
this case, even if the electron beam emittance is two order of
magnitude smaller than the wavelength, due to diffraction effects
one can still apply a Geometrical Optics approach, because $\max[2
\pi \epsilon \beta/(L_f \lambda), 1] \cdot \max[2 \pi \epsilon
L_f/ (\beta \lambda), 1] \gg 1$, i.e. the photon beam emittance is
much larger than the wavelength. As a result, dimensional analysis
suggests that bending magnet radiation may be treated exhaustively
in the framework of Geometrical Optics even for third generation
light sources.

\section{\label{sec:conc} Conclusions}

This work presents a theory of transverse coherence properties
from third generation light sources, valid while radiation evolves
in free-space. Besides being important for experiments involving
coherence that make no use of optical elements, it constitutes the
first step towards the solution of the image formation problem for
undulator sources (see \cite{OURP}) that will be a natural
follow-up to the present article.

We considered Synchrotron Radiation (SR) as a random statistical
process to be described using the language of statistical optics.
Statistical optics developed in connection with Gaussian,
stationary processes characterized by homogeneous sources.
However, for SR, there is no a priori reason to hold these
assumptions satisfied.

We showed that SR is a Gaussian random process. As a result,
statistical properties of SR are described satisfactory by
second-order field correlation functions. We focused, in
particular, on undulator sources. It should be noted here that
wiggler and bending magnets are still being used at third
generation facilities. However, as has been remarked in the
previous Section \ref{sub:geop}, these devices are characterized
by a much shorter formation length, which allows one to apply a
Geometrical Optics approach to describe them. Thus, in this case,
the formalism developed for second generation facilities can be
satisfactory taken advantage of. In other words, use of wigglers
and bending magnets is mainly related with applications requiring
higher photon flux (but not high coherent flux), compared with
analogous devices installed in second generation facilities. Our
choice of considering undulator sources is justified by the fact
that we are interested in transverse coherence properties of
radiation.

With this in mind, a frequency-domain analysis was used to
describe undulator sources from a mathematical viewpoint. As a
consequence of the frequency domain analysis we could study the
spatial correlation for a given frequency content using the
cross-spectral density of the system. This can be used to extract
information even if the process is non-stationary, and
independently of the spectral correlation function.

We gave a general expression for the cross-spectral density
dependent on six dimensionless parameters. Subsequently we assumed
small normalized detuning from resonance, thus obtaining a
simplified expression of practical relevance.

We simplified our expressions further in the case of third
generation light sources, based on a large horizontal emittance
(compared with the wavelength). In this case, the cross-spectral
density can be factored in the product of a horizontal and a
vertical factor.

Attention was subsequently drawn on the vertical cross-spectral
density $\hat{G}_y$, without loss of generality. We expressed
$\hat{G}_y$ in terms of one-dimensional convolutions between
universal functions and Gaussian functions, and studied different
asymptotic cases of interest.

In the case of a vertical emittance much smaller than the
radiation wavelength we derived the counter-intuitive result that
radiation is not fully coherent in the vertical direction. This
effect can be interpreted as an influence of a large horizontal
emittance on the vertical plane, and is related with the
particular non-Gaussian nature of the single particle field.

Subsequently, we studied  quasi-homogeneous cases of interest,
discussing the applicability of the VCZ theorem for Gaussian and
non-Gaussian quasi-homogeneous sources. Finally, we discussed the
accuracy of the quasi-homogeneous model.

It is interesting to spend a few words about relation of our
theory with numerical techniques. Computer codes have been written
(see e.g. \cite{CHUB,BAHR}) that deal with beamline design, based
on wave-optics techniques. Codes also begin to be used to treat
the case of partially coherent radiation: one of their final goals
is to solve the image formation problem starting from first
principles. Results may be obtained using numerical techniques
alone, starting from the Lienard-Wiechert expressions for the
electromagnetic field and applying the definition of the field
correlation function without any analytical work. Calculation of
the intensity at a single point on the image plane can be
approached by propagating wavefronts from different
macro-particles through the entire optical system, calculating
intensities and summing them up. This method is well-suited for
parallel processing, and relatively easy to implement. Yet, a
first-principle calculation of the field correlation function
between two generic points at the image plane involves very
complicated and time-expensive numerical evaluations. To be
specific, one needs to perform two integrations along the
undulator device and four integrations over the electron-beam
phase space distribution to solve the problem in free space. Then,
even in the simple case when the optical beamline can be modelled
as a single focusing lens, other four integrations are needed to
characterize coherence properties on the image plane, for a total
of ten integrations. The development of a universal code for any
experimental setup is then likely to be problematic. A more
conservative approach may suggest the use of computer codes based
on some analytical transformation of first principle equations
suited for specific experimental setups. From this viewpoint our
most general expressions (or alternatively, as it is being done,
expressions for the Wigner distribution function) may be used as
reliable basis for the development of numerical methods. Our
analytical theory allows treatment and physical understanding of
many asymptotes of the parameter space and their applicability
region with the help of a consistent use of dimensional analysis.
This physical understanding, together with the possibility of
using our asymptotic results as benchmarks for numerical methods,
will be of help to code writers.

In closing, as has been remarked in \cite{HOWE}: "[...] it is very
desirable to have a way to model the performance of undulator
beamlines with significant partial coherent effects, and such
modelling would, naturally, start with the source. The calculation
would involve the knowledge of the partial coherence properties of
the source itself and of how to propagate partially coherent
fields through space and through the optical components used in
the beamline. [...] it is important to recognize that, although
most of these calculations are, in principle, straightforward
applications of conventional coherence theory (Born and Wolf,
1980; Goodman, 1985), there is not much current interest in the
visible optics community.". These statements were formulated more
than ten years ago, when operation of third generation light
sources started. While it was immediately recognized that usual SR
theory was not adequate to describe them, no theoretical progress
was made in that direction. Our paper finally answers the call in
\cite{HOWE}.

\newpage

\section*{Acknowledgements}

The authors wish to thank Hermann Franz and Petr Ilinski for many
useful discussions, Massimo Altarelli, Jochen Schneider and Edgar
Weckert for their interest in this work.

\end{document}